\documentclass[11pt]{article}
\usepackage{color,amsfonts,amsmath,amssymb,mathrsfs,verbatim,epsf}

\def\inn{\in}

\def\ccdot{ \!\! \mbox{{\raisebox{0.5ex}{\makebox[1em]{$\centerdot$}}}} \!\!}

\def\TM{\mbox{\it TM}}

 \def\psibbar{
\stackrel{\raisebox{-2pt}[0pt][0pt]{\mbox{{\tiny ({\bf ---})}}}}
 {\raisebox{-0.0pt}[0pt][0pt]{\mbox{$\psi$}}}}
\def\phibbar{
\stackrel{\raisebox{+0pt}[0pt][0pt]{\mbox{{\tiny ({\bf ---})}}}}
  {\raisebox{-0.0pt}[0pt][0pt]{\mbox{$\varphi$}}}}
\def\Aamu{A^{\alpha}_{\ph{o}\mu}} 

\def\rrr{{\mathbb{R}}}

\def\b1{\mbox{\boldmath $1$}}

\def\bh{\mbox{\boldmath $h$}}
\def\bk{\mbox{\boldmath $k$}}

\def\bp{\mbox{\boldmath $p$}}
\def\bPP{\mbox{\boldmath $P$}}
\def\bsl{\boldsymbol}
\def\omp{\omega_{\bsl{p}}}

\def\bv{\mbox{\boldmath $v$}}

\def\hG{\hat{G}}

\def\bx{\mbox{\boldmath $x$}}

\def\lvtfep{L_8(\bv_{248})_{\mathrm{E}_8}=1}

\def\lpvng{L_p(\bv_n)_{\hG}=1}

\def\uo{\mbox{U}(1)}

\def\suth{\mbox{SU}(3)}

\def\ee{\mbox{E}_8}
\def\eeg{\mbox{E}_{8(-24)}}
\def\ese{\mbox{E}_7}
\def\eseg{\mbox{E}_{7(-25)}}
\def\esi{\mbox{E}_6}
\def\esig{\mbox{E}_{6(-26)}}

\def\SML{\mbox{SU}(3)_c \times \mbox{SU}(2)_L \times \mbox{U}(1)_Y}

\def\past{{}^{\,\ast\!}}
\def\ph{\phantom}

\def\lag{{\mathcal L}}

\def\mcM{{\mathcal M}}

\def\mcX{{\mathcal X}}
\def\mcY{{\mathcal Y}}

\def\pal{\partial}
\def\fh{\frac{1}{2}}

\def\ol{\overline}

\def\GN{G_{\! N}}

\def\cstr{c^{\alpha}_{\ph{\alpha}\beta\gamma}}

\def\Lsl{L\!\!\!\!\!\!\;{\mbox{\scriptsize$\boldmath{\diagup}$}}}
\def\Lsl{L\!\!\!\!\!\!\;{\mbox{\scriptsize${\diagup}$}}}
\def\vkin{\bv_{\mathrm{kin}}}

\def\setb{\setlength{\baselineskip}{0.625\baselineskip}}

\begin{document} 

%% take out double spacing (28/11/15)
{\setlength{\baselineskip}{0.625\baselineskip}

\begin{center}
       
     {\LARGE {\bf  \hspace*{-1cm} \mbox{Quantum Gravity from the Composition of Spacetime}
                     \\    \vspace{8pt}     
                                   Constructed through Generalised Proper Time }}

\bigskip
   \vspace{30pt}

% \mbox {{\Large David J. Jackson}}
%\footnote{email: david.jackson.th@gmail.com}  \\

  \mbox {{\Large David J. Jackson}} \\ 
  \vspace{8pt}  
  {david.jackson.th@gmail.com}  \\

 \vspace{30pt}  
  
%% \vspace{10pt}
  %\vspace{-7pt}
 
%\today
 { \large August 28, 2020 }

 \vspace{30pt}
 %\vspace{5pt}

{\bf  Abstract}

%\vspace{-10pt} 
 
\end{center}

%{\setlength{\baselineskip}{0.69\baselineskip}
%{\setlength{\baselineskip}{0.625\baselineskip}

  We describe a theory amalgamating quantum theory and general relativity through the 
identification of a continuous 4-dimensional spacetime arena constructed from the 
substructures of a generalised multi-dimensional form for proper time. In beginning with 
neither a matter field content nor the geometric structure of an extended 4-dimensional 
spacetime the properties of these entities, together with their mutual correspondence, are 
derived through the constraints implied in founding the theory upon the general form of 
proper time alone. The geometric structure of the spacetime identified exhibits a locally 
degenerate description in terms of an inherent ambiguity in the apparent matter field 
content, in a manner consistent with the Einstein field equation of general relativity 
while also incorporating an intrinsic indeterminacy that is characteristic of quantum 
phenomena. Here the quantum properties of all non-gravitational fields arise from this 
composition \textit{of} spacetime, rather than via a postulated set of rules applied 
\textit{in} spacetime to be extended and applied to the gravitational field or spacetime 
itself.
 The contrast with several other approaches to quantum gravity and the manner in which 
this framework can incorporate the essential properties of general relativity and quantum 
theory, both individually and in combination, will be elaborated. 
 While emphasis will be upon providing a unifying conceptual basis the development of the 
associated mathematical framework will also be described.

{

%\vspace{-0.1cm}
%\vspace{-17pt}
% 16/12/18 try to fit t.o.c. in first page...

%{\setlength{\parskip}{-2.0pt}
%\tableofcontents }

%\vspace{5pt}

% 13/05/19 ...or not

\pagebreak

\vspace*{1pt}
{\setlength{\baselineskip}{1.3\baselineskip}
\tableofcontents
\par}

% for footnote space
%\vspace{0.9cm}
%\vspace{15pt}
}

%\par} 

% put this at bottom to take out double line spacing for all
%\par}% \linespread{1.0} for main text (28/11/15)
%match '{\setlength{\baselineskip}{0.625\baselineskip}' above

\pagebreak
%\vspace{30pt}
{\setlength{\baselineskip}{0.983\baselineskip}
\section{Introduction}
\label{qugr1}

     Having been initially expounded in the early 20$^{\mathrm{th}}$ 
century quantum theory and general relativity have each developed 
independently into mathematically sophisticated and empirically 
well-tested theoretical structures. However the elucidation of a 
definitive framework unifying these two principal pillars of 
fundamental physics has remained elusive into the first decades of 
the 21$^{\mathrm{st}}$ century. The term `quantum gravity' refers 
generically to this quest of combining quantum theory and gravity 
within a single coherent and  mathematically consistent theoretical 
scheme.

   Several well-established approaches begin with a basis in  quantum 
theory and attempt to impose a method of `quantisation' upon the 
gravitational field or to the structure of 4-dimensional spacetime 
itself, as might be suggested by the term `quantum gravity'. The 
unifying principle in such frameworks is essentially that `everything 
is quantum', and in aiming to apply quantisation rules to all 
dynamical entities there may be no significant impingement upon the 
foundations of quantum theory itself.  

   On the other hand for the present theory the starting point and 
unifying basis is simply the continuum of proper time, as described 
by one real parameter, and the direct arithmetic generalisation of an 
infinitesimal interval to a multi-dimensional form. In identifying 
the geometric structure of 4-dimensional spacetime from substructures 
of  generalised proper time, consistently with general relativity, 
the phenomena of quantum theory are proposed to derive from an 
intrinsic local degeneracy in the apparent matter field composition 
of this spacetime. This unification scheme hence implies a greater 
concession on the side of quantum theory rather than general 
relativity as they are merged within this framework, in contrast to 
most other theories of quantum gravity. As well as providing a 
unifying basis for quantum gravity we shall argue that the new theory 
can also address long-standing conceptual questions concerning 
quantum theory alone.
 
  In the four subsections of section~\ref{qugr2} we first review the 
characteristic features of quantum theory and general relativity in 
themselves as well as the pertinent issues to be resolved in their 
combination. In section~\ref{qugr3} we review the motivation for 
basing a fundamental physical theory on generalised proper time
 and the immediate implications for the construction and geometry of 
4-dimensional spacetime independently of any quantum characteristics. 
The emergence of quantum and elementary particle phenomena through an 
implicit local degeneracy in the composition of the spacetime 
geometry will then be described in largely qualitative terms in 
section~\ref{qugr4}.  

 Progress towards a full mathematical structure for the description of 
particle  interaction processes
  and quantitative calculations connecting with the application of 
quantum field theory in the high energy physics laboratory will be 
described in more technical detail in section~\ref{qugr5}. That will 
provide a sufficient mathematical basis to account for the elementary discreteness of 
particle quanta as we shall argue in section~\ref{qugr6}. In 
section~\ref{qugr7} we return to the conceptual questions raised in 
section~\ref{qugr2} and further describe how the features of quantum 
theory and general relativity are amalgamated, with many specific 
issues potentially resolved, consistently in this theory. Finally we 
conclude in section~\ref{qugr8} with a summary of the main issues 
that have been addressed and the main areas for further development 
of the theory, including in connection with laboratory experiments.
 We begin by assessing the features of quantum theory and general 
relativity that must ideally be reproduced, reflected, resolved or 
reconciled in a unified theory.

\par}

\pagebreak
\section{Quantum Gravity: The Nature of the Problem}
\label{qugr2}

%\vspace{10cm}

\subsection{Quantum Theory Alone}
\label{qugr21}

 On the one hand the mathematical machinery of quantum theory has 
arguably been successfully tested much more thoroughly than that of 
general relativity, owing largely to the typical scale associated 
with quantum phenomena and the far greater ease in constructing 
corresponding laboratory experiments. On the other hand quantum 
theory seemingly also presents rather more significant conceptual 
difficulties and hence appears to be in greater need of a deeper 
understanding through a boarder theoretical context than general 
relativity. While Einstein's theory of gravity can be directly 
conceived of in terms of a geometric curvature of spacetime, quantum 
theory does not provide a fully coherent conceptual description of 
\textit{what is actually happening} in spacetime to accompany the 
corresponding practical calculations. These questions concerning the 
meaning and interpretation of quantum theory have been present from 
its inception in the 1920s and have been well-established since at 
least the 1980s (see for example~\cite{Rae} chapter~11). This 
long-standing major area of uncertainty in the basic nature of 
quantum theory itself suggests that a resolution may well require the 
context of a new conceptual framework, rather than  interpretational 
tweaks to the existing quantum formalism.   

  At the heart of these conceptual issues is the rift between quantum 
theory and classical physics, with
  the initial break opening up in 1900 with the origins of the `old 
quantum theory'  as marked by Planck's `quantum hypothesis'. Planck 
proposed that electromagnetic waves could only be absorbed or emitted 
in discrete bundles of energy $E = h\nu$, with $h$ a fundamental  
constant of nature and $\nu$ the oscillation frequency, rather than 
as a classical continuum. This hypothesis was introduced to elucidate 
Planck's radiation law, which in turn had been motivated to match the 
observed spectrum of blackbody radiation as a function of frequency 
and temperature (\cite{Pais} section~19(a)). 
 The introduction of photons as quanta of light, that is as actual 
physical particles of radiation with energy $E = h\nu$, by Einstein 
in 1905 was then able to account for the otherwise puzzling phenomena 
seen in the photoelectric effect (\cite{Pais} chapter~19), and also 
for the later observations of Compton scattering (see for 
example~\cite{Rae} section~1.2).

 The old quantum theory culminated in the 1923 proposal by de Broglie 
that the wave and particle properties associated with massless 
photons should be generalised to all material particles, in 
particular those with finite mass such as the electron. It was 
suggested that all particles might be associated with wave effects 
through the relations:
\begin{eqnarray}
   E & = & \hbar \omega \label{ehbaro}  \\
 \bp & = & \hbar \bk    \label{phbark}
\end{eqnarray}
  with $E$ the energy and $\bp$ the 3-momentum of the particle state, 
  while $\omega = 2\pi \nu$ is the angular frequency and $\bk$ is the 
wave 3-vector, with the corresponding de~Broglie wavelength $\lambda 
= 2\pi/\vert \bk \vert$, while $\hbar = h/2\pi$ is Planck's constant. 
The wave-like properties of electrons, which had been originally 
observed and thought of as purely particle phenomena,   were 
subsequently confirmed in 1927 through diffraction effects observed 
in the scattering off a crystal lattice (\cite{Rae} section~1.4, 
\cite{Pen} section~21.4). 
 This universal wave-particle duality, with discrete particle-like 
interactions associated with wave-like propagation, as encapsulated 
in the relations of equations~\ref{ehbaro} and \ref{phbark}
 -- marking an attempt to retain the familiar classical concepts of 
particles and waves albeit now associated with the same entity
 -- is the principal characteristic feature of the old quantum 
theory.

   The divergence from classical physics became even more pronounced 
for the `new quantum theory', introduced in particular by Heisenberg 
in 1925 and Schr\"{o}dinger in 1926, originally to account for atomic 
spectra but found to have much broader applicability. (We shall 
generally refer to this new development under the designation 
`quantum theory' or more specifically as `quantum mechanics').
 In the classical world it can generally be assumed that the act of 
making observations does not significantly influence the phenomena 
under investigation, while for quantum theory the properties of the 
system being studied cannot be distinctly separated from the 
observational apparatus. 
 Since no quantum system can ever be isolated and investigated purely 
in itself 
the dilemma remains in understanding precisely what it is that quantum 
theory tells us about the elementary nature of the physical universe. 

 One central feature of quantum theory that has been very well 
established is that of indeterminacy. Strict determinism, familiar in 
the macroscopic classical world, does not apply on the microscopic 
scale, for which events do not follow as a unique consequence of the 
immediately preceding state of the system, with an inherent 
randomness characteristic of quantum phenomena.
 Indeterminism is not in itself necessarily a conceptual 
\textit{problem}; it can simply be posited as a `brute fact' about 
the physical world. However, given in particular that for centuries 
before quantum theory it had been natural to assume the physical 
world to be deterministic and a reluctance to change this view
(`the Old One', Einstein was famously convinced, `does not play dice'  
\cite{Pais} section~25(a))
 it can still be asked \textit{why} the basic laws of physics should 
exhibit this uncertainty. There is also an intrinsic asymmetry in the 
observation of certain events at the expense of other possibilities 
that apparently \textit{could have been observed} with a known 
probability but that are not realised in our world, raising the 
question concerning the meaning of the potential alternative 
outcomes.

 In quantum mechanics the probability of an outcome is calculated in 
terms of a wavefunction $\Psi$ that describes the state of the system 
under study. Before a measurement is made the wavefunction evolves in 
time $t$ according to the Schr\"{o}dinger equation:
\begin{equation}
\label{Schro} 
   i \hbar \frac{\pal}{\pal t} \Psi \; = \; \hat{H} \Psi   
\end{equation} 
  where $\hat{H}$ is the Hamiltonian operator representing the total 
energy of the quantum mechanical system (see for example~\cite{Rae} 
equation~4.45 and chapter~4 generally for the basic postulates of 
quantum mechanics. Subsequently in this paper natural units will 
generally be employed with Planck's constant $\hbar = 1$ and also the 
speed of light $c=1$). 
  Even for a single-particle system the quantum mechanical 
wavefunction $\Psi(\bx,t)$, as a function of position $\bx$ in 
3-dimensional space, has characteristic non-local properties. This 
feature is exhibited for example in the archetypical double-slit 
experiment depicted for an electron source in figure~\ref{dslit}.

%\pagebreak % ***** temporary *****
% figure moved from here

 A particle such as an electron is typically conceived of as a 
\textit{local} entity, as attributed to the distinctive particle-like 
properties of a local emission at $S$ and a local detection at $I$ in 
figure~\ref{dslit}. On the other hand the electron wavefunction 
$\Psi(\bx,t)$ must be considered a \textit{non-local} entity in 
`passing through both slits at the same time', exhibiting a 
distinctly wave-like propagation as a solution for 
equation~\ref{Schro}, in order to describe the interference pattern 
observed on accumulating a number of hits in proportion to $\vert 
\Psi(\bx,t) \vert^2$ on screen $M$ in figure~\ref{dslit}. This 
apparent `wave-particle duality', retained from the old quantum 
theory, is a central enigmatic feature of quantum mechanical systems.
   
 % figure moved to here
 \vspace{-2pt}
 \begin{figure}[htbp]  
\centering
\epsfxsize=12.0cm
\leavevmode
\epsffile[0 0 1413 863]{aQfig1e}
\vspace{-8pt}
\caption{\setb  The likelihood of observing the interaction $I$ of a 
single electron emitted from the source $S$ at time $t_0$ after 
`passing through' the double-slit $D$ to the measurement screen $M$ 
at time $t_1$ is described by the probability distribution $\vert 
\Psi(\bx,t) \vert^2$ determined by the wavefunction $\Psi(\bx,t)$ 
across $M$ for this quantum mechanical \mbox{electron state.} 
\mbox{(See for example~\protect\cite{Rae} figure~1.3} for a description of 
this experiment with photons).}
\label{dslit}
\end{figure}   
   
  The non-local character of quantum mechanics is perhaps even more 
marked for a wavefunction describing the entanglement of a 
multi-particle system.
 For example the observed spin state of each of two electrons can be 
correlated between two measurements separated by an arbitrary 
distance in an EPR-type experiment (\cite{Pais} section~25(c), 
\cite{Pen} section~23.3, \cite{EPR}). The two particles apparently do 
not have a fully independent existence even if the two measurements 
are made outside the relative light cone and in principle light-years 
apart. The resulting correlations between the two measurements as 
observed in actual experiments agree with the predictions of quantum 
mechanics. While also consistent with special relativity in that no 
signal can be propagated faster than light speed, the correlations 
violate the inequality of Bell's theorem (\cite{Rae} equation~11.13, 
\cite{Bell}) and hence cannot be accounted for by a local and 
deterministic `hidden variable' theory (\cite{Rae} section~11.2) in 
which the two separated measurements are made entirely independently. 
As was the case noted above for \textit{indeterminacy}, the question 
can also be raised concerning how to make sense of such intrinsically 
\textit{non-local} phenomena as observed in the workings of the 
universe. 

  A further distinctive feature of quantum mechanics is that of the 
apparent collapse or \textit{reduction} of the wavefunction when a 
measurement takes place. This is a further non-local characteristic, 
quite unlike any property of a classical wave, in which the entire 
wavefunction and corresponding probability distribution apparently 
instantly reduces down to the state of the particular measured value. 
This is the case for example with $\Psi(\bx,t)$ `collapsing to a 
point' with the electron detection at $I$ in figure~\ref{dslit}, 
corresponding to the sudden return to a local particle description. 
This apparently instantaneous non-local reduction, with zero 
probability instantly inferred for all regions of $M$ excluding $I$, 
is cause for significant doubt in the wavefunction as furnishing a 
complete description of the physical world.

  The unitary Schr\"{o}dinger evolution {\bf U} associated with the 
Hermitian operator $\hat{H}$ of equation~\ref{Schro}, preserving a 
normalised sum total probability of one over a range of possible 
outcomes, is hence punctuated by the reduction of the wavefunction 
{\bf R}, as a non-unitary jump to the particular outcome of a 
measurement that is then itself alone assigned a probability of one.
 This dualism in the development of the wavefunction in alternating 
between {\bf U} and {\bf R}, with
 their very different properties, 
  is problematic from the perspective of the coherence of quantum 
theory as a single framework (\cite{Pen} sections~22.1 and 22.2). 
While {\bf U} is continuous and time-symmetric the abrupt collapse 
{\bf R} represents a discontinuous and time-asymmetric action. The 
irreversible change in a detector as a record is formed in a 
measurement under {\bf R}, as an apparent real action in the physical 
world, appears closer to physical reality than the evolution {\bf U} 
of the wavefunction $\Psi$ in describing the evolving array of 
probabilities for the outcome of an observation. 
As a mathematical expression to compute the probabilities of possible 
outcomes the wavefunction $\Psi$ can be considered to represent the 
experimenter's best `knowledge' of a quantum system before a 
measurement is made.
The differing properties of {\bf U} and {\bf R} are associated with 
the `measurement problem' regarding whether, why, when and how the 
wavefunction collapse ${\bf U} \to {\bf R}$ actually takes place, 
suggesting that the formalism of quantum mechanics, with any 
interpretation, may be an approximation to a new more uniform and 
coherent theory.

  In attempting to address the measurement problem a range of 
interpretations and approaches have been proposed (see for 
example~\cite{Rae} section~11.3, \cite{Pen} sections~29.1 and 29.2).
In order to make a measurement at all it can be considered 
essentially necessary to postulate that detectors obey classical 
laws, with the quantum mechanical wavefunction evolution of 
equation~\ref{Schro} hence not universal. 
 The simplest case in principle is to treat quantum and classical 
mechanics as two separate theories. 
 This kind of dualism is however not entirely dissimilar to that 
noted above for the {\bf U} and {\bf R} processes themselves.
 That it is aesthetically undesirable to have two independent 
theoretical descriptions of the universe, together with the fact that 
the nature and location of an apparent boundary between the quantum 
and classical world is unclear, motivates attempts to take the 
quantum evolution further, extending the application of 
equation~\ref{Schro} to macroscopic apparatus and even to the 
hypothetical cat of Schr\"{o}dinger. 
  Interpretations of how to best comprehend quantum mechanics then 
include the branching of `many worlds', a role for the observer's 
consciousness in subjective theories (see for example~\cite{Wign} 
chapters 13 and 14), or even a combination in a `many minds' 
proposal. Other approaches include `pilot wave' theory, 
schemes with modified dynamics and further
explicit mechanisms 
 such as `environmental decoherence' (see~\cite{Tegk} for an example 
combining many worlds with environment-induced decoherence).

  However it is the `Copenhagen interpretation', the earliest 
approach developed in the latter 1920s and associated with Niels Bohr 
and collaborators, that is still generally favoured as a pragmatic 
working guide to the assessment of laboratory experiments.
    That there is no compelling argument for the concrete language of 
macroscopic classical physics to apply consistently on the very 
different and unfamiliar scale of the microscopic world provided part 
of the motivation for the Copenhagen interpretation. While 
measurements themselves must be expressed in conventional classical 
terms observables simultaneously accessible at the classical level 
can be mutually exclusive, or complementary, at the microscopic 
scale.  

 Unlike the case for classical observables, which are represented 
directly by real numbers, the corresponding observables are 
represented in quantum mechanics by operators acting on the system 
wavefunction. In cases where the operators do not commute the 
corresponding eigenvalues, extracted from the action on the 
wavefunction, do not represent measurements that can be made 
independently. In quantum theory there is hence a `complementarity' 
between non-commuting operators that represent observations that 
cannot be made simultaneously and precisely. 
Apparatus designed to reveal one aspect of a system may prohibit the 
observation of another. This is the case for the relation between the 
predicted measurements of the position and the momentum of a particle 
according to the `Heisenberg uncertainty principle', which itself 
follows from the fundamental postulates of quantum mechanics 
(\cite{Rae} section~4.5).

   Hence properties that might be attributed to the quantum system 
are actually inseparably properties of the combined microscopic 
system and macroscopic classical equipment, which is configured to 
probe particular observables. While some form of apparatus is essential to make any observation at all, any measurement is a property of the 
whole setup. For example the correlations observed in an EPR-type 
experiment as alluded to above relate to the two electrons 
\textit{and also} the orientation of the two sets of measurement 
apparatus. It is not appropriate to speak of a non-local influence of 
one particle on the other since there \textit{are no} particle 
entities with an existence independent of the apparatus used to 
observe them.   

  This inseparability of a quantum system and the arrangement of the 
experimental apparatus is well supported in practice by a wealth of 
laboratory measurements. For example if the setup in 
figure~\ref{dslit} is \textit{changed} to detect \textit{which} of 
the two slits each electron from $S$ actually passes through then on 
accumulating a number of events the intermediate screen $D$ 
essentially casts a classical `shadow' and the interference pattern 
on $M$ disappears. The question remains however to elucidate the full 
nature and relation of quantum and classical systems and their 
apparent ill-defined boundary in a new unified theory. In particular 
with pertinent questions also hanging over all interpretations of 
quantum theory the quest for a deeper understanding regarding 
\textit{what is actually happening} in experiments such as that in 
figure~\ref{dslit} is still open for consideration.

  These features of quantum theory apply both for quantum mechanics, 
as considered above, and also for the development to the relativistic 
case of quantum field theory (QFT).
 Here we have initially focussed on the conceptual picture presented 
in non-relativistic quantum mechanics. While sidestepping the more 
technical mathematics of relativistic QFT, characteristic quantum 
properties are also empirically more accessible and have been more 
thoroughly investigated in the non-relativistic case.
  As for quantum mechanics the calculations of QFT for the high 
energy physics (HEP) environment of particle colliders have 
nevertheless also achieved complete success in matching the empirical 
observations. However the more technical aspects inevitably lead to 
more questions being raised concerning the calculational side of QFT, 
some of the basic aspects of which we summarise in the remainder of 
this subsection. 

  For the calculations in QFT the `interaction picture' may be 
adopted allowing the employment of free fields analysed into 
 $e^{\pm ip\cdot x}$  Fourier components where $p \!\cdot\! x$ is the 
Lorentz inner product (see for example~\cite{Pesk} chapter~4). For 
example a real scalar field, as a function of location $x$ in the 
flat 4-dimensional spacetime of special relativity, can be written as 
(\cite{Unifi} equation~10.13):
\begin{equation}
   \hat{\phi}(x)  =  \int \frac{d^3 \bsl{p}}{(2\pi)^3} 
\frac{1}{\sqrt{2\omp}} \,
          \Big( a(\bp) \, e^{-ip\cdot x} \, + a^{\dag}(\bp) \, 
e^{+ip\cdot x} \Big)
		     \label{kgosol2}
\end{equation}
 The wave 4-vector $p=(p^0,\bp)$ of each oscillation mode is  
\textit{directly} identified with the \mbox{4-momentum} $(E,\bp)$ on 
the left-hand side of equations~\ref{ehbaro} and \ref{phbark} (that 
is with $\hbar=1$). The energy $p^0 = E$ and the 3-momentum $\bp$ are 
related in the relativistic expression
 \mbox{$p^0 = \omp = \sqrt{\bp^2 + m^2}$} where $m$ is the mass of 
the corresponding particle states.
 The particle states are identified through the
 `quantisation' of the operator field $\hat{\phi}(x)$ which is 
achieved by imposing commutation relations
 for the operator coefficients attached to the Fourier modes in 
equation~\ref{kgosol2} with:
\begin{equation}
   \begin{array}{c}
    \lbrack a(\bsl{p}),  a^{\dag}(\bp') \rbrack \, = \, (2\pi)^3 \, 
\delta^3 (\bp - \bp') 
	   \\
    \lbrack a(\bp),  a(\bp') \rbrack =0, \qquad
	\lbrack a^{\dag}(\bp),  a^{\dag}(\bp') \rbrack =0 
	  \end{array} \label{aacomr}
\end{equation}
  as reviewed for (\cite{Unifi} equation~10.16).
 In acting as ladder operators on a Fock space of particle states 
 $a^{\dag}(\bp)$  and $a(\bp)$ are interpreted as particle
   `creation'  and
 `annihilation'  operators respectively. For example, adopting the conventions of (\cite{Unifi} equation~10.17),   a single particle state can be expressed as 
$\vert \bp \rangle \, = \, \sqrt{2\omp} \; a^{\dag}(\bp)\vert 0 
\rangle$, 
where $\vert 0 \rangle$ represents the vacuum state.
 Hence `wave-particle duality' is explicitly exhibited by the 
structure of the quantum field $\hat{\phi}(x)$  in 
equation~\ref{kgosol2} -- the integral over the Fourier modes $e^{\pm 
ip\cdot x}$ represents a linear combination of `wave' solutions for 
the Klein-Gordon equation while the operators $a^{(\dag)}(\bp)$ 
create and annihilate quanta of field excitations corresponding to 
`particles' (\cite{Pesk} section~2.4).

  The cross-section $\sigma$ representing the likelihood for a transition from the
particles of a given initial state $\vert i \rangle$ to those of 
a specific final state $\langle f \vert$ is directly proportional to 
$\vert \mcM_{fi} \vert^2$, where the transition amplitude $\mcM_{fi}$ 
is determined by the non-trivial part of the $S$-matrix in the 
scattering matrix element (\cite{Unifi}~equations~10.3--10.6):
\begin{equation}
   S_{fi}  =  \langle f \vert S \vert i \rangle
   \label{sfi}
\end{equation}
  The $S$-matrix describes the evolution of the initial state
 according to the possible interactions of the particular QFT.
 These are described by an interaction Hamiltonian $H_{\mathrm{int}}$ 
which is written down directly in terms of the interaction Lagrangian density 
$\lag_{\mathrm{int}}$ for the model being considered  (\cite{Unifi} 
equations~10.27):
\begin{eqnarray}
   H_{\mathrm{int}} & = & \int d^3 \bx \, 
    {\mathcal H}_{\mathrm{int}} \: = \: -\int d^3 \bx \, 
\lag_{\mathrm{int}}\; .
	\label{htlagx} 
   \\
  \mbox{For example} \quad
    \lag_{\mathrm{int}} & = & -g\hat{\phi}\hat{\mcX}^{\dag}\hat{\mcX}
	                    \, - \,  
g\hat{\phi}\hat{\mcY}^{\dag}\hat{\mcY}  
	 \label{lagfint}
\end{eqnarray}
  would represent a Lagrangian density for the field $\hat{\phi}(x)$ 
of equation~\ref{kgosol2} interacting with complex scalar fields 
$\hat{\mcX}(x)$ and $\hat{\mcY}(x)$, each associated with Fourier 
expansions involving their own creation and annihilation operators, 
and where $g$ is a coupling constant (\cite{Unifi} equations~10.14, 10.15 
and 10.23).

\pagebreak

 With the interaction taking place at time $t\sim 0$ and the initial 
state before the transition process expressed as 
$\vert i \rangle = \vert \Psi(t=-\infty)\rangle$ the evolution of the 
state corresponding to the Schr\"{o}dinger equation~\ref{Schro} here 
in the interaction picture for the QFT can be written as:
\begin{equation}
\label{Schroq} 
   i \frac{\pal}{\pal t} \vert \Psi(t)\rangle 
     \; = \; H_{\mathrm{int}}(t) \vert \Psi(t)\rangle  
\end{equation} 
  The unitary operator $U$ describing the evolved state 
  $\vert \Psi(t) \rangle  \equiv U(t,-\infty)\vert i \rangle$ at time 
$t$ hence satisfies the same relation:
\begin{equation}
     i \frac{\pal}{\pal t} U(t,-\infty) = 
                 H_{\mathrm{int}}(t)U(t,-\infty) 
    \label{uevolve}
\end{equation}
 Although the background is the Minkowski spacetime of special 
relativity it can be seen from the above equations that, as for 
non-relativistic quantum mechanics, progression in time still plays a 
privileged role compared with dependence on the three dimensions of 
space in QFT.   
  
  Given the unitarity of the evolution in time an initial 
normalisation $\langle i \vert i \rangle =1$ is preserved
 through to $t=+\infty$ after the interaction. This
 implies (\cite{Unifi} equation~11.45):
\begin{equation}
 \label{fsumunit}
   \sum_f \vert \langle f \vert \Psi(\infty) \rangle \vert^2 = 1
\end{equation} 
 corresponding to a conserved summed probability of one over the full 
range of possible final state outcomes.
 Given this full evolution to 
 $\vert \Psi(\infty) \rangle  \equiv U(+\infty,-\infty)\vert i 
\rangle$, which incorporates \textit{all} potential final states that 
can be attained from the initial state via the interaction 
Hamiltonian,
 the individual transition probability to a particular final state $\langle f 
\vert$ and the determination of the corresponding event cross-section 
then depends upon the matrix element: 
\begin{equation}
    S_{fi} =  \langle f \vert \Psi(\infty) \rangle 
         = \langle f \vert U(+\infty, -\infty) \vert i \rangle 
	  \label{sfifi}
\end{equation}

  The $S$-matrix in equation~\ref{sfi} is hence defined in terms of 
this full evolution from the initial time $t=-\infty$ through to the 
final time $t=+\infty$. Solving equation~\ref{uevolve} for a general 
interaction Hamiltonian by iteration, taking into account the 
operator expansion in the interaction picture (\cite{Unifi} 
equations~10.31--10.35), leads to the general form:
\begin{equation}
  \label{smatrix}
   S = U(+\infty,-\infty) = T \left[
   \exp \left( -i\int_{-\infty}^{+\infty} dt \,    
     H_{\mathrm{int}}(t) \right)
	    \right]
\end{equation}
  This expression incorporates an algebraic rearrangement with a  
time-ordered product $T$ of operators while the exponential 
represents a formal expansion as a sum over terms of increasing order 
in $H_{\mathrm{int}}$.
 When sandwiched between $\langle f \vert$ and $\vert i \rangle$ 
states equation~\ref{smatrix} describes all possible chains of 
mediating interactions 
 and determines the overall process probability.
 If the coupling, such as $g$ in equation~\ref{lagfint}, is 
sufficiently small then the first few terms of the perturbative 
expansion will generally provide the required accuracy for the 
corresponding cross-section calculation.

 In expanding the $S$-matrix between given $\langle f \vert$ and 
$\vert i \rangle$ states the algebra of the annihilation and creation 
operators, such as for $a(\bsl{p})$ and $a^{\dag}(\bp)$  in 
equation~\ref{aacomr}, 
  effectively act to exchange $e^{\pm ip\cdot x}$
 Fourier components between fields that appear in the same term of 
the interaction Lagrangian, such as between $\hat{\phi}(x)$ and
 $\hat{\mcX}^{(\dag)}(x)$ in the first term for the model of 
equation~\ref{lagfint}. Factors of creation and annihilation 
operators and $e^{\pm ip\cdot x}$ modes composed together under 
 3-momentum integrals (from fields such as $\hat{\phi}(x)$ in 
equation~\ref{kgosol2}) together with 4-dimensional spacetime 
integrals (from substituting equation~\ref{htlagx} into 
equation~\ref{smatrix}) 
 result in 4-momentum conserving $\delta$-functions for these field 
exchanges 
(see for example \cite{Unifi} equations~10.45 and 10.46), as associated with the 
`vertices' of a Feynman diagram.
 As an element of these calculations the
 time-ordering in equation~\ref{smatrix} implies the introduction of 
$\theta$-functions that effectively augment intermediate 3-momentum 
integrals to apparent `4-momentum' integrals (\cite{Unifi}~equations 
10.64--10.71) leading
  to field `propagator' terms between the vertices  (\cite{Unifi} 
equations~10.38--10.42),
 as also depicted in a Feynman diagram (see for example \cite{Unifi} 
figure~10.4, and also figure~\ref{extofey}(b) in 
subsection~\ref{qugr52} here).

  All possible sequences of field interactions, as expressed through the 
$T$-ordering and corresponding Feynman diagrams, contribute to the 
matrix element $S_{fi}$ in equation~\ref{sfifi}.  In the Standard 
Model for particle physics the field content and Lagrangian structure 
are largely constructed by hand (see for example \cite{Teub} 
chapter~5 equation~5.1, \cite{Unifi} section~7.2) as guided by 
empirical observations and as consistent with requirements such as 
local gauge symmetry. The coupling parameters for the Standard Model 
(such as in the Lagrangian terms of \cite{Teub} equations~5.30--5.33, 
\cite{Unifi} equations~7.39--7.42) are sufficiently small for 
perturbation theory to apply for practical calculations, well within 
the typical statistical errors encountered for measurements in HEP experiments.

While the postulated rules for calculating event probabilities, which 
extend well beyond the above brief review of core elements, have been 
unfailingly tested in a vast range of experiments the question of an 
underlying origin for these mathematical tools of quantum theory, 
beyond their considerable pragmatic value, remains open.
 As a closely related issue while the notion of `particles' has a 
natural role in summarising laboratory data a definitive conceptual 
correlate of \textit{physical} particle states is not explicitly provided by QFT and 
is not needed in performing QFT calculations with results that 
accurately match the observations. Using this machinery QFT is 
pragmatically geared towards mathematical techniques which determine 
measurable cross-sections and decays rates via the $S$-matrix, as we 
have reviewed above, rather than the accuracy of a conceptual 
description of the physical structure of the microscopic world. Hence 
the question of the true nature of particle states, as well as their 
position relative to fields as fundamental entities, is left largely 
open in QFT.    

 For QFT calculations the question can be raised concerning not only 
the basis of these mathematical methods employed to determine event 
probabilities but also regarding the appearance of infinities which 
frequently arise in performing the relevant integrals despite the unitarity of the state evolution and  
equation~\ref{fsumunit} (see for example \cite{Unifi} figure~10.9 and 
equation~10.86).
 This issue itself requires the sophisticated 
  techniques of renormalisation, which involve an element of 
calibration against measured parameters (\cite{Pesk} Part II opening discussion). 
 Physical charges and masses for example are then distinct from the 
`bare' coupling parameters of the original Lagrangian that are 
allowed to absorb the renormalisation.

 That the requirement of these and other mathematical methods seem 
somewhat
 inappropriate for a fundamental physical theory suggests that a 
further underlying understanding of the origin of quantum theory, as 
a basis for characteristic features such as indeterminism as well as quantitative 
likelihood calculations, would be desirable. 
  This would imply that QFT as applied in the HEP laboratory is a 
low-energy \textit{effective} theory, a limit of a boarder and more 
coherent new theory. (This is the opinion expressed in \cite{Wein} 
volume I chapter 12 opening, as quoted in \cite{TimeE} section~6, and 
also for example in \cite{Wall} section~3 where `X' denotes the 
presumed new theory, see also~\cite{Pesk} section~22.5).  
 The formalism of QFT `is the way it is' through the need to reconcile 
quantum mechanics with special relativity (\cite{Wein} volume I 
chapter 2 opening). However since the geometry of spacetime is not 
flat QFT is inherently incomplete and, in the context of this paper, 
what is really needed is a theory to reconcile quantum phenomena  
directly with general relativity -- from which QFT and quantum 
mechanics might be recovered in the appropriate limits.

  One motivation for this subsection has been to summarise basic 
elements of QFT calculations for later reference, in particular in 
subsection~\ref{qugr52}.
  The other main purpose has been to emphasise the considerable 
uncertainty that still remains concerning both the meaning of 
calculations in quantum theory and regarding how to extend the 
corresponding formalism to connect with that of classical physics in 
the laboratory, even \textit{before} considering a further extension 
of the domain of quantum theory to cover also gravitational 
phenomena.
 Given the existing technical and conceptual issues with quantum 
theory in itself directly attempting to apply `quantisation' rules to 
gravity or to the structure of spacetime may in this sense be 
somewhat overreaching.
 On the other hand the contention of the present theory is that a 
full understanding of `quantum gravity' may in fact provide a 
clarification of the nature and domain of quantum phenomena 
themselves, affording a deeper basis for comprehending a more 
complete and coherent conceptual picture for quantum theory alone as 
well as in combination with general relativity.

%\pagebreak
\subsection{General Relativity Alone}
\label{qugr22}

     Compared with quantum theory Einstein's theory of gravity 
carries a far more literal and direct picture of the physical world 
and with somewhat less conceptual uncertainty in itself. The 
application of the  theory of general relativity can be considered to 
consist of two compatible parts: the Einstein field equation relating 
the geometry of 4-dimensional spacetime $M_4$ to the distribution of 
matter and equations of motion for matter through spacetime. In the 
Einstein field equation the Einstein tensor $G^{\mu\nu}(x)$, 
describing the geometry of the extended spacetime through a function 
of spacetime derivatives of the symmetric metric tensor 
$g_{\mu\nu}(x)$ with
 $x \in M_4$, is equated directly with the energy-momentum tensor 
 $T^{\mu\nu}(x)$:
\begin{equation}
 \label{Eineq}
    G^{\mu\nu} = -\kappa T^{\mu\nu}
\end{equation} 
  where $\mu,\nu = 0,1,2,3$ are indices of general coordinates in 
spacetime and $-\kappa$ is a normalisation constant (the geometric 
constructions and conventions of Riemannian geometry behind this 
equation are reviewed in~\cite{Unifi} sections~3.3 and 3.4, as 
discussed in particular for equations~3.53 and 3.71--3.74 therein).
 The structure of general relativity follows largely as a consequence 
of the `equivalence principle', according to which the laws of 
physics converge to those of special relativity in the limit of small 
spacetime volumes (see for example~\cite{Rind} sections~1.13, 8.4 and 
14.2, \cite{Carr} chapter~4, \cite{WeinGR} sections~3.1 and 4.1).

   By the equivalence principle local coordinates can always be 
constructed about any point $x \in M_4$ for which the metric 
$g_{\mu\nu}(x)$ locally takes the form of the Lorentz metric
 \mbox{$\eta_{ab} = \mbox{diag}(+1,-1,-1,-1)$}, with $a,b = 0,1,2,3$ 
local inertial frame coordinate indices, defining a light cone 
structure throughout the extended spacetime $M_4$. Equations of 
motion for fields and particles then describe the causal propagation 
of matter within the light cone structure, with material bodies 
postulated to pursue geodesic trajectories in the absence of any 
non-gravitational forces, essentially generalising the inertial 
motion of Newton's first law.
  
   In order to empirically verify the theory solutions of the field 
equation need to be determined and compared with actual observations 
in the world, as we review here. 
 The employment of the Einstein field equation~\ref{Eineq} to 
determine the gravitational field described by the metric tensor 
$g_{\mu\nu}(x)$ can in principle be tested through measurements of 
the geodesic deviation in the motion of two nearby bodies,  with the 
spacetime curvature being proportional to the observed deviation. 
 However since the field equation places mutual restrictions on the 
spacetime geometry $g_{\mu\nu}(x)$ as well as the matter sources 
described by $T^{\mu\nu}(x)$ it is not straightforward to extract a 
solution for 
  the metric tensor.
 With $G^{\mu\nu}(x)$ a non-linear function involving second-order 
derivatives of $g_{\mu\nu}(x)$ the difficulty in solving the field 
equation is also further complicated in that $T^{\mu\nu}(x)$ is 
itself in general a function of the metric $g_{\mu\nu}(x)$ as well as 
material entities.

   Hence with the distribution of matter itself  dynamically 
intertwined with the spacetime geometry through which it propagates  
    it is generally not possible to begin with a given source term on 
the right-hand side of the field equation~\ref{Eineq}. Further, a 
coordinate system is required in order to even specify the components 
of $T^{\mu\nu}(x)$.
 A possible procedure is to begin with a given general coordinate 
system $\{x^{\mu}\}$ in which an  arbitrary metric function 
$g_{\mu\nu}(x)$ is adopted. This will uniquely yield a specific form 
for the energy-momentum tensor $T^{\mu\nu}(x)$ through 
equation~\ref{Eineq}, and iterations over  $(g_{\mu\nu}(x), 
T^{\mu\nu}(x))$ pairs might be performed in an attempt to converge 
upon a particular physical application (\cite{Rind} section~14.2). 
However most such arbitrary metrics $g_{\mu\nu}(x)$ generate 
unphysical properties for $T^{\mu\nu}(x)$, for example with regions 
of negative energy density (see also \cite{Carr} discussion of 
equation~4.72), and hence restrictions are usually adopted to ensure 
more realistic forms of energy-momentum such as based on known matter 
compositions. 

 For more direct specifications of $T^{\mu\nu}(x)$ guided by 
empirical phenomena 
 an implicit restriction is still provided by the Einstein field 
equation itself through the four relations of the twice-contracted 
Bianchi identity (\cite{Unifi} equation~3.71, \cite{Carr} 
equation~3.96;
   with $\nabla_{\:\!\!\!\mu}$ the covariant derivative and the 
summation convention for repeated indices adopted throughout this 
paper):
\begin{equation}  
  \nabla_{\:\!\!\!\mu}G^{\mu\nu} = 0
   \label{Bian}
\end{equation}  
  The purely geometric origin of the Bianchi identities is discussed 
in (\cite{MTW} chapter~15).
 The geometric identity of equation~\ref{Bian}, via 
equation~\ref{Eineq},
imposes four constraints \mbox{$\nabla_{\:\!\!\!\mu} T^{\mu\nu}=0$} 
on the energy-momentum tensor.
 This vanishing divergence in turn implies that only  
 six of the ten field equations, in components of the symmetric 
tensors $G^{\mu\nu}(x)$ and $T^{\mu\nu}(x)$, are independent.
  Hence the ten components of the symmetric metric $g_{\mu\nu}(x)$ 
are not determined uniquely by equation~\ref{Eineq}, but rather four 
degrees of freedom remain for arbitrary coordinate transformations. 
That is,
 a solution for $g_{\mu\nu}(x)$ in one coordinate system 
$\{x^{\mu}\}$ will be a solution  in any other $\{x^{\mu'}\}$, as 
related by a set of four functions $\{x^{\mu'}(x^{\mu})\}$, and hence 
there are four unphysical degrees of freedom in $g_{\mu\nu}(x)$. 
 The field equation~\ref{Eineq} is hence sufficient to constrain the 
six degrees of freedom required to define $g_{\mu\nu}(x)$ up to an 
equivalence class $(M_4,g)$ of physically indistinguishable 
geometries on the manifold $M_4$ as related by general coordinate 
transformations (see for example \cite{Unifi} discussion of  
figure~3.6 and after equation~3.105).

  The Einstein equation is then not physically underdetermined and 
supplies a sufficient number of equations for the unknown quantities 
to allow for example a well-posed `initial value problem' for the 
metric $g_{\mu\nu}(x)$ (\cite{Carr} chapter~4, \cite{HawkEl} 
chapter~7).
 Typically in other branches of physics 
 once the \textit{initial} data concerning the location and motion of physical entities 
is given it is the \textit{future} motion, or evolution of the system 
in time, that is predicted by the equations of the theory. 
 The initial value problem for general relativity, also known as `the 
Cauchy problem', consists in using the Einstein equation~\ref{Eineq} 
to determine the metric $g_{\mu\nu}(x)$ in the proximity of a 
\mbox{3-dimensional} spacelike hypersurface upon which $g_{\mu\nu}(x)$ and 
$\pal g_{\mu\nu}(x)/ \pal x^0$ (with $x^0$ a timelike coordinate 
parameter) are given. 

  The identity of equation~\ref{Bian} implies constraints on the 
initial data such that
 only the spatial components of the metric are required if the 
`harmonic gauge' condition $\square x^{\mu} = 0$, with $\square = 
\nabla^{\nu}\nabla_{\!\nu}$, is employed for the coordinates.
 This coordinate condition is of course \textit{not} generally 
covariant, rather the purpose of such a choice is to remove some of 
the ambiguity in the metric due to the general covariance of the 
Einstein equation (\cite{WeinGR} sections~7.4 and 7.5). 
 The full 4-dimensional geometry can then be uniquely determined by 
evolving forwards to obtain a well-defined solution (barring the 
presence of `singularities', as will be discussed below).
  More general solutions can also be found by incorporating initial 
conditions including the specification of matter fields obeying 
suitable equations of motion with similar methods (\cite{HawkEl} 
section~7.7).
 In all cases numerical techniques will in general be required to 
iterate forwards in small steps of time.
  
 However most solutions in general relativity are obtained in 
spacetime $M_4$ directly as full 4-dimensional geometries given the 
input of more general `boundary conditions'.
 This expresses more generally the fact that differential equations 
of motion alone do not give the full picture and in order to make 
observable predictions it is necessary to specify such a set of boundary 
conditions for the fields and physical entities under study.
  As well as initial state data the boundary conditions might include 
an assumption of specific symmetry structures imposed on the 
spacetime geometry or further particular properties of the 
energy-momentum. 
 Generally a simple form for the energy-momentum, such as the vacuum 
case with $T^{\mu\nu}(x) = 0$ or for a perfect fluid with density and 
pressure terms (\cite{Unifi} equation~5.37 and 12.2), as well as 
simplifying symmetry assumptions, for which preferred coordinates can 
be employed such that fewer parameters are required to fully describe 
$g_{\mu\nu}(x)$,
 are needed to find exact solutions in general relativity. 

  Exact solutions for the Einstein equation in the vacuum case 
include the Schwarzschild solution (\cite{Unifi} equation~5.49), 
which assumes spherical spatial symmetry and can be applied to 
determine the orbit of planets (where deviation from the Newtonian 
theory in accounting for the observed perihelion advance of the 
planet Mercury
 provided the first test general relativity \cite{Pais} 
section~14(c).4).
 On the other hand matter treated as a perfect fluid, under the 
symmetry assumptions of large scale homogeneity and isotropy for the 
cosmic spacetime, provides the basis for a range of exact solutions 
for cosmological models for the evolution of the universe (as 
described for \cite{Unifi} equation~12.5).

 In the absence of a high degree of symmetry or other simplifying 
assumptions in general exact solutions are hard to find. In the case 
for example of three bodies of comparable mass mutually `orbiting' 
each other numerical methods are needed to find an approximate 
solution, as might be expressed through an initial value problem. 
   In practice numerical approximations might be required for many 
situations, as would be the case for more general cosmological 
solutions for example.
  
  As a common feature, however they are obtained, 
 all solutions in general relativity consist of a metric description 
for a \textit{complete} extended 4-dimensional spacetime geometry.
  This complete spacetime is needed for the study of cosmology, while 
only a limited region of the manifold may be of physical interest in 
many cases, such as in the analysis of planetary motion using the 
Schwarzschild solution. 
   All solutions also depend upon the boundary conditions assumed for 
the Einstein field equation. A change in these conditions, such as 
the introduction  of large scale inhomogeneities in the matter 
distribution for the cosmological case, will result in a different 
solution. 

   However, while exact calculations might prove intractable, from 
the conceptual point of view in general relativity there is no 
difficulty in conceiving of the existence of precise, unique, 
deterministic and consistent solutions for a vast range of given 
initial and boundary conditions, with one exception. 
 With gravity being attractive,
for the Schwarzschild solution adapted to study a massive body 
collapsing under its own weight, beyond the point at which known 
physical processes are capable of arresting the inward fall of 
matter, the field equations lead to the formation of a point of 
infinite density and curvature, namely a central `singularity' (see 
for example~\cite{HawkEl} chapters 8--10). A singularity is no longer 
part of a smooth spacetime manifold but rather represents a
 discontinuity in the
 structure of spacetime for which the field equations themselves no 
longer hold.
 The nature of such a black hole singularity, or an initial 
singularity associated with the Big Bang in cosmology, is perhaps the 
main conceptual difficulty with general relativity in itself. It can 
however be interpreted as an environment explicitly beyond the domain 
of general relativity alone, implying the need for further 
understanding of microscopic physics, and in particular of quantum 
phenomena in conjunction with enormous gravitational fields, in order
 to rigorously analyse the extreme spacetime regions in the vicinity 
of such singularities.

   At the other extreme, in the case of very weak gravitational 
fields, the simplified  field equations of `linearised general 
relativity' provide an approximate method that may be sufficient for 
some practical applications (\cite{Carr} chapter~6, \cite{FosN} 
chapter~5). In this manner the complications due to the non-linearity 
of the full Einstein equation~\ref{Eineq} can be avoided, with linear 
equations in general being much easier to solve. For a weak 
gravitational field the extended metric function $g_{\mu\nu}(x)$ can 
be decomposed into    
 the fixed Lorentz metric $\eta_{\mu\nu}$ of flat Minkowski spacetime 
together with a small perturbation $h_{\mu\nu}(x)$:
\begin{eqnarray}
   g_{\mu\nu} & = & \eta_{\mu\nu} + h_{\mu\nu} \label{glin} \\
   \mbox{with:} \quad \eta_{\mu\nu} & = & \mbox{diag}(+1,-1,-1,-1)
   \quad \mbox{and} \quad \vert h_{\mu\nu} \vert \ll 1    
      \label{nhlin}
\end{eqnarray}
  Even for weak gravity the smallness of the magnitude of the 
components $h_{\mu\nu}(x)$ will depend upon the global coordinates 
employed, but a convenient choice can be made from coordinate systems 
consistent with equations~\ref{glin} and \ref{nhlin}. The smallness of each component 
$\vert h_{\mu\nu} \vert \ll 1$ means that expressions to first order 
in $h_{\mu\nu}(x)$ will be sufficient, with for example the general 
relation $g^{\mu\nu}g_{\nu\sigma} = \delta^{\mu}_{\ph{\mu}\sigma}$ 
and equation~\ref{glin}
   implying that the inverse metric can be written as
    $g^{\mu\nu}  =  \eta^{\mu\nu} - h^{\mu\nu}$, 
	with $h^{\mu\nu} = \eta^{\mu\rho}\eta^{\nu\sigma}h_{\rho\sigma}$,
	 to first order.
	 
 Under this approximation general relativity can be effectively 
treated as a linear field theory for the symmetric tensor field 
$h_{\mu\nu}(x)$ in the flat spacetime of special relativity. The 
Einstein field equation~\ref{Eineq} becomes (\cite{Carr} 
equation~6.8, \cite{FosN} following equation~5.1.4; with care needed 
in comparing different references owing to the various sign 
conventions, although the only significant differences with 
reference~\cite{FosN} are here in the signs of $\kappa$ and 
$\square$. Here we also use either upper or lower indices for $G^{\mu\nu}$ and $T^{\mu\nu}$ according 
to convenience):
\begin{equation}
  \frac{1}{2} (-\pal_{\lambda}\pal_{\nu}h^{\lambda}_{\ph{\lambda}\mu}
   -\pal_{\lambda}\pal_{\mu}h^{\lambda}_{\ph{\lambda}\nu}
   +\pal_{\mu}\pal_{\nu} h + \square h_{\mu\nu} 
   -\eta_{\mu\nu}\square h
   +\eta_{\mu\nu}\pal_{\rho}\pal_{\sigma} h^{\rho\sigma})
  \, = \,  - \kappa T_{\mu\nu}
\label{Einlin} 
\end{equation}
  with $h(x) = h^{\nu}_{\ph{\nu}\nu}(x)=\eta^{\mu\nu}h_{\mu\nu}(x)$
   and here $\square = \eta^{\mu\nu}\pal_{\mu}\pal_{\nu}$ for this 
flat spacetime approximation. 
 On adopting a choice of harmonic coordinates with $\square 
x^{\mu}=0$ the above expression simplifies 
 (via \cite{Carr} equation~6.17) to:
\begin{equation}
  \square h_{\mu\nu} 
   -\frac{1}{2}\eta_{\mu\nu}\square h
  \, = \,  - 2\kappa T_{\mu\nu}
\label{Einlinh} 
\end{equation}
 For the trace-reversed perturbation 
 $\bar{h}_{\mu\nu}(x) := h_{\mu\nu}(x) - 
\frac{1}{2}\eta_{\mu\nu}h(x)$ the above equation can be written as simply 
(\cite{Carr} equation~6.22, \cite{FosN}~5.1.19):
\begin{equation}
  \square \bar{h}_{\mu\nu} 
  \, = \,  - 2\kappa T_{\mu\nu}
\label{Einlint} 
\end{equation}

   An important application of linearised general relativity, 
identified under the further simplification of the vacuum case with 
$T_{\mu\nu}(x) = 0$, concerns the propagation of gravitational waves. 
For this example the metric perturbation as a solution for the linear 
wave equation: 
\begin{eqnarray}
   \square \bar{h}_{\mu\nu} & = & 0 \label{gwaveq} \\
   \mbox{takes the form:} \quad\;  
  \bar{h}_{\mu\nu} & = & \mbox{Re}(A_{\mu\nu}e^{ik\cdot x})
  \label{waveh}
\end{eqnarray} 
 with $k\! \cdot\! x = \eta_{\mu\nu}k^{\mu}k^{\nu}$.
 Here the wave 4-vector $k=(k^0,\bk)$, with $k^0$ the angular 
frequency and the spatial part $\bk$ describing the direction of 
propagation, consists of just three constant parameters due to the 
null condition
 $k^{\mu}k_{\mu}=0$, by virtue of satisfying equation~\ref{gwaveq},
 corresponding to a gravity wave travelling at the speed of light 
through the approximately flat spacetime.

 On choosing a \mbox{$k$-dependent} subgauge of the possible range of 
harmonic coordinates, known as the transverse traceless or TT gauge, 
for which $\bar{h}_{0\nu} = 0$ 
 and $\bar{h}=\bar{h}^{\nu}_{\ph{\nu}\nu}=0$ (and hence 
$h_{\mu\nu}=\bar{h}_{\mu\nu}$), the symmetric amplitude matrix 
$A_{\mu\nu}$ in equation~\ref{waveh} also takes a simplified form. In fact under the 
constraint of the TT gauge the original ten possible constant real 
components of $A_{\mu\nu}$ reduce down to just two parameters that 
characterise the wave. On taking $k = (k^0, 0, 0, k^3)$, for 
propagation in the $x^3$ direction with $k^0 = k^3$, the solution for 
equation~\ref{gwaveq} in equation~\ref{waveh} can be written as 
(\cite{Carr} equation~6.53, \cite{FosN} equations~5.2.7 and 5.2.8):
\begin{equation}
     h_{\mu\nu} \, = \, 
     (A_{11} \:\! e^1_{\mu\nu} \, + \, A_{12} \:\! e^2_{\mu\nu})
	      \cos k \!\cdot\! x 
  \label{wavepol}
\end{equation} 
\begin{equation}
     \mbox{with:} \quad
    e^1 = \left( \begin{array}{rrrr}
	     0\;\; & 0 & 0 & \;\; 0  \\
		 0\;\; & 1 & 0 & \;\; 0  \\
		 0\;\; & 0 &-1 & \;\; 0  \\
		 0\;\; & 0 & 0 & \;\; 0  
		 \end{array} \right) \quad \mbox{and} \quad
    e^2 = \left( \begin{array}{rrrr}
	     0\;\; & 0 & \;\; 0 & \;\; 0  \\
		 0\;\; & 0 & \;\; 1 & \;\; 0  \\
		 0\;\; & 1 & \;\; 0 & \;\; 0  \\
		 0\;\; & 0 & \;\; 0 & \;\; 0  
		 \end{array} \right)	\qquad 
  \label{waveee}
\end{equation}
  That is, all $A_{\mu\nu}$ components are zero apart from $A_{11} = 
-A_{22}$ and $A_{12} = A_{21}$.
  Equation~\ref{waveee} is the gravitational analogue of the 
polarisation states of an electromagnetic wave as will be noted for 
equations~\ref{ecoeff} and \ref{waveem} in subsection~\ref{qugr51}.
  The amplitude in equation~\ref{wavepol} is here a linear 
combination of the two polarisation matrices $e^1$ and $e^2$ which 
describe the manner in which test particles will be displaced in the 
transverse directions as the gravitational wave passes
 (\cite{Carr} equations~6.65--6.69 and accompanying figures, \cite{FosN} table~5.1).
   Given the weakness of gravity dramatic events are required to 
observe gravitational waves, for which the predictions of general 
relativity have been verified in recent years~\cite{Ligo}.

  Notwithstanding the range of applications and tests of general 
relativity, including those reviewed above, open questions still 
remain regarding the structure of the theory itself. On the large 
scale the nature and role of a cosmological constant $\Lambda$ in the 
possible augmentation of the Einstein equation~\ref{Eineq} to the 
form (\cite{Unifi} equations~3.84 and 12.1, \cite{Carr} 
equation~4.74):
\begin{equation}
   \label{einlamt}
    G^{\mu\nu} + \Lambda g^{\mu\nu} = - \kappa T^{\mu\nu}
\end{equation}
 is yet to be fully understood. A significant impact for the extra 
term in driving the observed accelerating expansion of the universe
 has been identified for some of the cosmological models alluded to earlier, 
however the source of this apparent `dark energy' is unknown.

  On a much smaller scale there is also an open issue concerning whether 
the geometric property of torsion might play a role. In such an 
augmentation of general relativity to Einstein-Cartan theory the 
linear connection $\Gamma^\rho_{\ph{\rho}\mu\nu}(x)$  in general also has 
an antisymmetric component in the lower indices (\cite{Unifi} equation~3.60) and cannot be 
expressed as a unique function of the metric $g_{\mu\nu}(x)$ alone
(unlike the symmetric Levi-Civita connection, \cite{Unifi} equation~3.53, for the torsion-free case). 
 The question concerning the specific nature of the source and structure of torsion 
can then also be raised for this extension to general relativity.

  Finally in this subsection we make some observations concerning 
locality in general relativity. As a theory of the gravitational 
field in spacetime general relativity is considered a `local' theory. 
In fact it can be expressed in terms of a local Lorentz symmetry in a 
manner analogous to the structure of local gauge theories (as 
reviewed in \cite{Unifi} near the end of section~3.4 with reference 
to~\cite{Uti} here). On the other hand considered as a theory of the 
structure of spacetime itself, including the geometry of spacelike 
hypersurfaces which cut through the non-causal directions of the 
continuous array of light cones, there is a distinctly `non-local' 
aspect to general relativity. 
  That is, unlike a geometrical disturbance such as that of 
equations~\ref{glin} and \ref{wavepol} \textit{in} spacetime, in representing by 
definition a coherent arena even outside the light cone structure 
spacetime \textit{itself} is an intrinsically non-local entity.

  Even in specifying an initial 3-dimensional spacelike geometry as a 
boundary condition for a solution to equation~\ref{Eineq} this 
structure is itself manifestly non-local in nature. It could be 
argued that such an initial condition is merely the result of the 
earlier evolution of the spacetime geometry. However in evolving back 
in time as far as possible to the era of the Big Bang the initial 
singularity at cosmic time $t=0$ is \textit{itself} conceived of as 
an infinitely extended spacelike hypersurface (see for example 
\cite{Pen} figure~28.17(c)). The `start-up problem' (discussed in 
\cite{Unifi} towards the end of section~12.3 with reference to 
figures~12.3 and 12.4 therein) alludes to this intrinsic 
non-locality.  
 Essentially
 with all field solutions for equation~\ref{Eineq} covering a full 
4-dimensional spacetime $M_4$, which 
 \textit{carries} the assembly of light cones within which matter 
evolves in a causal manner, the spacetime manifold $M_4$ itself is 
not something that \textit{evolves within} that causal structure, and 
is rather of a necessarily external and non-local nature. 

For example given the Hubble expansion rate $H(t)$ at  cosmic time 
$t$ two bodies locally at rest with respect to comoving cosmic 
coordinates and separated by a physical spatial proper distance $d(t) 
= a(t)\Delta \Sigma$, where $a(t)$ is the cosmological scale factor 
and $\Delta \Sigma$ is the comoving `coordinate distance' from the 
spatial part of the Robertson-Walker line element (\cite{Unifi} equation~12.5), will 
be mutually receding at a speed of $v = H(t)d(t)$  (\cite{Unifi} 
equation~12.12).
  Hence at the present cosmic time $t_0$, 
  with \mbox{$H_0 = H(t_0) \simeq 
70\,$km$\,$s$^{-1}\,$Mpc$^{-1}$~\cite{PDG20}},  for $d(t_0) \gtrsim 14 
\times 10^9$ light-years this mutual recession speed exceeds that of 
light, and can be arbitrarily larger for yet more distant objects. 
   With a present day particle horizon of $R_p(t_0)\simeq 46 \times 
10^9$ light-years (discussed in \cite{Unifi} after equation~12.24) 
this
also implies that the most distant objects that we can see are `now', 
at cosmic time $t_0$, receding from us at more than the speed of 
light.
 However there is no violation of the limiting speed of light within 
the local light cone structure, with nothing actually 
\textit{travelling} faster than light.

  Spacetime is the arena of events in the world and while exhibiting 
the properties of an extended continuum it is not itself subject to 
the causal relations defined \textit{by} the spacetime structure. 
 This is very much unlike the formalism of quantum theory, as 
described in the previous subsection, which is embedded within the local 
structure of a background space and time. While QFT can be applied in 
the laboratory setting under the assumption of a locally flat 
spacetime, as consistent with the equivalence principle of general 
relativity, the latter theory concerns the properties of the 
spacetime arena itself and is hence somewhat \textit{external} to the 
application of quantum theory.
   This is a key observation that will help guide the construction of 
a framework for quantum gravity within the context of a theory based 
upon generalised proper time. 
 Since there are also environments for which quantum and 
gravitational effects will both be significant the question of how 
they might be coherently combined in such a unified theory for the 
general case will need to be addressed.

%\pagebreak
\subsection{Quantum Theory with Gravity}
\label{qugr23}

   While there are well-known conceptual questions concerning quantum 
theory in itself, as reviewed in subsection~\ref{qugr21}, further 
issues arise in attempts to incorporate general relativity in a 
unified theory of quantum gravity (see for 
example~\cite{Kief,Carl1}). 
 The most direct approach would be to treat the metric deviations 
$h_{\mu\nu}(x)$ in equation~\ref{glin}, away from the flat
  metric $\eta_{\mu\nu}$ (or from a more general classical solution 
\cite{Kief} equation~16, for which linearised gravity can also be adopted as noted in \cite{Carr} after 
equation~6.3), as the field to be quantised.
 Gravitational radiation, in the form of equations~\ref{wavepol} and 
\ref{waveee}, would then be expected to be composed of `gravitons' as 
 massless (since $k^{\mu}k _\mu=0$), spin-2 (since the polarisation 
modes are invariant under $180^{\circ}$ transverse rotations, \cite{Carr} discussion following equation~6.69)  quanta of the 
gravitational field.
  Such gravitons would
 satisfy equations~\ref{ehbaro} and \ref{phbark}, similarly as for 
the case of photons as massless spin-1 quanta of the electromagnetic 
field.

 However for such attempts at a quantum theory of gravity new 
infinities appear at each order of perturbation and the theory is 
non-renormalisable (\cite{Kief} section~2). This problem arises as the 
gravitational coupling constant in equation~\ref{Eineq} has negative 
mass dimension, $[\kappa] = [\mbox{mass}]^{-2}$,
 unlike the case for electromagnetism and the interactions of the Standard 
Model of particle physics generally (\cite{Pesk} section~22.5, as 
also discussed in \cite{Unifi} after equation~10.86).
 This difficulty is central to the technical challenge in obtaining a 
consistent quantisation of gravity. In this subsection we review 
further issues to be addressed for any unification of quantum theory 
with general relativity.

  A seemingly more fundamental inconsistency is in the conception of 
time in quantum theory compared to that in general relativity.
  In quantum theory time is an independent background parameter, 
having roots in the Newtonian concept of time as 
       singled out and treated differently to the spatial dimensions,
 through which states evolve as described by the Schr\"{o}dinger 
equation~\ref{Schro} as also for QFT in equation~\ref{Schroq}, with 
the Hamiltonian operator generating infinitesimal translations of the 
quantum state through time.   
  In general relativity however the focus is on full 4-dimensional 
\mbox{spacetime} solutions for equation~\ref{Eineq} as reviewed in 
the previous subsection, with time itself a dynamical variable in 
this background independent theory.

  A specific time parameter could be defined in spacetime through the 
choice of a particular 3-dimensional spatial foliation. However such 
a structure is not invariant under diffeomorphisms, while classical 
general relativity is foliation-independent with different choices of 
3-space foliations equally permitted in describing the same region of 
4-dimensional spacetime, as might be employed in obtaining solutions 
to the `initial value problem' described in the previous subsection 
for example. 
 Further, in the canonical approach to quantising gravity, based upon 
a foliation of 3-dimensional spatial hypersurfaces, the classically 
constructed Hamiltonian exhibits constraints which at the quantum 
level imply 
 $\hat{H}\Psi = 0$ with $\Psi$ representing the quantum state of the 
whole universe (\cite{Kief} equations~20--22, \cite{Ander} equations~3 and 4). By 
comparison with equation~\ref{Schro} this is 
  an apparently timeless equation for which nothing happens, seemingly very much at odds with 
observations and non-trivial to interpret. 
  Under quantum fluctuations involving superpositions of possible 
3-space geometries, in a `foamy' spacetime, the concepts of causality 
and locality, as well as time, can also be ill-defined.

The technical difficulty in uniquely specifying a time evolution for 
the quantum state in general relativity, by consistently combining 
the one-dimensional and 4-dimensional aspects of the two theories, 
with its several related facets, is known as the `problem of time' in 
quantum gravity (\cite{Ander,Kuch,Isham}, \cite{Kief} section~4).
 Strategies for dealing with the problem of time range from 
internally identifying time as part of a background structure before 
quantising, hence maintaining a Schr\"{o}dinger picture of quantum 
theory~\cite{Brown}, to considering the concept of time itself to 
have a secondary or emergent phenomenological status in a 
fundamentally timeless theory~\cite{Rove1}.

  A further apparent stark incompatibility between quantum theory and 
general relativity is in the notion of vacuum energy. The Einstein 
equation~\ref{Eineq} can be augmented to include a
cosmological term as described for equation~\ref{einlamt}.
 This $\Lambda g^{\mu\nu}(x)$ term can instead be placed on the 
right-hand side of the equation, and 
 is present even in the absence of ordinary matter $T^{\mu\nu}(x)$, 
with the constant $\Lambda$ interpreted as the energy density of the 
vacuum.
 However the QFT expectation is for a vacuum energy of
   $\Lambda \sim m^4_P$, which  (with the Planck mass being
     $m_P \sim 10^{19}\,$GeV) is more than $10^{120}$ times too large 
to be consistent with observations (as discussed for
 \cite{Carr} equation~4.75). This `cosmological constant problem' for 
$\Lambda$ also motivates the need for a much better understanding of 
the relation of quantum theory to gravity, and in a manner that may 
relate to the origin and density of the apparent `dark energy' as 
inferred from observations of the large scale structure of the 
universe.   

  At the opposite end of the distance scale
  the nature of spacetime singularities, associated with black holes 
and also the Big Bang, is perhaps the most significant conceptual 
blind spot to a full understanding of general relativity in itself, 
as noted in the previous subsection. A compatible role for quantum 
theory in a consistent quantum gravity framework might be expected to 
help resolve this issue (\cite{Kief} section~5). In the meantime it 
is perhaps unreasonable to suggest that `general relativity predicts 
its own downfall' in singularities, since in approaching such a state 
the properties of matter generally, let alone a full theory of 
quantum gravity, are presently largely unknown for arbitrarily high 
densities and may act as a limiting factor to the collapse.

  In any case it is well known that matter can collapse down beyond 
the event horizon leading to the formation of a black hole
from the perspective of an external observer. While being well beyond 
the realms of empirical observation (barring the existence of 
primordial black holes of an appropriate mass) an understanding of 
Hawking radiation emitted from the vicinity of the event horizon 
and the properties of black hole thermodynamics (\cite{Carl1} 
section~V, \cite{Carl2}) is perhaps the most, and only, seemingly 
reliable theoretical application involving quantum theory together with general 
relativity. 
However rather than a full quantum gravity theory these results 
depend upon adapting the techniques of standard QFT, developed for 
flat spacetime, for a curved spacetime background. Hence, while such 
black hole properties are
 theoretically well-established having been derived in different 
ways, to some extent these theoretical phenomena can only represent 
an approximation to the complete physical picture, particularly for 
the extreme spacetime curvature involved in the final stages of black 
hole evaporation. 

  Associated with black hole thermodynamics is the `information 
paradox' in which `information' falling into a black hole is 
apparently ultimately lost in the thermal radiation of the 
evaporation.
   This corresponds to an apparent transition from an initial pure 
quantum state of matter entering the black hole to a final mixed 
state in the evaporation, an evolution which violates unitarity and 
hence contradicts quantum mechanics~\cite{Mat1,Mat2}. In place of the 
information being lost the possibility of it 
escaping through hidden correlations in the Hawking radiation or of 
being stored in the singularity, in principle for it to leak out 
through the radiation or in the final evaporative explosion, is 
sometimes proposed as a resolution. Another possibility is for 
unitarity to indeed be violated 
and the information to actually be lost in the extreme environment of 
a black hole, begging the question of the need for a complete theory 
of quantum gravity (see also \cite{Pen} section~30.8). After all, as 
it stands the `paradox' is a result of attempting to combine standard 
features of quantum theory and general relativity in an incomplete 
picture, providing a theoretical demonstration and input of 
\textit{requirements} to be met by a full quantum gravity theory. 
This provisional paradox is not in itself necessarily a problem 
unless or until a complete quantum gravity theory describing the 
nature of black hole singularities and evaporation is proposed that 
still cannot resolve this issue.

   Since the information paradox relates to unitarity in quantum 
theory there is a suggestion that the resolution may involve a 
reformulation, or new conception, of quantum phenomena for a general 
curved spacetime (see for example \cite{Carl2} section~10.1).
 It is also proposed in (\cite{Mat2} section~1) that `bypassing the 
paradox needs a basic change in our understanding of how quantum 
effects operate in gravity'. Here we consider that the identification 
of a novel basis for quantum theory in a consistent amalgamation with 
gravitation 
  might also address other issues discussed in this subsection. In 
the appropriate flat spacetime limit the calculations of standard QFT 
would need to then be reproduced, while in turn ideally the 
long-standing questions concerning the conceptual basis for quantum 
theory discussed in subsection~\ref{qugr21} might also be reassessed 
from this new perspective.

%\pagebreak
\subsection{Priority of `Quantum' or `Gravity'}
\label{qugr24}

    A typical approach to quantum gravity has been to take the scope 
of quantum theory as in some sense all-embracing and immutable to the 
extent of inevitably subsuming also a theory of gravity,
 as for example in the direct approach to quantising the 
gravitational field described in the opening of the previous 
subsection.
 Indeed the name `quantum gravity' itself to some degree carries the 
connotation of subsuming gravity under quantum theory (\cite{Pen} 
section~30.1).
 This can be motivated in part by the ambition of a broad unification 
under the guiding principle that `everything is quantum', as we 
alluded to in section~\ref{qugr1}, consistent with `an appeal to the 
unity of physics' (\cite{Carl3} section~2). While significant 
technical challenges remain
 this is arguably the case for string theory, in seeking to unify all 
interactions with gravitons identified as quantum excitations of 
closed strings (\cite{Kief} section~2.3, \cite{Carl1} section~III), 
and also for loop quantum gravity, in attempting to quantise the 
structure of spacetime itself (\cite{Kief} section~2.2.2, 
\cite{Carl1} section~IV).

     The success of such an approach in consistently quantising 
gravity would indeed probably lead to a more unified worldview, but 
would still beg the question of the origin of the quantisation 
formalism  itself and may not even touch upon many of the conceptual 
issues with quantum theory reviewed in subsection~\ref{qugr21}. There 
would also remain the further unification question of subsuming the 
specific symmetry and multiplet structure of the Standard Model of 
particle physics, which is empirically well-established at the 
microscopic scale dominated by quantum phenomena. 
  The ambition of accounting for the Standard Model 
 is faced with the `landscape problem' in string theory~\cite{Doug} 
while not being addressed at all in loop quantum gravity (\cite{Rove} 
section~III opening).
 With string theory having been initially developed in a context very 
different to quantum gravity there remains also more generally an 
open issue concerning for what physical applications string theory 
might actually be of most value (see for example~\cite{Duff} 
section~5, \cite{Alva} section~8 opening).

   If it had proved possible to quantise gravity, with say a similar 
level of technical difficulties as encountered in quantising the 
electromagnetic or other internal gauge interactions, it is still 
perhaps unlikely that such a theory would lead to phenomenology that 
could be empirically tested. This is due to the weakness of gravity 
on the laboratory scale and the feebleness of individual gravitons as 
the hypothetical quanta of the gravitational interaction. A theoretically consistent 
picture of black hole singularities and thermodynamic properties 
might be attained, addressing for example the information paradox 
discussed at the end of the previous subsection as an encouraging 
feature, but would be beyond the realms of observational 
confirmation. 

   Predictions concerning the nature of the initial singularity and 
the very early evolution of the universe emerging from the Big Bang 
might conceivably have observable implications in cosmology,
 although a decisive test 
   could prove to be very challenging 
(see for example \cite{Carl1}  section~VI). 
 In terms of possible observable cosmological effects and signals there are 
proposals linking quantum gravity models with inflation, the dark 
sector, gravitational waves and gamma-ray bursts (see respectively 
for example \cite{Ansel,Berg,Calc,Jacob} and references therein). 
 While there are also proposals for laboratory experiments capable of 
determining whether gravity is actually quantised again there are 
significant difficulties to be addressed (see for 
example~\cite{Haine,Faure,Howl,Kamp}). 
 On the other hand it can be argued that given the inability to 
directly test theories of quantum gravity to date other factors have 
led to the prominence of a small number of possible approaches as 
``today's working scientists fight for and achieve objectivity by 
meeting intersubjective standards for consensus'' as a measure of 
scientific validity~\cite{Gilb}.

  Despite these empirical challenges a theoretically successful 
consistent quantisation of gravity may have given a false sense of 
security or completeness in achieving a unification of quantum theory 
with general relativity. This in turn might have focussed more 
attention back on the conceptual basis of quantum theory itself, 
including the features of indeterminacy, non-locality and 
wavefunction reduction reviewed in subsection~\ref{qugr21}. The 
application of quantum theory, with these features, would still rely 
on a set of \textit{postulates} and seemingly ad hoc rules for their 
employment, such as in utilising the Copenhagen interpretation, all 
remaining to be better understood. Effectively quantisation of 
gravity in principle \textit{can only} address the issue of combining 
quantum theory with gravity, while the underlying `quantum' features 
of the world seemingly still need to be accepted as `brute facts' 
about the universe, as a somewhat unsatisfactory loose end. 
 The view could be taken not to raise questions concerning these 
features of quantum theory, or even to consider them to be 
unanswerable, essentially perpetuating a `shut-up-and-calculate' 
approach (\cite{Tegk} section~IV.A).

   On adopting a less positivist perspective
   at heart it is a question of what it is `that really exists'. From 
the outset quantum mechanics has been motivated by observations and 
developed pragmatically to make calculations for comparison with 
laboratory measurements, incorporating a theoretical construction to 
determine the relative probability of events. As we reviewed in 
subsection~\ref{qugr21} this involves the continuous unitary evolution {\bf 
U} of a wavefunction $\Psi$ of seemingly no direct physical 
significance, together with irreversible changes at instants of 
wavefunction reduction {\bf R} which do seem to have a real objective 
meaning in corresponding to physical measurements
  (although these statements are a matter of interpretation).
 The conceptual problems are apparently amplified in attempting to 
extend the wavefunction description to incorporate macroscopic matter 
in a uniform manner, leading to the debate over the fate of 
Schr\"{o}dinger's cat for example (see also \cite{Pen} sections~29.7--29.9). 
 This already strongly suggests that quantum theory is 
\textit{not universal}, diminishing the motivation for attempting the 
further extension towards the quantising of gravity or the structure 
of spacetime itself.

 Thought experiments involving an assumed extrapolation of quantum 
properties beyond the domain of empirical evidence, such as in 
constructing a superposition of gravitational fields associated with 
a superposition of differing locations for a macroscopic `lump' of  
matter as a variation of Schr\"{o}dinger's feline experiment (see for 
example \cite{Pen} figure~30.20),  
  are sometimes used as the basis for arguments that gravity must 
ultimately itself be subject to a formalism of quantisation. However 
it can be
 `shown that all these arguments are of a heuristic value and that they 
do not lead to the quantisation of gravity by a logical conclusion' 
(\cite{Kief} section~1). The need to quantise gravity is also 
questioned for example in \cite{Hugg,Wuth}. 
   
   In general relativity a \textit{free} particle of matter follows a 
geodesic path through a curved spacetime, with a trajectory hence 
determined by gravity alone. This conception of gravity is then 
\textit{not} as a `force' as for the Standard Model gauge 
interactions, but rather as merely a direct consequence of the geometric 
curvature of the extended spacetime itself. The gravitational field 
describes this \textit{external} geometry of spacetime as the arena 
for all events, and is hence quite unlike the \textit{internal} gauge 
fields such as associated with electromagnetism and
 particle interaction events.  
   This significant distinction is itself reason to question the aim 
of quantising gravity in accord with the strong, weak and 
electromagnetic Standard Model interactions.
 That there is no experimental or observational evidence suggesting 
any discreteness or quantum behaviour of the gravitational field or 
spacetime, as well as the theoretical difficulties in submitting  
gravity to quantisation in any wholly satisfactory manner, all adds 
to the doubt over the \textit{need} to quantise gravity. 
There remains of course however the ambition of a coherent theory of 
`quantum gravity' from which the familiar structures of quantum 
theory and classical general relativity as applied in practice both 
consistently derive as limiting cases.
  
   The fact that it has proven technically far harder to quantise 
gravity compared with the other forces of nature, together with
 the differing characteristics of the gravitational field and also
 a desire to address the internal conceptual issues in the quantum 
realm, suggests that rather than adapting, or tinkering with, the 
formalism of quantum mechanics there is an opportunity to consider a 
more decisive shift towards a new theoretical framework that might 
\textit{subsume}, and provide a better understanding for the origins 
of, quantum theory itself. A more coherent quantum theory together 
with a consistent theory of quantum gravity
  might then be sought by placing the initial focus on gravity, which 
in itself exhibits fewer conceptual problems as we discussed in 
subsection~\ref{qugr22}.

 In assigning the priority more on the side of gravitation in such a 
theory of quantum gravity consistency with general relativity might 
be assured while also seeking to reproduce all observed quantum 
phenomena in the appropriate limits. 
This would be analogous to the advance from Newton's mechanics and 
law of universal gravitation in the conceptual leap to special and 
general relativity, providing a more complete and unifying picture 
from which the Newtonian worldview can be recovered in the 
non-relativistic limit.
  As well as being fully relativistic such a unified quantum gravity 
theory should be well-defined and lacking in arbitrariness, while 
also being consistent with all observations in general from the 
largest scale surveyed in cosmology to the smallest scale studied in 
the laboratory.
 Ideally an explanation for the origin of the specific structures of 
the Standard Model of particle physics should also be incorporated.

  In this paper we describe such a unifying framework. Unlike string 
theory or loop quantum gravity, which originate directly with a high 
level of technical mathematical sophistication, here the approach is 
to begin rather with a simple and intuitively natural conceptual 
picture that can be expressed in elementary mathematical terms.
Rather than beginning with the notion that `everything is quantum', 
as considered in the opening of this subsection, here the unifying 
principle is essentially that `everything happens in time'.
 This foundation will be argued to be clearly and uniquely defined 
and upon which a more complete and sophisticated mathematical 
formalism for the theory can then be developed.  
 The manner in which the theory can be considered a further 
generalisation from general relativity, rather than with the priority 
on extending the quantum domain, is described in (\cite{Gener} 
subsection~5.1).

  In the following section we review the motivation and basis of this 
approach founded upon `generalised proper time'.  From 
section~\ref{qugr4} of this paper we describe how through the 
identification of an extended spacetime structure and an associated 
matter field content this theory can convolve the properties of 
general relativity with quantum theory in a manner addressing many of 
the conceptual issues reviewed in this section.

%\pagebreak
\section{Generalised Proper Time as a Unifying Basis}
\label{qugr3}

    In general relativity the Einstein tensor $G^{\mu\nu}(x)$, as 
described for equation~\ref{Eineq}, is a function of the spacetime 
derivatives of the metric tensor $g_{\mu\nu}(x)$ in 
\mbox{4-dimensional} spacetime. In a system of general coordinates 
$\{x^{\mu}\}$ a local infinitesimal interval of proper time $\delta 
s$ can be expressed as 
 $(\delta s)^2 = g_{\mu\nu}(x)\delta x^{\mu}\delta x^{\nu}$. By the 
equivalence principle of general relativity (as alluded to following 
equation~\ref{Eineq}) a set of local coordinates $\{x^{a}\}$ for a 
local inertial reference frame can be identified at any point for 
which this interval can be expressed as:
\begin{equation}
  \label{sfourd}
  (\delta s)^2 \; = \; \eta_{ab}\delta x^{a}\delta x^{b}
\end{equation}
  with the Lorentz metric $\eta = \mbox{diag}(+1,-1,-1,-1)$ and $a,b 
= 0,1,2,3$.
 Local Lorentz transformations then relate the possible choices of 
such local inertial coordinate frames.

   Theories with extra spatial dimensions typically conceive of our 
familiar world of
  \mbox{4-dimensional} spacetime as an embedded submanifold within
   an extended higher-dimensional bulk space that may or may not be 
compactified down to the preferred extended 4-dimensional spacetime
 (\cite{Gener} opening of subsection~2.2 and references therein).
 In all cases at a local level such theories augment the structure of 
equation~\ref{sfourd} to that of an $(n>4)$-dimensional spacetime for 
which a proper time interval then takes the form
  $(\delta s)^2 \: = \: \hat{\eta}_{ab}\delta x^{a}\delta x^{b}$ now
  with  $a,b = 0,\ldots,(n-1)$ and $\hat{\eta} = 
\mbox{diag}(+1,-1,\ldots,-1)$ the augmented local Lorentz metric for 
$n$-dimensional spacetime.

   For the present theory we further generalise this local form for 
proper time on noting that as well as relaxing the restriction to 
$n=4$ we can also drop the assumption of a quadratic $p=2$ structure 
and augment equation~\ref{sfourd} to:
\begin{equation}
 \label{salpha}
  (\delta s)^p  \; = \; \alpha_{abc\ldots}\delta x^a 
                            \delta x^b \delta x^c \ldots
\end{equation}
  This expresses the possibility for a homogeneous polynomial power 
$p>2$, with indices $a,b,c,\ldots$ running over the full set of $n$ 
components and with the coefficients \mbox{$\alpha_{abc\ldots} \inn 
\{-1,0,1\}$} further generalising the local Lorentz metric structure. 
This further augmentation is permitted since a $p=2$ 
\textit{quadratic} structure is \textit{only} required for the 
original four components of the external spacetime, including the 
three components that we actually \textit{perceive} as 
\textit{spatial} dimensions, as we shall discuss further below.

  The above 
expression for infinitesimal intervals can be written more 
conveniently in terms of the generally finite components
 $v^a := \frac{\delta x^a}{\delta s} 
          {\big{\vert}}_{\mbox {\tiny $\delta s \! \to \! 0$}}$ of an 
$n$-vector $\bv_n \in \rrr^n$ on dividing both sides of 
equation~\ref{salpha} by $(\delta s)^p$ and defining:
\begin{equation}
  \label{lpvn}
  L_p(\bv_n)_{\hat{G}} 
  \; := \; \alpha_{abc\ldots} \frac{\delta x^a}{\delta s}
   \frac{\delta x^b}{\delta s}\frac{\delta x^c}{\delta s}\ldots
   \Big\vert_{\delta s \to 0} \; = \;
    \alpha_{abc\ldots}v^a v^b v^c \ldots \; = \; 1
\end{equation}
  Here again $p$ is the homogeneous polynomial power and $n$ is the 
total number of components while $\hat{G}$ is the full symmetry 
group, generalising from the Lorentz transformations of 
equation~\ref{sfourd} for this generalised form of proper time.

  It is a very evident fact about the universe that we observe the 
material flux of the physical world in \textit{space} as well as 
through \textit{time}, with matter and events presented to us through 
an extended 3-dimensional spatial arena. At the most elementary local 
level the basis for the Euclidean properties of space is encapsulated 
in the three spatial components of equation~\ref{sfourd} which
 incorporates an infinitesimal interval of 3-dimensional space 
$\delta \Sigma$ expressed as 
 $(\delta \Sigma)^2 = (\delta x^1)^2+(\delta x^2)^2+(\delta x^3)^2$
  (entering equation~\ref{sfourd} with an overall minus sign from the 
Lorentz metric signature convention). 
 
 However the assumption of this consistency with the Pythagorean 
theorem, together with its direct \textit{spatial} interpretation, 
can be dropped for any extra components, beyond the 4-dimensional 
form of equation~\ref{sfourd}.
  This is the case since \textit{we do not perceive} the additional 
components in the form of \textit{any} spatial or geometrical 
structure.
 Hence the generalisation to the \mbox{$n$-dimensional} form of 
equations~\ref{salpha} and \ref{lpvn} is permitted \textit{provided} 
there is a term in the expansion of the right-hand side that contains 
a factor of the 4-dimensional form $\eta_{ab}\delta x^a\delta x^b$ of 
equation~\ref{sfourd}. While in this theory we begin with the 
\mbox{one-dimensional} continuum of time $s \in \rrr$, which exhibits 
a direct elementary substructure for an infinitesimal interval 
$\delta s$ via the basic arithmetic composition of 
equation~\ref{salpha},
 such a factor of  $\eta_{ab}\delta x^a\delta x^b$  can be identified 
with the local metric $\eta_{ab}$ and local coordinates 
$x^a \in \rrr^4$ forming a local basis for the
geometric structure that we \textit{do} perceive as the
 continuum of an external 4-dimensional spacetime manifold $M_4$.   
   
  The necessary extraction of the local
   \textit{external}
   Lorentz symmetry acting on the  subcomponents 
   $\bv_4 = (v^0,v^1,v^2,v^3) \in \TM_4$ projected onto the 
4-dimensional  tangent space from the full form of 
equation~\ref{lpvn} both breaks the full symmetry $\hat{G}$ and 
fragments the full set of
 $\bv_n \in \rrr^n$ components.  Subgroups identified within the 
residual \textit{internal} symmetry $G$ can be associated with  gauge 
fields $A(x)$ on $M_4$ while the fragmented components of $\bv_n(x)$ 
transforming under irreducible representations of the broken 
symmetry:
\begin{equation}
 \mbox{Lorentz} \times G \, \subset \, \hat{G}
 \label{gbreak}
\end{equation}
  are associated with further matter fields on $M_4$.
  
  For this theory, without supposing any preconceived or pre-existing 
external \mbox{4-dimensional} spacetime manifold, the geometric 
construction of the extended arena $M_4$, in extrapolating  beyond the 
local infinitesimal form of $\eta_{ab}\delta x^a\delta x^b$, cannot 
be separated from the constraints implied in equation~\ref{lpvn}
 and the resulting associated matter fields described for 
equation~\ref{gbreak}.
 That is, rather than \textit{beginning} with spacetime $M_4$ and 
\textit{then} introducing matter fields and proposing physical laws, 
here all physical properties of \textit{matter} as observed in 
\textit{spacetime} derive directly through their \textit{mutual} 
origin in the more basic foundation of generalised proper 
\textit{time}. 
  That the underlying simplicity in being founded upon proper time 
alone implies stringent constraints on the forms of matter in 
spacetime obtained, hence with the potential for significant 
explanatory and predictive power,
 is a key point for the present theory. 
This construction necessarily implies a connection between the 
external geometry on $M_4$, as described by the Einstein tensor 
$G^{\mu\nu}(x)$, and the internal gauge fields $A(x)$ as well as the 
fragmented components of $\bv_n(x)$, as expressed respectively and 
generically as:
\begin{eqnarray}
 G^{\mu\nu} & = & f(A)  \label{gefa}  \\
 G^{\mu\nu} & = & f(\bv_n)  \label{gefv}
\end{eqnarray}  
 Here $f\equiv f^{\mu\nu}$ represents tensor functions of the field 
components.
   The specific form of these equations mutually constrains both the 
possible geometry \textit{of} spacetime and the apparent forms of 
matter observed \textit{in} spacetime, with the energy-momentum 
tensor $T^{\mu\nu}(x)$ for all matter essentially \textit{defined} 
through the Einstein equation~\ref{Eineq} as will be described for 
equation~\ref{gfromavt}.

  As alluded to near the end of subsection~\ref{qugr24}, with the 
initial priority placed more upon the gravitational than the quantum 
side, the historical connection of this theory with general 
relativity and early unified field theories is presented in 
(\cite{Gener} subsections~1.2 and 5.1, \cite{Ufield}), with the 
underlying simplicity of this approach in deriving from the `one 
dimension' of proper time alone as encapsulated in `one simple equation' discussed in  (\cite{Gener} 
subsection~5.2). The conceptual picture regarding the relation 
between the substructure of this one dimension of time and the 
structure of 4-dimensional spacetime and the matter it contains is 
further expounded in~\cite{Struct}. The basic motivation behind 
equations~\ref{salpha} and \ref{lpvn} is described further in 
(\cite{Gener} subsections~2.1 and 2.2), in particular through 
comparison with the theories based on extra spatial dimensions. 

  A specific form for equation~\ref{gefa} for the general case of 
equation~\ref{lpvn}, for a gauge field associated with the internal 
symmetry $G$ of equation~\ref{gbreak}, is obtained via links with 
Kaluza-Klein theories~\cite{KKone} as will be reviewed here for 
equations~\ref{fofa} and \ref{gchift}. 
  The symmetry breaking projection of the subcomponents $\bv_4 \in 
\TM_4$ out of equation~\ref{lpvn}, again for the general case, 
impacts the spacetime geometry as represented by equation~\ref{gefv} 
and in turn, via the implied energy-momentum, can  be interpreted as 
the basis for the `origin of mass'. The projected $\bv_4 \in \TM_4$ 
components are correspondingly associated with the role of the Higgs 
field in the Standard Model, 
as will be described more explicitly for 
equations~\ref{gwarph}--\ref{lpvnb} also in subsection~\ref{qugr51} and utilised in subsection~\ref{qugr62}.

  The unique development of the theory with specific mathematical 
forms for equation~\ref{lpvn} through to $\hG = \esi$ (with $p=3$, 
$n=27$) and $\hG = \ese$ (with $p=4$, $n=56$) leads to a symmetry 
breaking structure for $\bv_n(x)$ fragments making direct connections 
with a series of esoteric properties of the Standard Model of 
particle physics, as described extensively in (\cite{Unifi} 
chapters~6, 8 and section~9.2, \cite{Novel} \mbox{sections~4--6}, 
\cite{TimeE} section~4, \cite{Gener} subsection~3.1). These 
connections include the identification of spinor structures, strong 
$\suth_c$ singlets and triplets, electromagnetic $\uo_Q$ fractional 
charges and an intrinsic left-right asymmetry; while further elements 
of electroweak theory can be identified that further justify the 
association of the projected $\bv_4 \in \TM_4$ components with the 
Higgs sector (as described in \cite{Unifi} section~8.3).
 In this manner significant elements of the Standard Model are 
derived from generalised proper \textit{time} far more directly than  
can be obtained in models with extra \textit{spatial} dimensions (see 
for example \cite{Gener} section~6 first two pages). 

   On the technical mathematical side an intrinsic role for the 
octonions, the largest division algebra, in the construction of the 
$\esi$ and $\ese$ levels of forms for proper time and their 
corresponding relevance for particle physics is expounded 
in~\cite{Octo}. This progression through the exceptional Lie groups 
leads to a prediction of an octonion-based full $\hG = \ee$ 
construction for equation~\ref{lpvn}, in principle with \mbox{$p=8$ 
and $n=248$} and hence of the form (\cite{Octo} equation~9):
\begin{equation}
 \label{lvto}
 \lvtfep
\end{equation}
 with an anticipated symmetry breaking pattern capable of reproducing 
the full Standard Model three-generation multiplet structure 
(\cite{Unifi} section~9.3, \cite{TimeE}, \cite{Gener} 
subsection~3.2). Constraints on the structure of the mathematical form 
of this application for the largest exceptional Lie group
 $\ee$, for which properties of the octonion algebra are anticipated 
to be essential, 
  in turn lead to preliminary predictions for physics beyond the 
Standard Model in the Higgs, neutrino and dark sectors as described 
in (\cite{Gener}~section~4, \cite{Ufield} section~6).

      In this paper we focus on how the amalgamation and direct 
generalisation of equations~\ref{gefa} and \ref{gefv} leads to a 
framework that can be interpreted as a theory of quantum gravity, 
reviewing and building upon the presentation in \mbox{(\cite{Unifi} 
chapters~10 and 11)}. The initial discussion in the following section 
will make the case for convolving the essential empirical properties 
of general relativity and quantum theory together within this 
conceptual context of constructing matter in 4-dimensional spacetime 
through a generalised form for proper time. A key test will be the 
quantitative reproduction of laboratory phenomena in a quantum field 
theory limit, with a mathematical basis for this direction presented 
in sections~\ref{qugr5} and \ref{qugr6}. We then return to the 
conceptual issues raised in section~\ref{qugr2} and describe the 
extent to which they can be addressed within the new framework in 
section~\ref{qugr7}, before concluding with further comments in the 
final section.

%\pagebreak
\section{Degeneracy of Solutions for 4-Dimensional Spacetime}
\label{qugr4}

\subsection{Identifying Gravitational and Quantum Phenomena}
\label{qugr41}
     
	 Neither equation~\ref{gefa} nor equation~\ref{gefv} implies a 
unique geometry $G^{\mu\nu}(x)$ for an extended 4-dimensional 
spacetime $M_4$. Rather an endless range of possible solutions for a 
continuous spacetime geometry $G^{\mu\nu}(x)$ as a function of 
continuous gauge fields in  components of $A(x)$ or of continuous 
fields in subcomponents of $\bv_n(x)$ can be identified (an example 
will be given in equations~\ref{tmnem}--\ref{wavecom} for an 
electromagnetic plane wave). The prime and essential requirement is 
the construction of an extended and coherent arena in space, as well 
as in time, as a continuous and smooth external manifold through 
which we perceive events in the world.
 However this overriding global demand for the 
    external spacetime does \textit{not} impose a condition of 
continuity upon the internal fields.
   Hence with
    no such \textit{a priori} requirement for fields in components of 
$A(x)$ and $\bv_n(x)$ to be everywhere continuous there are vastly 
more solutions for a continuous external spacetime permitted on 
combining equations~\ref{gefa} and \ref{gefv} in the generalised 
form: 
\begin{equation}
 \label{gfromavt}
  G^{\mu\nu} \: = \: f(A,\bv_n)   \: =: \:  -\kappa T^{\mu\nu}   
\end{equation}  
   Through this expression the energy-momentum tensor $T^{\mu\nu}(x)$ 
is also \textit{defined}, as alluded to after equation~\ref{gefv}  
(see for example \cite{Unifi} equations~5.32 and 15.1, \cite{Struct} 
equation~9).

  This construction manifestly incorporates the Einstein field 
equation~\ref{Eineq} in representing full 4-dimensional  solutions 
for a continuous spacetime geometry as for the classical theory of 
general relativity. On the other hand the central function 
$f(A,\bv_n)$ in equation~\ref{gfromavt} will be characterised by a 
\textit{degeneracy} of local  field functions in subcomponents of $A(x)$ and $\bv_n(x)$ 
that describe the \textit{same} local geometric structure 
$G^{\mu\nu}(x)$ and corresponding gravitational field $g_{\mu\nu}(x)$ 
metric solution for the spacetime manifold $M_4$.
 While the properties of the global 4-dimensional spacetime solutions 
encapsulate general relativity, the local degeneracy in internal field contributions  
is proposed to underlie the phenomena of quantum theory for all 
non-gravitational fields.
 The mathematical constraints on this degeneracy, largely deriving 
from equation~\ref{lpvn} as a direct generalisation from the one dimension 
of proper time, and their main consequences will be described in 
sections~\ref{qugr5} and~\ref{qugr6}.

At the level of an infinitesimal interval the one-dimensional 
continuum of time intrinsically contains the arithmetic substructure 
of equations~\ref{salpha} and \ref{lpvn}. This inherent substructure 
of time itself in turn underlies the \textit{construction} and 
\textit{degenerate composition} of 4-dimensional spacetime, with a 
geometric structure and mutually associated matter fields described 
in equations~\ref{gefa} and \ref{gefv} as combined in 
equation~\ref{gfromavt}.
The local degeneracy is a natural consequence of the large degree of 
redundancy entailed in building an extended 4-dimensional spacetime
 from substructures of the $n$-dimensional form for proper time in 
equation~\ref{lpvn}, which provides far more than a minimal means of 
constructing the spacetime manifold.
Here quantum theory is not something that happens \textit{in} 
spacetime, rather quantum phenomena are an intrinsic feature 
resulting from identifying spacetime \textit{itself} through 
substructures of the general form of time of equation~\ref{lpvn}
with an inherent multiplicity of possible solutions for equation~\ref{gfromavt}.

 The distinctive probabilistic nature of quantum events here 
essentially arises through the relative `number of ways' that a given 
local spacetime geometry can effectively be \textit{composed out of} 
the internal  $A(x), \bv_n(x)$ field components.  
  By comparison the nature of classical probabilities concerns the 
relative `number of ways' that events can happen \textit{in} 
spacetime.
Hence while bringing general relativity and quantum theory together 
in the single framework of generalised proper time, the role of a  
degeneracy count in determining the likelihood of spacetime   
solutions for equation~\ref{gfromavt} also brings
 the notion of probability in the quantum realm  much closer to that 
in classical physics. This is in contrast with the probabilities 
obtained with the standard postulates of quantum theory as based on 
the squared modulus of a complex wavefunction or amplitude
  (a means of linking the two methods of calculation will be 
developed for equation~\ref{pdddds}).

   We noted in subsection~\ref{qugr21} that the difficulty in 
interpreting quantum theory in terms of physically real entities in 
space and time strongly suggests that the world is not objectively 
built out of wavefunctions propagating in a background spacetime 
arena. 
 The associated conceptual problems are here considered to arise 
through introducing such entities \textit{into} spacetime and then 
imposing quantisation `postulates' to match empirical observations. 
In the present theory we do not begin by positing 4-dimensional 
spacetime as a pre-existing arena for physical entities, rather here 
the founding basis is simply a one-dimensional progression in time.

 With the original ordered one-dimensional progression in time 
underlying the 4-dimensional spacetime solutions for 
equation~\ref{gfromavt} the degeneracy of the latter implies a 
corresponding causal progression of inherently indeterministic events 
characterised by probabilities for local exchanges in matter fields 
that shape the overall solution for the extended spacetime geometry. 
This temporal accumulation of probabilistic events is proposed to 
underlie the distinguished role for time in quantum theory as 
reviewed for equations~\ref{Schro}, \ref{Schroq} and \ref{uevolve}. 
  In this manner with an intrinsic degeneracy of the matter fields in subcomponents of
$A(x)$ and $\bv_n(x)$ built into
 the construction of the spacetime manifold the corresponding 
apparent quantum phenomena are sewn into the fabric of the 
4-dimensional spacetime continuum at the most elementary level.

  Hence in principle this construction provides a coherent framework 
for combining the distinguished role for one-dimensional temporal 
evolution of quantum systems together with the global nature of 
4-dimensional spacetime solutions for equation~\ref{Eineq} in general 
relativity. In being free in principle from conceptual contradictions 
this framework then provides a basis for developing the full 
mathematical expressions and incorporating the full range of  
empirical phenomena associated with both the quantum and 
gravitational realms for these two central theories of fundamental 
physics as developed in the $20^{\mathrm{th}}$ century.

On the largest scale our own universe is considered to consist of 
\textit{one} of \textit{very many} possible solutions for a full 
extended 4-dimensional spacetime with a continuous and smooth 
geometry consistent with equation~\ref{gfromavt} which on the 
macroscopic scale reproduces the phenomena of classical general 
relativity. As we perceive events in spacetime this single solution 
is infused with a local degeneracy or `slippery ambiguity' in field 
components of $A(x)$ and $\bv_n(x)$, describing the \textit{same} 
local geometry, leading to processes that given all preceding 
observations are indeterministic and of a probabilistic nature, as 
becomes evident as we delve into the microscopic substructure of 
matter.  
 That there is no contradiction between the Einstein field 
equation~\ref{Eineq} and events in spacetime with an intrinsic 
element of indeterminacy will be discussed explicitly in the opening 
paragraphs of subsection~\ref{qugr73}.
 It is this local degeneracy in the identification of the external 
spacetime geometry through equation~\ref{gfromavt} that underlies the 
apparent `quantisation' of all non-gravitational fields associated 
with the various subcomponents of $A(x)$ and $\bv_n(x)$.

  The geometry of the 4-dimensional spacetime manifold $M_4$ 
\textit{defines} the light cone structure through which the causal 
progression of events as perceived \textit{in} spacetime is 
channelled, with no signals travelling faster than the speed of light 
through spacetime, as discussed towards the end of 
subsection~\ref{qugr22}.
 On the other hand in representing a global extended solution for the construction of spacetime itself, consistent with the structures of 
Riemannian geometry, equation~\ref{gfromavt} exhibits an 
intrinsically \textit{non-local} nature. This feature will be 
associated with the apparent non-locality of quantum phenomena 
reviewed in subsection~\ref{qugr21}, as we shall argue in 
subsections~\ref{qugr73} and \ref{qugr74}, where the correlate of 
`wavefunction reduction' in the present theory will also be 
described.

    In obtaining solutions for $G^{\mu\nu}(x)$ and the metric 
$g_{\mu\nu}(x)$ from equation~\ref{gfromavt} partial information will 
be provided in practice by our knowledge of macroscopic 
energy-momentum $T^{\mu\nu}(x)$, in particular in the terrestrial 
environment for which the approximations of a flat spacetime limit 
can be adopted. On the small scale in the laboratory this will 
include the measurement apparatus, which itself constitutes a fixed 
element of $T^{\mu\nu}(x)$ in spacetime and acts as a `boundary 
condition' to a full geometric solution for  equation~\ref{gfromavt}, 
closely analogous to the employment of boundary conditions in general 
relativity as discussed in subsection~\ref{qugr22}. Here the 
\textit{potential} for exchanges in components of the fields $A(x), 
\bv_n(x)$ and the corresponding likelihood of such interactions, 
consistent with the constraints and local degeneracy, will act as 
further significant boundary conditions for equation~\ref{gfromavt}.
  The setup of the apparatus will hence determine the set of boundary conditions and in turn constrain the nature of the observations that can be made in an experiment.

  Equation~\ref{gfromavt} can also be written more compactly as:
\begin{equation}
 \label{gavt}
  G^{\mu\nu} \: = \:   -\kappa T^{\mu\nu}(A,\bv_n)   
\end{equation}  
  to emphasise the implicit field composition of the identified 
energy-momentum tensor. At the macroscopic level an effective 
coarse-grained expression for $T^{\mu\nu}(x)$ will generally be 
sufficient to describe the corresponding continuous spacetime 
geometry $G^{\mu\nu}(x)$ (see for example \cite{Unifi} equation~5.35 
for a `dust cloud' and equation~5.37 for a `perfect 
fluid'). While $G^{\mu\nu}(x)$ is everywhere smooth and continuous on 
all scales in this theory, at the microscopic level the explicit form 
for $T^{\mu\nu}(x)$ in terms of components of $A(x)$ and $\bv_n(x)$ 
and their possible exchanges and interactions will become evident, 
giving rise to the apparent elementary structure of matter. The 
specific structure of this composition will be determined and 
constrained by the specific form of proper time for 
equation~\ref{lpvn}; with the pattern of symmetry breaking  over the 
local spacetime $M_4$ leading directly to the elementary physical 
properties of matter and the interactions through which it is 
observed. 

As noted towards the end of section~\ref{qugr3} there is a unique 
progression in expressions for equation~\ref{lpvn} beyond a local 
4-dimensional spacetime form, through the cases of $\hG = \esi$ and 
$\hG = \ese$, which indeed inevitably yield structures strongly 
resembling those of the Standard Model of particle physics. 
 While the familiar structures of the Standard Model are not 
\textit{imposed} in this theory they naturally provide a guide for 
the features to look for in the symmetry breaking of the
 $\hG = \esi$ and $\hG = \ese$, and ultimately $\hG = \ee$, forms for 
proper time. Similarly while we do not \textit{impose} any postulates 
of quantum theory here such a standard formalism and its practical 
application can help inform the exploration of the associated 
mathematical structures that arise naturally in the present theory in attempting to 
account for the corresponding empirical phenomena.
 At the same time there will be an opportunity to acknowledge novel 
insights that might be gained on adopting the new perspective.

  Both sides of equation~\ref{gavt} refer to the \textit{same} overall structure, with 
the Einstein tensor
 $G^{\mu\nu}(x)$ describing the spacetime geometry as a function of 
the metric and the associated energy-momentum tensor $T^{\mu\nu}(x)$ 
accommodating both the classical and quantum properties of matter.
To also `quantise' the elements of gravity $G^{\mu\nu}(x)$ on the 
left-hand side of equation~\ref{gavt} would be effectively to 
quantise the \textit{same} object twice in two different ways, and 
with no natural mechanism to justify the quantisation of 
gravitational phenomena.  
  Equations~\ref{gfromavt} and \ref{gavt} describe how \textit{all} 
matter is identified through the composition and construction of the 
4-dimensional spacetime geometry on all scales, with the 
identification of the gravitational field $g_{\mu\nu}(x)$ enveloping 
all matter and essentially \textit{prescribing} the quantisation of 
matter through the composite structure of the corresponding apparent 
energy-momentum tensor $T^{\mu\nu}(x)$.

  Gravity is hence a property of all matter while quantum effects, in 
being subsumed under the geometric structure, are limited in scope 
and scale rather than being of universal applicability.
Quantum phenomena are not events that happen \textit{in} spacetime 
but are rather fused within the structure of general relativity as an 
intrinsic feature of the manner in which 4-dimensional spacetime 
\textit{itself} is constructed. 
 Equations~\ref{gfromavt} and~\ref{gavt} can be interpreted both as 
the \textit{origin} of quantum theory and also as effectively 
\textit{limiting} quantum phenomena to the microscopic realm of the 
physical world.

 Such quantum effects, far removed from the scale of everyday life, 
were only uncovered through laboratory experiments dating from the 
late $19^{\mathrm{th}}$ and early $20^{\mathrm{th}}$ centuries. These 
phenomena were subsequently described by quantum mechanics and then 
quantum field theory, which will both be considered here as effective 
theories applicable under the appropriate limiting approximations.
 While the internal conceptual issues with quantum theory reviewed in 
subsection~\ref{qugr21} have never been fully resolved on attempting 
to extend the quantum realm to macroscopic or gravitational 
phenomena, here we very much go the other way and propose to address 
these questions on effectively subsuming quantum theory under general 
relativity and an extended continuous spacetime geometry as 
constructed through the general form of proper time.    
 The implications of this framework regarding the conceptual 
interpretation of quantum theory both in the QFT and non-relativistic 
limits, for example in connection with Schr\"{o}dinger's cat and the 
double-slit experiment, will be elaborated in section~\ref{qugr7}. 

  The tangible spacetime form of an elementary particle, as observed 
in laboratory experiments, will arise from the microscopic 
substructure of solutions to equation~\ref{gavt}. In  
sections~\ref{qugr5} and \ref{qugr6} we shall describe a  means of 
developing a rigorous mathematical account of these elementary 
phenomena, in particular in connection with the quantum field theory 
limiting approximation of particle states interacting in a flat 
spacetime arena. In the meantime in the following subsection we first 
describe the \textit{concept} of an elementary particle as suggested 
by the present theory.

%\pagebreak
\subsection{Identifying Particle States and HEP Processes}
\label{qugr42}

  For the present theory the properties of the various types of 
elementary particles emerge from the field components of $A(x)$ and 
$\bv_n(x)$ and
 the array of possible interactions and exchanges between them that underlie the myriad of 
locally degenerate solutions for the 4-dimensional spacetime geometry 
$g_{\mu\nu}(x)$ through equation~\ref{gfromavt}. The characteristics 
of a particular state
such as an electron will derive from the small number of specific 
field components involved and the corresponding permitted 
interactions associated with the apparent emission and detection of 
the state.
   Particle properties hence depend upon the properties
  of the underlying subcomponents of the $A(x),\bv_n(x)$ fields which 
are in turn determined, and their possible exchanges constrained, by 
the breaking of an explicit $n$-dimensional expression for 
equation~\ref{lpvn} over 4-dimensional spacetime.
 As reviewed in section~\ref{qugr3} significant progress has been made for this theory in accounting for the empirically observed particle multiplet structure from the elementary symmetry breaking patterns,  as will be utilised 
for equation~\ref{lpvnb} in the following subsection.

 A particle state is not here conceived of as an entity to 
be created and \textit{then} manipulated, as for a classical body, 
but rather exists as an element in the context of a full 
4-dimensional spacetime solution for equation~\ref{gfromavt} in 
constructing the manifold $M_4$.
 While hence not resembling a `particle' in a `billiard-ball' sense, 
 a somewhat amorphous `electron' state for example nevertheless corresponds to a distinct 
objective \textit{type} of impression in the geometry of spacetime 
itself.
Such particles do not then correspond to the continuous trajectories 
of independent `material' entities, rather the field components and 
interactions through which a particle state is both composed and 
observed are enveloped under the continuous geometric solution for 
$G^{\mu\nu} = f(A,\bv_n)$ in equation~\ref{gfromavt}.

  While the geometry described by $G^{\mu\nu}(x)$ and $g_{\mu\nu}(x)$ 
is everywhere smooth and continuous in spacetime $M_4$ the 
characteristic properties of elementary and composite particle states 
correspond to the characteristics of discrete topologies for the 
possible set of geometric solutions, together with their particular 
kinematic properties implied through $T^{\mu\nu}(x)$ in 
equations~\ref{gfromavt} and \ref{gavt},  that emerge in the 
near-vacuum limit (\cite{Unifi}~discussion of figure~15.2). 
 The Bianchi identity of equation~\ref{Bian} \textit{implies} the 
conservation of energy-momentum for all systems including such 
`particle' phenomena, as we shall also describe for 
equation~\ref{emcon} in the next subsection,  imparting an `electron' 
for example with an element of seemingly independent kinematic 
behaviour. 
  The consistency with which such particle phenomena can be 
reproduced, including the robust measurable invariant mass $m_e$ of 
the electron and the observed spectrum of elementary particle 
 masses generally, will require an explanation in the 
full mathematical development of the theory.

    It will also be necessary to account for the quantum properties 
of elementary particles in a quantitative manner, reproducing the 
successes of quantum theory reviewed in subsection~\ref{qugr21}. This 
should include an explanation of the basic particle `quanta' 
relations of equations~\ref{ehbaro} and \ref{phbark} and an account 
of the  $\vert \Psi \vert^2$ probability postulate as for example 
relating to the probable location to detect an electron in an 
experiment such as that in figure~\ref{dslit}. In quantum mechanics 
such a probable position of an electron can only be described given 
the details of the experimental apparatus used to determine it. This 
latter feature is very much incorporated into the present theory for 
which solutions to equation~\ref{gfromavt}  describe the extended 
geometry of the particle system immersed within a macroscopic 
environment. Our knowledge of the macroscopic world can be considered 
as a `boundary condition' in determining such a solution, as alluded 
to in the previous subsection.
  Hence for example an electron together with the corresponding 
experimental apparatus are collectively and inseparably part of the 
\textit{same} solution for a spacetime geometry and collectively determine the 
probabilities for detecting the electron at different locations.

  As we shall describe in subsection~\ref{qugr73} this framework also 
incorporates systems involving several or many particles. This 
includes the case of `indistinguishable particles', exhibiting the 
behaviour and statistical properties of bosons or fermions 
(themselves considered in subsection~\ref{qugr72}), again always 
relating to probabilities for the topologies of solutions for the 
spacetime geometry permitted through equation~\ref{gfromavt} as 
associated with the observation of apparent discrete particle 
effects. The observed properties can be counter-intuitive if an 
independent existence is attributed to such particle entities. This 
is in fact not only the case for the apparent non-local quantum 
entanglement associated with multi-particle systems, but also for the 
wave-like interference effects and non-local wavefunction reduction 
in the quantum mechanical description of a single particle state, as discussed for 
figure~\ref{dslit} in subsection~\ref{qugr21} for example.
  While for the present theory a \textit{single} enveloping geometry for 
equation~\ref{gfromavt} will apply in all cases, including for entangled states, in this subsection 
we are considering principally the concept of individual particles 
and their local interactions, both with each other and with 
experimental apparatus, such as registered by the click of a Geiger 
counter or by particle detectors more generally. 

  The identification of particle processes in the laboratory 
corresponds to the tangible form of a `microscopic' particle system 
as embedded in a macroscopic solution for equation~\ref{gfromavt}. 
One striking observation regarding such processes concerns the 
displaced vertices of particle decays seen in high energy physics 
experiments. Here the signatures of particle interactions can be 
reconstructed on a macroscopic scale of millimetres through recording 
devices situated several centimetres to several metres away from the 
purely inter-particle interactions.
 This is seen for example in processes at an electron-positron 
collider of the type:
\begin{equation}
  e^+e^- \, \to \, Z^0 \, \to \, b\bar{b}  
 \label{eezbb} 
\end{equation}
 with such an event pictured in figure~\ref{sldbb} (the same event is 
shown in \cite{Unifi} figure~10.1). 
     
% temporary:	 
%\pagebreak	 
	 
\begin{figure}[htbp]
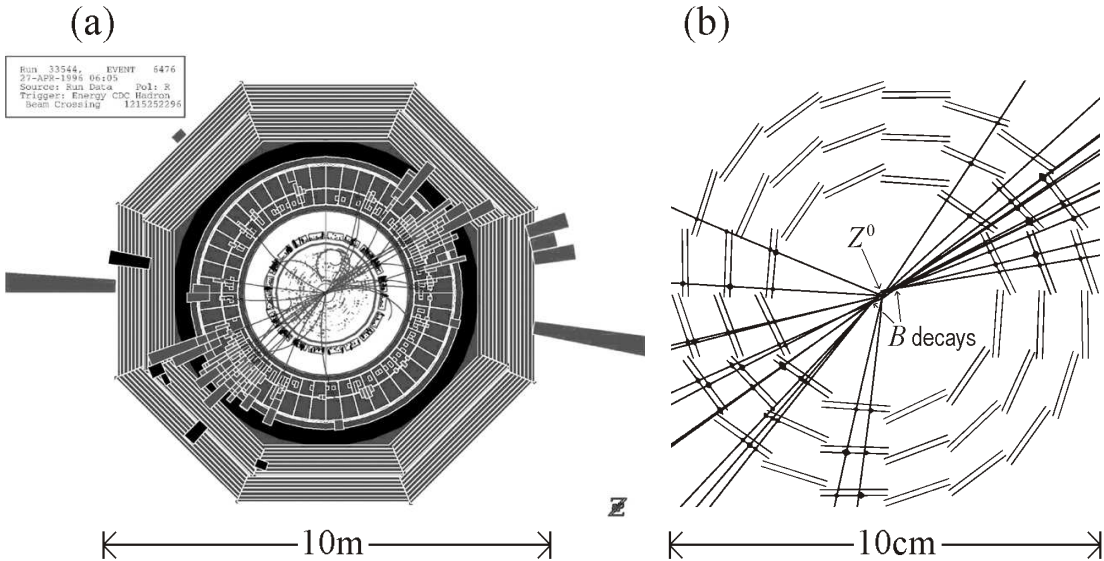
  
\centering
\epsfxsize=14.4cm
\leavevmode
\epsffile[0 0 1625 64]{aQfig2tope}
 %\hspace*{-48pt}
\vspace{-7pt} 
\hspace*{-40pt}
\epsfxsize=8.5cm
\leavevmode
\epsffile[0 0 1049 788]{sldevente} 
\hspace{0.1cm}
\epsfxsize=5.7cm
\leavevmode
\epsffile[0 0 577 606]{aQfig2be} 
%\hspace*{-48pt}
\hspace*{0pt}
\epsfxsize=14.4cm
\leavevmode
\epsffile[0 0 1458 51]{aQfig2bote} 
\caption{\setb  (a) A $Z^0 \to b\bar{b}$ event recorded by the SLD 
collaboration~\protect\cite{SLDweb} as identified (b)~via $B$ decay 
vertices displaced from the central $Z^0$ interaction point, as 
reconstructed through the extrapolation of charged particles inwards 
from the tracking devices, including the three layers of the `vertex 
detector' (\protect\cite{VXD3} figure~3) as pictured. (A~further zoom in by a  
factor of ten is depicted in figure~\protect\ref{sldbbv}(a) in 
subsection~\protect\ref{qugr72}).}
\label{sldbb}
\end{figure}

  In QFT the probability for the production of $b$-quarks in such a 
process can be calculated as well as the average lifetimes of the 
resulting $B$-hadrons, which are reflected in the typical $B$ decay 
lengths of several millimetres  observed, but there is no clear 
description of \textit{what is actually happening} in the extended 
spacetime region containing these phenomena such as reconstructed in 
figures~\ref{sldbb}(b) and \ref{sldbbv}(a). For the present theory 
the fact that the $e^+e^- \to Z^0 \to B$-hadrons production and decay 
topology can be \textit{seen} in spacetime suggests that something of 
a genuinely geometric nature \textit{is} physically happening there, 
and can be considered as evidence for solutions to 
equation~\ref{gfromavt} for the spacetime geometry with a  
\textit{topology} channelling the particle and quantum properties 
pertaining to this high energy physics environment, as will be 
discussed further for figure~\ref{sldbbv} in subsection~\ref{qugr72}.  

    Along with attempts to combine quantum theory and general 
relativity in a single consistent framework, one other major 
challenge in fundamental physics is to explain the specific 
classification of elementary particle types as described by the 
Standard Model and as investigated in experiments such as that in 
figure~\ref{sldbb}. As alluded to in the opening of this subsection 
 direct and explicit success in accounting for a series of esoteric 
features of the Standard Model  \textit{has} already been achieved 
for the present theory, which performs in this respect much better 
than models with extra spatial dimensions for example, as noted 
towards the end of section~\ref{qugr3} (and extensively demonstrated 
in the references therein). This ability to reproduce structures of 
the Standard Model through the direct symmetry breaking structure of 
natural and unique mathematical expressions for the generalised form 
of proper time in equation~\ref{lpvn}, subsuming a 4-dimensional 
Lorentzian form,  itself provides a significant vindication of this 
approach. 

  These specific mathematical forms for equation~\ref{lpvn}  
determine the constraints and possible elementary field interactions 
underlying microscopic solutions for equation~\ref{gfromavt}. This 
will set limits on the possible geometries and topologies for these 
solutions and on the corresponding range of elementary and composite 
particle types and their properties that can be observed in 
laboratory experiments, such as that in figure~\ref{sldbb}. However a 
full mathematical understanding of the quantum and gravitational 
aspects of extended spacetime solutions will also be needed for 
complete calculations of HEP processes. The aim would then be to 
reproduce the results of QFT for cross-section and lifetime 
calculations, and in principle to go further by determining the complete  
elementary particle classification structure and mass spectrum for 
example as well as by
 providing an account of what is actually physically 
\textit{happening} in spacetime for such events.

 The development of 
this mathematical structure will be guided by the clear conceptual 
picture of the present theory, as we explore in the following 
sections as outlined here.
  In subsection~\ref{qugr51} we describe the mathematical constraints 
implied in constructing matter fields and 4-dimensional spacetime 
together from the general form of proper time in equation~\ref{lpvn}, 
and also give a preliminary account of a propagating particle state. 
  In subsection~\ref{qugr52} a means of determining the likelihood of 
particle interactions will be developed and connected with the 
corresponding calculational tools of QFT. In the light of that 
analysis a more complete picture of the nature of particle quanta 
for the present theory will be further developed in 
section~\ref{qugr6}.

%\pagebreak
\section{Mathematical Expression and Connection with QFT}
\label{qugr5}

\subsection{Constraints on Solutions and Ricci Curvature Waves}
\label{qugr51}

  So far in this paper the focus has been upon describing a 
conceptual picture within which the essential phenomena of quantum 
theory and general relativity might plausibly be combined in a 
coherent manner. This picture concerns the composition of an extended 
4-dimensional spacetime $M_4$ as deriving from the unifying basis of 
generalised proper time, which itself involves a basic analysis of 
the arithmetic substructure of the one dimension of time alone. While 
non-trivial elements of the Standard Model have been identified 
directly in the local symmetry breaking structure, as briefly 
summarised in section~\ref{qugr3}, ultimately further rigorous 
mathematical expression will also be needed to describe the dynamics 
in the extended spacetime and the quantum gravity element of the 
theory in order to make full contact with the physical world.

  This is very much in contrast with approaches such as string theory 
and loop quantum gravity which essentially set out head-on in 
directly confronting the significant technical mathematical 
challenges that are faced in attempting to apply a prescription of 
quantisation to gravitation or the spacetime structure itself. While 
continuing to address the technical difficulties that have arisen 
these programmes have also been unable to make any notable original 
contact with the empirical world, and have not accounted for the 
specific structures of the Standard Model for example. Further, as we 
also alluded to in subsection~\ref{qugr24}, attempts to quantise 
gravity essentially leave the conceptual issues associated with 
quantum theory itself, as reviewed in subsection~\ref{qugr21}, 
largely untouched.

  For the present theory both the gravitational field and spacetime 
manifold remain smooth, continuous and `unquantised', and in 
providing an underlying basis for the quantisation of other fields, 
through the composition of spacetime itself, the conceptual issues of 
quantum theory can to some extent be addressed, as will be considered 
further in section~\ref{qugr7}. While we shall also discuss the 
conceptual picture corresponding to the extreme case of black holes, 
explicit calculations are likely to prove the most challenging in 
such an environment and would in any case most probably be beyond the 
reach of empirical confirmation. Hence here the aim will be to 
describe the initial steps towards reconstructing and extending upon 
the results and predictive power of quantum field theory as applied 
and investigated in the laboratory setting. 

  While suitable limiting approximations might be applied for the 
near-flat spacetime of this environment both the quantum and 
gravitational elements of the present theory will be needed. 
 QFT itself provides a pragmatic set of calculational tools yielding 
results to be compared with empirical measurements, as reviewed in 
part in subsection~\ref{qugr21} for 
equations~\ref{kgosol2}--\ref{smatrix}.
 However, as noted towards the end of subsection~\ref{qugr21} and in 
subsection~\ref{qugr42} in the discussion of figure~\ref{sldbb}, QFT 
does not
 furnish a manifestly clear conceptual picture for the physical 
reality underlying the observed phenomena.
 One aim for the present theory will be to provide such an underlying 
substantial basis for the formalism of QFT itself. 
This basis should both \textit{explain} the true nature of HEP 
phenomena and reproduce the results of the corresponding QFT calculations in the 
appropriate limit. In providing a fuller picture of HEP phenomena the 
mathematical structure of the new theory might then be employed to 
reach beyond QFT to determine specific particle parameters or to 
predict new phenomena, allowing for the theory to be tested in 
laboratory experiments.

  The significant degree of empirical success noted above that has 
already been achieved for the present theory in the identification of 
Standard Model structures derives through an analysis of the 
infinitesimal composition of the \mbox{one-dimensional} continuum of 
time, via explicit mathematical expressions for equation~\ref{lpvn} 
through to $\hG = \ese$ and the corresponding local symmetry breaking 
pattern over spacetime $M_4$. Here we are specifically  interested in 
the microscopic composition of the extended \mbox{4-dimensional} 
continuum of spacetime itself, as identified through the components 
of generalised proper time, and the implications this has for 
empirical observations on the macroscopic scale corresponding to 
solutions for equation~\ref{gfromavt}.
 That is, while equation~\ref{lpvn} concerns the composition of 
one-dimensional time and the arithmetic ways an interval $\delta s$ 
can be decomposed into subcomponents within the symmetry $\hG$, 
leading to structures identified in the symmetry breaking that resemble the Standard Model, the natural 
extension to the construction of equation~\ref{gfromavt} concerns the composition of 
4-dimensional spacetime and the degeneracy of ways to describe the 
same geometry of $M_4$ from substructures of proper time, in manner 
that will be associated with the phenomena of quantum theory.
 Essentially we shall need to identify both field types and 
interaction terms corresponding to the Lagrangian employed as input 
to a quantum field theory, in particular that for the Standard Model 
of particle physics~\cite{Teub}, and interpret the corresponding 
calculations of QFT in the context of the present theory.

   Specifically the Standard Model includes Lagrangian terms of the 
form of equations~\ref{lagsff}--\ref{lagdir} described below. The 
Yang-Mills or kinetic terms for non-Abelian gauge fields are of the 
form
 (equation~\ref{lagkff} here, with the field strength 
$F^{\alpha}_{\ph{\alpha}\mu\nu}(x)$ defined in equation~\ref{fofa}, 
see also \cite{Unifi} equation~3.94, \cite{Teub} equation~5.15):
\begin{equation}
  \lag_{\mathrm{YM}} \sim
  F^{\alpha}_{\ph{\alpha}\mu\nu}F_{\alpha}^{\ph{\alpha}\mu\nu}
  \label{lagsff}
\end{equation} 
  Lagrangian mass terms
 containing factors in a mass matrix $M_{ij}^f$, as a product of 
Yukawa couplings $Y_{ij}^f$ and the vacuum value of the Higgs doublet field 
$\Phi$,
  for three generations ($i,j = 1,2,3$) of left and right-handed 
fermions $f^i_{L,R}$ are of the gauge invariant form
 (\cite{Unifi} equations~7.69 and 7.70, \cite{Teub} equations~5.29 and 6.1):
\begin{equation}
 \label{lagmas}
   \lag_{\mathrm{M}} \sim Y_{ij}^f\,\ol{f}^{\,i}_L \Phi f^j_R
\end{equation}
 This term applies specifically for the quark sector, while in the lepton 
sector there are generally modifications dependent on assumptions 
regarding the nature of right-handed neutrino states (\cite{Gener} 
subsection~1.1 and references therein). 
 Gauge couplings
  to leptons and quarks represented by 4-component Dirac spinors $f(x)$ 
   are introduced through 
   Lagrangian kinetic terms of the form (see for example \cite{Unifi} 
equations~3.96--3.97 and 7.39--7.40, \cite{Teub} 
equations~5.30--5.33):
\begin{equation}
  \label{lagdir}
   \lag_{\mathrm{D}} \sim \ol{f} \gamma^{\mu}D_{\mu}f
\end{equation}
 via the gauge covariant derivative $D_{\mu}$ and $4 \times 4$ 
$\gamma$-matrices.  
  In the following we describe how each of these particular kinds of 
term in equations~\ref{lagsff}--\ref{lagdir} is accounted for in the 
present theory.  

  The breaking of the full symmetry $\hG$ of equation~\ref{lpvn} 
leads to an internal symmetry $G$ in equation~\ref{gbreak} and 
associated gauge fields $A^{\alpha}_{\ph{a}\mu}(x)$ which perturb the 
spacetime geometry as alluded to in equation~\ref{gefa}. 
 The components of the corresponding field strength tensor 
$F^{\alpha}_{\ph{\alpha}\mu\nu}(x)$ can be written in the standard 
way (\cite{Unifi} equation~3.38):
\begin{equation}
  F^{\alpha}_{\ph{\alpha}\mu\nu} \, = \,  
\partial_{\mu}A^{\alpha}_{\ph{a}\nu} - 
	         \partial_{\nu}A^{\alpha}_{\ph{a}\mu} + \cstr 
A^{\beta}_{\ph{a}\mu} A^{\gamma}_{\ph{a}\nu}
\label{fofa}
\end{equation}
  where $\{\alpha,\beta,\gamma\}$ are Lie algebra indices and $\cstr$ 
the corresponding structure constants, while
  partial derivatives $\pal_{\mu}$ here replace the covariant 
derivatives $\nabla_{\!\mu}$ in the external curved spacetime since 
the symmetric Levi-Civita linear connection
 $\Gamma^{\rho}_{\ph{\rho}\mu\nu}(x)=
   \Gamma^{\rho}_{\ph{\rho}\nu\mu}(x)$ is employed.
   
   The field strength term on the right-hand side of 
equation~\ref{lagsff}
  is incorporated for the present theory
  in a purely geometric manner through 
  a scalar field $\tilde{R}(x)$ defined on $M_4$ via a
   scalar curvature identified on the principal fibre bundle
 $P \equiv M_4 \times G$ (\cite{KKone} equation~90).  
 The explicit form for equation~\ref{gefa}
 is derived under the stationarity of the action 
 $\delta \mbox{\raisebox{1.5pt}{$\int$}}\! \tilde{R} = 0$ on $M_4$ 
under variations of the metric  $\delta g_{\mu\nu}(x)$ in a manner 
analogous to Kaluza-Klein theories leading to 
  (\cite{KKone}
  equations~90--93, \cite{Unifi} equation~5.20, as reviewed for 
\cite{Gener} equation~25): 
\begin{equation}
 \label{gchift}
  G^{\mu\nu} \:  = \: 
  2\chi \left(  - F^{\alpha \mu}_{\ph{\alpha\mu} 
\rho}F_{\alpha}^{\ph{\alpha}\rho\nu}
	                -\frac{1}{4} g^{\mu\nu} \, 
F^{\alpha}_{\ph{\alpha}\rho 
\sigma}F_{\alpha}^{\ph{\alpha}\rho\sigma} \right)   \: =: \:  -\kappa 
T^{\mu\nu}   
\end{equation} 
 where $\{\mu,\nu,\rho,\sigma\}$ are spacetime indices
  and $\chi$ is a normalisation constant. 
 The expression for the Einstein tensor $G^{\mu\nu}(x)$ for the 
external geometry in equation~\ref{gchift} then effectively 
incorporates cubic and quartic `self-couplings' in the gauge field 
$A(x)$
 via equation~\ref{fofa}.
  The origin of these couplings is then similar to that in the 
  Standard Model, except with the field strength entering explicitly 
through a geometric context with a scalar curvature $\tilde{R} \sim 
F^2$ rather than being included specifically through scalar 
  Lagrangian terms of the form $\lag \sim F^2$ in 
equation~\ref{lagsff}.
  That is,  an effective matter Lagrangian
  $\lag \sim F^2$ is implicitly incorporated simply by extending the 
structure of the vacuum geometry of general relativity from the base 
space $M_4$ to the principal fibre bundle space $P \equiv M_4 \times 
G$ identified for the internal symmetry $G$ in equation~\ref{gbreak} 
 to obtain equation~\ref{gchift}.

   In extracting the local 4-dimensional Lorentz metric structure of 
equation~\ref{sfourd} from the full form of proper time in equation~\ref{lpvn} onto $M_4$ via the 
projection of components \mbox{$\bh(x) \equiv \bv_4(x) \in \TM_4$} onto the 
external tangent space,  variations in the magnitude 
\mbox{$h(x) = \vert \bv_4(x) \vert = 
(\eta_{ab}v^av^b)^{\frac{1}{2}}$}, with $\bv_4 = \{v^a\} =\{\delta 
x^a/\delta s\}$ for $a=0,1,2,3$, 
 imply time dilation effects on the 4-dimensional manifold $M_4$ 
associated with a 
 warping of the spacetime geometry.
If this were to be the only distortion from a flat Minkowski 
spacetime the metric would take the conformal form 
 (\cite{Unifi} equation~13.2):
\begin{equation}
  \label{gwarph}
 g_{\mu\nu}(x) = \frac{1}{h^2(x)}\eta_{\mu\nu}
\end{equation}
  (note that this is of course a different `$h$' to that in 
equations~\ref{Einlin} and \ref{Einlinh}).
  This geometric deformation is here associated with the Standard 
Model Higgs and the origin of mass, as alluded to for 
equation~\ref{gefv} and reviewed for (\cite{Gener} equations~22--24), 
and in terms of the Einstein tensor takes the explicit form (see also \cite{Unifi} equation~13.4):  
\begin{equation}
 \label{gmnconh}
  G^{\mu\nu} \: = \:
  -3 h^{-2} \pal_{\! \rho}h \, \pal^{\rho}\:\!\! h \, g^{\mu\nu} 
  -2 h^{-1} \pal^{\mu} \pal^{\nu} \:\!\! h  
  +2 h^{-1} \square h \, g^{\mu\nu} 
  \: =: \:  -\kappa T^{\mu\nu}
\end{equation} 
    
	In turn the residual components of $\bv_n(x)$, which already 
resemble leptons and quarks for the $\hG = \ese$ level of 
equation~\ref{lpvn} as summarised in (\cite{Gener} equation~35--36), 
gain mass by impinging upon these `Higgs' $\bh(x)$ components.
  This interaction takes place through the fragmented terms of 
equation~\ref{lpvn} as partitioned into a sum of parts individually 
invariant under the broken symmetry of 
  equation~\ref{gbreak} (\cite{Gener} equation~29):  
\begin{equation}
 \label{lpvnb}
    \Lsl_p(\bv_n)_{\mathrm{Lorentz}\times G} \, = \, 
		  \sum (\mbox{invariant parts}) \,= \, 1  
\end{equation}   
  Some of these terms can be interpreted as containing factors in 
each of a Yukawa coupling, the `Higgs' field $\bh(x)$ and the lepton 
or quark components that hence gain mass (as discussed in 
\cite{Gener} from midway through subsection~4.2 for equations~29 and 41 
therein).
  The structure of such terms is
 closely analogous to the construction of the 
  Standard Model Lagrangian mass terms of equation~\ref{lagmas}.

   The constancy of equation~\ref{lpvnb} in the extended spacetime 
$M_4$ implies the vanishing of the gauge covariant derivative and 
hence with the scalar contraction:
\begin{equation}
 \label{dlpvnb}
    \gamma^{\mu}\! D_{\mu} \, 
   \Lsl_p(\bv_n)_{\mathrm{Lorentz}\times G} \, = \, 
		  \sum \gamma^{\mu}\! D_{\mu} \,
		   (\mbox{invariant parts}) \,= \, 0  
\end{equation}
 A suitable representation of the Dirac $\gamma$-matrices, as well as 
the gauge covariant derivative $D_{\mu}$, and the exact form of this 
expression will depend upon the properties of the full form for 
equation~\ref{lpvn} and the corresponding spinor structure deriving 
from the symmetry breaking. In particular, as noted towards the end 
of section~\ref{qugr3}, the construction of the predicted full form 
of proper time for $\hG = \ee$ in equation~\ref{lvto} is expected to 
make significant use of octonion-valued components as is the case for 
the intermediate $\esi$ and $\ese$ levels.
 The general structure of equation~\ref{dlpvnb} will 
  introduce terms coupling the gauge fields $A(x)$, such as the 
electromagnetic field, to fragments of $\bv_n(x)$, including lepton and 
quark spinor components (as discussed for \cite{Unifi} 
equations~11.33--11.35). This is
 analogous to the introduction of the coupling of
  gauge and spinors fields 
  in the Standard Model through equation~\ref{lagdir}.

 The heavy gauge bosons of electroweak theory are proposed to gain 
mass through an impingement of the associated gauge fields upon the 
projected external components $\bh \equiv \bv_4 \in \TM_4$ (as 
described for \cite{Unifi} equations~8.54--8.57, 8.70 and 8.71). This 
interaction is identified in the terms of the corresponding gauge 
covariant derivative action on the partitioned expression of equation~\ref{lpvnb}, which will vanish 
identically similarly as for equation~\ref{dlpvnb}.
 The origin of these gauge boson mass terms is analogous
  to that via impingement on the Higgs components in electroweak 
symmetry breaking in the Standard Model (\cite{Unifi} 
equations~7.55--7.62).

  In general relativity with the Einstein equation~\ref{Eineq} 
derived via the Einstein-Hilbert action the energy-momentum 
$T^{\mu\nu}(x)$ on the right-hand side is \textit{defined} in terms 
of a matter Lagrangian $\lag$ (\cite{Unifi} equations~3.79--3.85). 
The Einstein equation might then be typically interpreted as 
describing how the metric $g_{\mu\nu}(x)$ responds to the matter 
$T^{\mu\nu}(x)$, with the form of the matter Lagrangian $\lag$ being 
largely guided by empirical input. In contrast, for the present 
theory, here energy-momentum $T^{\mu\nu}(x)$ is \textit{defined} 
directly through equation~\ref{gfromavt}, with restrictions on the 
possible form of the spacetime metric $g_{\mu\nu}(x)$ solutions as 
expressed by the above constraint 
equations~\ref{gchift}--\ref{dlpvnb} in turn determining the possible 
forms of matter $T^{\mu\nu}(x)$ observed. Hence these constraints, 
associated with a specific form for equation~\ref{lpvn} and symmetry 
breaking structure, effectively replace the matter Lagrangian here.

  Through equations~\ref{gchift} and \ref{gmnconh}, and more 
generally equation~\ref{gfromavt}, all forms of energy-momentum 
$T^{\mu\nu}(x)$ are defined through solutions for the extended 
4-dimensional spacetime geometry $G^{\mu\nu}(x)$, with gravity 
subsuming all phenomena. In turn the contracted form of the Bianchi 
identity in equation~\ref{Bian}, expressing the vanishing divergence  
of the Einstein tensor, imposes a geometric constraint upon all 
energy-momentum:
\begin{equation}
\label{emcon}
  \nabla_{\:\!\!\!\mu}G^{\mu\nu} = 0 \quad \Longrightarrow \quad 
  \nabla_{\:\!\!\!\mu}T^{\mu\nu} = 0 \quad \Longrightarrow \quad 
  \pal_{\mu}T^{\mu\nu} = 0
\end{equation}
 with the latter expression holding to a good approximation in 
suitable global coordinates 
  in the flat spacetime limit. The Bianchi identity hence 
\textit{implies} energy-momentum \textit{conservation} in the 
familiar form $\pal_{\mu}T^{\mu\nu} = 0$ (as discussed in 
\cite{Unifi} section~5.2 opening), while 
$\nabla_{\:\!\!\!\mu}T^{\mu\nu} = 0$ can be more generally 
interpreted as `the conservation of energy in the presence of a 
gravitational field' (see for example~\cite{Carr} equation~4.23).

 This conservation applies everywhere in 4-dimensional spacetime on 
all scales, from the macroscopic arena of for example the solar 
system down to the microscopic solutions corresponding to phenomena 
studied in the laboratory. While still possessing a smooth and 
continuous geometric structure processes at the microscopic scale 
have a different character in that the consequences of the local 
field degeneracies underlying equation~\ref{gfromavt} and 
corresponding discrete particle and probabilistic quantum effects 
become manifest. 
  In the near-vacuum environment of particle interactions observed in 
experiments approximations relevant to approaching the flat spacetime 
limit will be applicable. This will allow for example a linearised 
form of general relativity, as reviewed for 
equations~\ref{glin}--\ref{Einlint}
 in subsection~\ref{qugr22}, to be employed for all equations 
involving $g_{\mu\nu}(x)$ and the spacetime geometry, as well as the 
approximation on the right-hand side of equation~\ref{emcon}.

As described above for the present theory the constraints of 
equations~\ref{gchift}--\ref{emcon} replace Lagrangian terms, such as 
equations~\ref{lagsff}--\ref{lagdir}, to determine the possible local 
field interactions (see \cite{Unifi} equation~11.29 and table~15.1 
for an earlier survey of the constraints). 
 The question then concerning how this collection of constraints and 
interaction terms in equations~\ref{gchift}--\ref{emcon}, which
 determine the permitted local degeneracy for possible solutions
 and shape the geometry of spacetime as expressed through 
equation~\ref{gfromavt},   can be connected with the probabilities to 
observe particle processes in the HEP laboratory will be considered 
mainly in the following subsection.

  While equation~\ref{gfromavt} incorporates a \textit{local} 
degeneracy for $A(x),\bv_n(x)$ matter field solutions for a given 
$G^{\mu\nu}(x)$ geometry within the constraints, the propagation of 
field values and geometric distortions into the \textit{broader} 
expanse of spacetime will depend upon the specific field content and 
corresponding equations of motion for matter fields permitted within 
the constraints. 
 Hence rather than deriving these equations of motion via a 
Euler-Lagrange equation as for 
 Lagrangian field theory  (\cite{Unifi} section~3.5), here they will 
also be determined by the constraints described in 
equations~\ref{gchift}--\ref{emcon} (see for example \cite{Unifi} 
sections~5.2 and 11.1) as we consider below. 

 For example equation~\ref{gchift} incorporates the standard 
expression for the energy-momentum of the electromagnetic field in a 
curved spacetime:
\begin{equation}
     \frac{-1}{\kappa} G^{\mu\nu} \,  = \,  
	\frac{2\chi}{\kappa} \left(
	 F^{\mu}_{\ph{\mu}\rho}F^{\rho\nu}
	      + \frac{1}{4} g^{\mu\nu} \, F_{\rho\sigma}F^{\rho\sigma}
		  \right) \, =  \, T^{\mu\nu}  
		     \label{tmnem}    
\end{equation}	
 Here $F_{\mu\nu}(x) = \partial_{\mu}A_{\nu}(x) - 
	         \partial_{\nu}A_{\mu}(x)$    
	from equation~\ref{fofa} is the electromagnetic field tensor and 
$A_{\mu}(x)$ the corresponding Abelian gauge field.
 Consistent with the discussion before equation~\ref{Bian} in 
subsection~\ref{qugr22} this expression is associated with a 
non-negative energy density and indeed represents a known empirical 
source of energy-momentum. 
   The Bianchi identity of equation~\ref{emcon} as applied to equation~\ref{tmnem} can be shown to 
itself imply Maxwell's equation for the propagation of the 
electromagnetic field over macroscopic distances, as can be written 
under the Lorenz gauge condition $\pal^{\mu}\! A_{\mu}(x) = 0$,
 here considering the flat spacetime limit, 
 as  (\cite{Unifi}~equations~5.29, 5.30 and 11.2):
\begin{equation}
 \label{maxafree}
   \square A_{\mu}=0 
\end{equation} 
  This contrasts with the standard derivation of Maxwell's equation 
via the Euler-Lagrangian approach based on the stationarity of the 
action integral for equation~\ref{lagsff} under variations in the 
gauge field $\delta A(x)$ for the electromagnetic case (see discussion of 
\cite{Unifi} equation~3.93). 
  
   The solutions include the case of the propagation of a plane wave 
in the electromagnetic field which, 
    on assuming an approximately flat spacetime for which a Minkowski 
coordinate frame can be adopted, takes the form:
\begin{eqnarray}
 A_{\mu}(x) & = &  A_{\mu}(\bk)  \cos k\!\cdot\! x
                  \label{acos} \\ 
     & = & A_r(\bk)\;\!\varepsilon_{\mu}^r(\bk)  
			  \cos k\!\cdot\! x
	              \label{ecoeff}
\end{eqnarray} 
  on writing $A_{\mu}(\bk)=  A_r(\bk)\;\! \varepsilon_{\mu}^r(\bk)$
  for $r = 0,1,2,3$,
   where the $\varepsilon^r$ are polarisation vectors
  (these equations are equivalent to \cite{Unifi} 
equations~11.3--11.6 with $A_r(\bk)$ real and within a conventional 
factor of two). 
  As described for (\cite{Unifi} equations~11.2--11.11) 
equations~\ref{acos} and \ref{ecoeff} provide a solution to 
equation~\ref{maxafree} for
 $(k)^2= k^{\mu}k_{\mu} = 0$, while the Lorenz gauge condition 
implies
  $k^{\mu} \varepsilon_{\mu}^r= 0$. Taking the propagation of the 
electromagnetic wave in the $x^3$ direction there are two possible 
polarisation states corresponding to the transverse vectors:
\begin{equation}
    \varepsilon^1 = (0,\;\, 1,\;\, 0,\;\, 0)
	 \quad \mbox{and} \quad
	 \varepsilon^2 = (0,\;\, 0,\;\, 1,\;\, 0) 
  \label{waveem}
\end{equation}
 These states can be compared with the two polarisation modes for 
gravitational waves described for equations~\ref{wavepol} and 
\ref{waveee}. 

  On substituting equation~\ref{ecoeff} into equation~\ref{tmnem} the 
energy-momentum of an electromagnetic wave for either polarisation 
state $r=1$ or $r=2$ is (as pictured in \cite{Unifi} figure~11.1 for 
equation~11.12 there):
\begin{eqnarray}
 T^{\mu\nu} & = & \frac{2\chi}{\kappa} \, \vert A_r(\bk) \vert^2 \, 
          k^{\mu}k^{\nu} \:\! 
	  \sin^2 k \!\cdot\! x  \nonumber \\
	  & = & \frac{\chi}{\kappa} \, \vert A_r(\bk) \vert^2 \, 
          k^{\mu}k^{\nu} 
		\left( 1 - \cos 2k \!\cdot\! x  \right) \nonumber  \\
	  & = &	 \frac{\chi}{\kappa} \, \vert A_r(\bk) \vert^2 \, 
          k^{\mu}k^{\nu}  \, - \,
		  \frac{\chi}{\kappa} \, \vert A_r(\bk) \vert^2 \, 
          k^{\mu}k^{\nu}  \:\! \cos 2k \!\cdot\! x
	   \label{twave}
\end{eqnarray} 
  The first term in the bottom line of equation~\ref{twave} is then 
of the form of a relativistic  `pressureless perfect fluid' or `dust 
cloud' with $(k)^2=0$ and constant energy density  
$\frac{\chi}{\kappa}  \vert A_r(\bk) \vert^2$, while the second term 
describes an accompanying oscillation also travelling at the speed of 
light.

 Here we are interpreting $T^{\mu\nu}(x)$ as being defined and 
derived \textit{through} a solution for the geometry $G^{\mu\nu}(x)$ in equation~\ref{tmnem}, as the 
electromagnetic case of equation~\ref{gchift}. Hence the oscillation 
in the electromagnetic field of equation~\ref{acos} is     
   directly associated with a tiny distortion in the spacetime 
geometry itself, underlying equation~\ref{twave}, with: 
\begin{equation}
 G^{\mu\nu} = \lambda \, 
          k^{\mu}k^{\nu}  \, - \,
		  \lambda \, 
          k^{\mu}k^{\nu}  \, \cos 2k \!\cdot\! x
  \label{gwave}
\end{equation}
 where $\lambda = - \chi \vert A_r(\bk) \vert^2$ is a constant.
 For any electromagnetic field in equation~\ref{tmnem} the 
corresponding scalar curvature vanishes $R=0$, with the Ricci 
curvature then simply 
 \mbox{$R^{\mu\nu} = G^{\mu\nu}$} (as discussed in \cite{Unifi} after 
equation~5.28), and the Weyl curvature is also zero for an 
electromagnetic plane wave solution (as discussed in \cite{Unifi} 
after figure~11.1).    
  Hence an electromagnetic wave is accompanied 
 by a spacetime distortion in the form of a wave of Ricci curvature 
in the gravitational field.

   While the Einstein tensor $G^{\mu\nu}(x)$ encapsulates the 
spacetime curvature, the gravitational field in the form of the 
metric $g_{\mu\nu}(x)$ conveys a more direct sense of the actual 
warping of the spacetime geometry, at least for minor deviations from 
the flat case and in a suitable choice of coordinates.
Given that the corresponding geometric deformation of spacetime in 
equation~\ref{gwave} is extremely small for the propagation of such a 
`matter wave', a linearised approximation to the Einstein field 
equation can be adopted, as described in subsection~\ref{qugr22}, to 
determine the metric $g_{\mu\nu}(x)$ structure. In this case the two 
terms in equation~\ref{gwave} can be treated as providing independent 
contributions to the metric geometry. While the first term $\lambda 
k^{\mu}k^{\nu}$ is homogeneous and independent of $x$, the second 
term describes an oscillation, and with zero Weyl curvature in both 
cases. Since conformal distortions from a flat spacetime have no Weyl curvature we can 
adopt a trial solution for the second term
in equation~\ref{gwave} as $g_{\mu\nu}(x)=\eta_{\mu\nu}(1+\beta \cos 2k \!\cdot\! x)$, that is in the form of equations~\ref{glin} and \ref{nhlin} as:
\begin{equation}
   g_{\mu\nu} = \eta_{\mu\nu} + d_{\mu\nu}
   \quad \mbox{with} \quad d_{\mu\nu} = 
           \beta \, \eta_{\mu\nu}\cos 2k \!\cdot\! x 
	\quad \mbox{and} \quad	\vert \beta \vert \ll 1  
	\label{gconfu} 
\end{equation}   
  To first order in $\beta$ the geometric distortions
  $d_{\mu\nu}(x)$, via the left-hand side of equation~\ref{Einlin}, 
generate 
  $G^{\mu\nu} = -4\beta k^{\mu}k^{\nu}\cos 2k \!\cdot\! x$, and hence 
provide a solution for the second term in equation~\ref{gwave} 
 for $\beta = \lambda/4$ (see also discussion of \cite{Unifi} 
equation~11.13).
 
  The solution for the purely gravitational wave perturbation 
$h_{\mu\nu}(x)$ in equations~\ref{wavepol} and \ref{waveee} can be 
contrasted with 
  the above solution for the geometry $d_{\mu\nu}(x)$ associated with 
the propagation of an electromagnetic wave, with the respective 
forms:
\begin{equation}
    h_{\mu\nu} \! = \! \left( \begin{array}{cccc}
	     0\,\, & 0 & 0 & 0  \\
		 0\,\, & A_{11} & A_{12} & 0  \\
		 0\,\, & A_{12} & \! -A_{11} \; & 0  \\
		 0\,\, & 0 & 0 & 0  
		 \end{array} \right) \! \cos k \!\cdot\! x
		   \quad \mbox{and} \quad
    d_{\mu\nu} \! = \! \left( \begin{array}{rrrr}
	     \beta & 0 & 0 & 0  \\
		 0 & -\beta & 0 & 0  \\
		 0 & 0 & -\beta & 0  \\
		 0 & 0 & 0 & -\beta  
		 \end{array} \right) \! \cos 2k \!\cdot\! x 
  \label{wavecom}
\end{equation}
 These solutions can be considered a `vacuum wave' and a  `matter 
wave' respectively, propagating with $k$ and $2k$ denoting the 
unrelated wave 4-vectors. In the vacuum case the gravity wave 
$h_{\mu\nu}(x)$ is purely Weyl curvature with two real parameters 
$A_{11},A_{12}$, on the other hand the matter wave in the spacetime 
geometry $d_{\mu\nu}(x)$, associated with an electromagnetic wave, is 
purely Ricci curvature with $\beta$ the single real parameter.

  While linearised general relativity has been employed in both cases 
constructing a linear superposition of gravitational waves with 
varying wave vector $k$ is non-trivial owing to the $k$-dependence of 
the coordinate choice employed for the above form of $h_{\mu\nu}(x)$, 
that is the TT gauge as described for 
equations~\ref{waveh}--\ref{waveee}
 (although a looser \mbox{$k$-independent} harmonic gauge can be 
employed in which case six, rather than two, real parameters are 
needed, as shown for \cite{Carr} equation~6.35).
 However the deduction of the matter wave  $d_{\mu\nu}(x)$ above was 
free of any $k$-dependence and hence a superposition could be formed 
directly with a single amplitude parameter $\beta(\bk)$ for each 
contributing mode. Such a superposition with a small spread in the 
angular frequency $k^0$, always with $k^{\mu}k_{\mu}=0$, could be 
constructed for the description of a wave-packet localised in 
3-dimensional space, as might be associated with a particle state -- 
namely a photon of light for this electromagnetic case.  
 The construction of wave-packets more generally will be discussed for equations~\ref{fpack} and \ref{wpack} towards the end of the following subsection.

  Ultimately a fuller analysis will be required to determine the 
solution $g_{\mu\nu}(x)$ for such a matter wave-packet, also taking 
into account the localised contribution from the first term on the 
right-hand side of equation~\ref{gwave}.
 Although still predominantly describing Ricci curvature a more 
complete particle state solution associated with a matter wave-packet might 
be expected to also generate Weyl curvature components.  
 Care will also be needed not to be misled by the superfluous effects 
relating to freedom in the choice of coordinates, with the physical 
curvature on $M_4$ determined through the geodesic deviation of 
hypothetical `test particles', similarly as for the case of a gravity 
wave discussed after equation~\ref{waveee}. Artificial  coordinate 
effects will need to be considered for example in comparing the 
equivalence or otherwise of different solutions for $g_{\mu\nu}(x)$, 
as will be relevant for the analysis of local solutions exhibiting a 
degenerate composition in $A(x), \bv_n(x)$ field components 
considered in the following subsection.

  However here we consider the conformal undulations of 
$d_{\mu\nu}(x)$ in equations~\ref{gconfu} and \ref{wavecom} to 
represent a real physical element of the impact on the spacetime 
geometry directly connected with a propagating oscillation in the 
electromagnetic field $A(x)$ via equation~\ref{gchift} for the case 
of equation~\ref{tmnem}. While these disturbances with 
$k^{\mu}k_{\mu}=0$  
 will ultimately be associated with particle states travelling at 
light speed and hence of zero invariant mass, more general massive 
states with $k^{\mu}k_{\mu}> 0$, relating to other matter fields such 
as deriving from the components of $\bv_n(x)$, 
will correspond to a more general propagation of Ricci curvature.
  In particular while Maxwell's equation~\ref{maxafree} here derives 
from the geometric structure of equation~\ref{tmnem}, the Dirac 
equation (\cite{Pesk} equation~3.31, \cite{Unifi} equation~3.99)  should similarly at least be 
consistent with equation~\ref{emcon} for the spacetime geometry 
corresponding to the energy-momentum of a Dirac field (as presumably 
closely related to the standard expression, \cite{Pesk} 
equation~19.150).
 Solutions for the spacetime curvature might then also be associated 
with spinor states such as leptons and quarks or with the equations 
of motion for other fields (and for example hadronic states composed 
of quarks), including also for the non-relativistic limit.

In general at a time before and a time after an interaction the 
initial and final states respectively can be considered as free 
particles propagating according to the appropriate equations of 
motion. The equation of motion associated with a Dirac spinor must then also be 
related to that associated with the electromagnetic field
 in equation~\ref{maxafree} as a solution for equation~\ref{tmnem}  
 (and for the more general non-Abelian internal symmetry  case 
associated with equation~\ref{gchift},
  see discussion of \cite{Unifi} equation~5.21 and \cite{KKone} 
equation~94)
 by the consistency of the constraints on the continuous spacetime 
geometry $G^{\mu\nu}(x)$ through interactions involving both spinor states and gauge 
bosons, such as for the case of Compton scattering. This is 
analogous to the standard flat spacetime properties involved in 
considering the Dirac equation as the `square root' of the 
Klein-Gordon equation in a consistent manner (see for example 
\cite{Pesk} section~3.2). 
 These observations will be relevant for the discussion following 
equation~\ref{phieeb} in the next subsection, where a close interrelation between the 4-vectors
  $A^{\mu}(x)$ and $\ol{\psi}(x)\gamma^{\mu}\psi(x)$, associated with gauge and spinor fields respectively, will then be considered in an interaction process.  

  The electromagnetic plane wave of equations~\ref{acos} and \ref{ecoeff} (and 
equation~\ref{acosee} in the following subsection) and polarisation 
modes of equation~\ref{waveem} determine the nature of the particle 
states resulting from the standard quantisation of the 
electromagnetic field as massless spin-1 photons. Similarly given the 
structure of a classical gravitational wave in 
equations~\ref{wavepol} and \ref{waveee}, as recalled above for 
$h_{\mu\nu}(x)$ in equation~\ref{wavecom}, a successful quantisation 
of the gravitational field would be expected to predict the existence 
of particle states in the form of massless spin-2 `gravitons', as we 
noted in the opening of subsection~\ref{qugr23}.

  However for the present theory of `quantum gravity' the 
gravitational field remains unquantised, with gravity waves 
corresponding to essentially classical disturbances of Weyl curvature 
propagating as `vacuum waves' in the spacetime geometry with no \textit{reason} to be 
quantised. On the other hand `matter waves' of Ricci curvature as undulations in the gravitational field 
associated with an implicit matter composition, as described for the 
electromagnetic field above for $d_{\mu\nu}(x)$ in 
equations~\ref{gconfu} and \ref{wavecom} as well as for other matter fields, will be 
`quantised' through the constraints on these constructions. That is 
the \textit{mechanism} of `quantisation'  itself and the origin of 
all matter particle states will be identified through the constraints of 
equations~\ref{gchift}--\ref{emcon} in obtaining solutions for 
equation~\ref{gfromavt}, in terms of the gravitational field 
\textit{together with} a matter field composition, as consistent with  
an underlying basis in
  equation~\ref{lpvn} as the general form of proper time. In particular 
  the limitations on possible solutions within the constraints 
implied for this theory can be shown to
    elucidate the origin of the basic particle `quanta' relations 
of equations~\ref{ehbaro} and \ref{phbark}, as we shall argue in 
subsection~\ref{qugr61}.

   Terms in the constraints of equations~\ref{gchift}--\ref{emcon},
  through the intrinsic local degeneracy in the matter field 
composition of spacetime, 
    will effectively act as local selection rules for possible local 
matter field exchanges and hence for the corresponding possible 
particle interactions.  These selection rules in turn will determine 
the relative probabilities for field transitions and particle 
interactions leading to macroscopically distinct events, as might be 
recorded by an experiment such as that in figure~\ref{dslit} or 
\ref{sldbb}. The set of terms in these constraints, deriving through 
the construction of 4-dimensional spacetime from a basis in 
generalised proper time alone for the present theory, is proposed to 
replace the role of the single scalar Lagrangian function for the 
Standard Model, with contributions from 
equations~\ref{lagsff}--\ref{lagdir}, as we have proposed in this 
subsection.

 The aim will be to avoid any \textit{postulates} of field 
`quantisation' in considering QFT as providing a pragmatic 
computational rather than a conceptually accurate match to 
microscopic physical phenomena.  
  On the other hand, given the very significant quantitative success 
of QFT we can be guided by the corresponding calculational tools 
which we ultimately aim to account for. That is, while there is no 
imposed `quantisation' as such for the present theory, it will still 
be    
necessary to understand the link between the underlying conceptual 
motivation here and mathematical structures of a standard quantum 
theory approach, at least in the appropriate limit. 
 We describe how such a connection might be established in the following 
subsection.

%\pagebreak
\subsection{Calculating Event Probabilities in Particle Physics}
\label{qugr52}

 The aim of this subsection is to describe how calculations might be 
performed in the present theory to determine the likelihood of 
particle processes in HEP experiments such as that in 
figure~\ref{sldbb}. 
 While in making measurements methods of classical statistics are 
employed to analyse the data collected,  
   there are also elements of the standard cross-section and event 
rate expressions for particle collider experiments that have a 
classical statistical nature, as discussed in (\cite{Unifi} 
following equation~10.7). These in particular involve the `number of 
ways' an interaction can occur given the characteristics of the 
initial $i$ and final $f$ states, while the transition amplitude 
$\mcM_{fi}$ between these states is a distinctly quantum mechanical 
object, as reviewed here for equations~\ref{sfi}--\ref{smatrix}. For 
the present theory the goal is to account for this latter object also 
in the manner of a `number of ways', now in terms of a local relative 
degeneracy count for building the spacetime geometry itself, at the 
most elementary level through equation~\ref{gfromavt} within the 
constraints of equations~\ref{gchift}--\ref{emcon}. This construction 
will hence also be more similar to that for a classical probability 
as noted in the discussion following equation~\ref{gfromavt}. The aim will then be to 
determine observables including interaction cross-sections and 
particle lifetimes for given processes, such as identified in 
figure~\ref{sldbb}, directly in terms
of the possible exchanges between locally indistinguishable matter field compositions 
underlying the same local
 spacetime geometry.

  With the intent ultimately to connect with QFT calculations, we 
note a common feature between the interaction picture construction 
described for the operator field of equation~\ref{kgosol2} in 
subsection~\ref{qugr21} and the description in the previous 
subsection of a `matter wave' of Ricci curvature based on 
equation~\ref{acos} for the electromagnetic field, as a solution to 
equation~\ref{tmnem}, is an analysis in terms of a wave 4-vector, $p$ 
and $k$ respectively, describing the oscillation modes. While the 
real gauge field oscillation of equation~\ref{acos}
can be written out as:
\begin{equation}  
   A_{\mu}(x) \, = \,  A_{\mu}(\bk)  \cos k\!\cdot\! x
               \,  = \, A_{\mu}(\bk)\mbox{Re}(e^{+ik\cdot x})
			 \, = \,   
   \frac{1}{2} A_{\mu}(\bk)(e^{+ik\cdot x}
    + e^{-ik\cdot x}) 
 \label{acosee}
\end{equation}
  in subsection~\ref{qugr51} we did not use the complex 
subcomponents. However there are many areas of physics, not only in 
QFT, in which the analysis of real oscillations in terms of complex modes, such as
$e^{\pm ik\cdot x}$ above, 
 is employed owing to their useful algebraic properties
  (see also discussion in \cite{Unifi} section~11.4 pages 342--343).
 
  Here algebraically \textit{both} the $e^{+ik\cdot x}$ and 
$e^{-ik\cdot x}$ components form parts of the full functional form 
contributing to the solution in equation~\ref{acosee} for 
equation~\ref{tmnem}. In turn \textit{each} of the $e^{\pm ik\cdot 
x}$ parts can be individually replaced, within the constraints of the 
theory, by an alternative equivalent $e^{\pm ik\cdot x}$ matter field 
mode contribution from subcomponents of the fields $A(x),\bv_n(x)$  that maintains the same local geometry of spacetime 
as determined through equation~\ref{gfromavt}.
 The functional form of the $e^{\pm ik\cdot x}$ modes will be found 
to be convenient here owing to the algebraic properties of 
exponentials under multiplication, as they can be composed as factors 
in terms of the constraints such as equation~\ref{dlpvnb}, as will be 
consider for equation~\ref{dlpvnbex} below.

The  complex 
$e^{+ik\cdot x}$ and $e^{-ik\cdot x}$ parts can be represented through  independent local exchanges 
  of $A(x),\bv_n(x)$ subcomponents,  leading to local hybrid 
combinations of these fields.
 While such a hybrid construction is 
mathematically permitted locally a macroscopically propagating 
wave-packet solution, corresponding to a physical particle state as 
considered in the previous subsection, will involve a coherent 
description in terms of specific real field components, that is with matching $e^{\pm ik\cdot x}$ contributions  such as for 
equation~\ref{acosee} above, consistent with the equation of motion 
deduced for that particular field, such as equation~\ref{maxafree} for the case of the electromagnetic field.       
 
  The probability of a transition between different propagating 
particle solutions for equation~\ref{gfromavt} will then be 
proportional to the local degeneracy, or count, of the number of 
local field exchanges in the separate $e^{\pm ik\cdot x}$ component 
functions describing a transition linking the initial and final 
particle states through compositions of an equivalent local spacetime 
$g_{\mu\nu}(x)$ metric geometry. The numerical counts as a sum of the 
`number of ways'  to decompose the field exchanges
 consistently within the constraints
 in terms of $e^{+ik\cdot x}$ and $e^{-ik\cdot x}$ Fourier components 
are denoted  $D_+$ and $D_-$ respectively.  These degeneracy counts 
are represented by nested time integrals, in  the continuum limit, 
over the possible respective field exchanges underlying the solutions 
for equation~\ref{gfromavt}.

 The general expression for the degeneracy count $D_+$ for an 
interaction process to take place between an initial state at time $t=0$ and 
a final state at time $t=T$ can be written as (see also \cite{Unifi} 
equations~11.40, 11.41 and 11.44):
\begin{equation}
  D_+ \;  =  \; \sum_{n=1}^{\infty} \; \int_0^T dt_1 \int_0^{t_1} 
dt_2\ldots \int_0^{t_{n-1}} dt_n 
           \, R_{\mathrm{con}}(t_1)
		      R_{\mathrm{con}}(t_2) \ldots
			  R_{\mathrm{con}}(t_n) 
  \label{degplus}
\end{equation}
 Here $R_{\mathrm{con}}(t_i)$ represents allowed field 
\textit{R}edescriptions as permitted by the \textit{con}straint  
equations for the $e^{+ik\cdot x}$ components.  
 At $n^{\mathrm{th}}$-order a chain $R_{\mathrm{con}}(t_1)
  \ldots  R_{\mathrm{con}}(t_n)$ can be constructed
  with all $n$ factors set to $R_{\mathrm{con}}(t_i) = 1$, 
  as opposed to any $R_{\mathrm{con}}(t_i) = 0$, 
 for each possible consistent sequence of field exchanges linking 
given initial and final states, as will be demonstrated in the 
example below of equations~\ref{ext1}--\ref{ext4} and 
figure~\ref{extofey}(a).  
  Equation~\ref{degplus} incorporates an implicit sum over all such possible consistent chains at each order.

 A very similar expression can be constructed for the $D_-$ 
degeneracy count corresponding to the allowed exchanges of the 
$e^{-ik\cdot x}$ components. 
 The  probability $P_{fi}$ for a particular process involving  
transitions in the matter field components underlying 
equation~\ref{gfromavt} between the initial and final states will 
then be of the form:           
\begin{equation}
  P_{fi} \; \propto\; D_+D_-
\label{ppdd}
\end{equation}
 To describe an actual probability between 0 and 1 the right-hand 
side of this expression would need to be normalised by the degeneracy 
count for all processes leading to all possible final states.

  All of the constraint equations~\ref{gchift}--\ref{emcon} stem from 
equation~\ref{lpvn} for generalised proper time as expressed through 
the construction of the 4-dimensional spacetime $M_4$. For example 
equation~\ref{dlpvnb} will consist of a string of terms each composed 
of fragmented components from the symmetry breaking of 
equation~\ref{lpvn} over $M_4$, such as of the form:
\begin{equation}
 \label{dlpvnbex}
    \gamma^{\mu}\! D_{\mu} \, 
   \Lsl_p(\bv_n)_{\mathrm{Lorentz}\times G} \, = \, 
	\ldots + \ol{\psi}\,\gamma^{\mu}\! A_{\mu}\:\!\psi \, + \,              
            \ol{\varphi}\,\gamma^{\mu}\! A_{\mu}\:\! \varphi 
		  		+ \ldots  \,= \, 0  
\end{equation}
 In general in this sum of invariant parts there will be other terms as well as further factors from 
the fragmented $\bv_n(x)$ components in the above terms.
  A full understanding of the role of the octonion algebra in the full $\hG=\ee$ form of 
equation~\ref{lvto} and the symmetry breaking structure under 
equation~\ref{gbreak}, for $G$ identified with the Standard Model 
gauge group, may be needed to fully determine the specific structure 
of equation~\ref{dlpvnbex}.  
 Here $\psi$ represents a Dirac spinor under the external Lorentz 
symmetry, as constructed from octonion-valued components of 
$\bv_n(x)$, with $\ol{\psi}$ its conjugate 
 consistent with the appropriate $\gamma$-matrix representation as 
noted after equation~\ref{dlpvnb}. With a similar construction for 
$\varphi$ and $\ol{\varphi}$
 such spinors with properties resembling leptons and quarks have been  
identified at the $\hG = \ese$ level for equation~\ref{lpvn}, as 
described for (\cite{TimeE}~figure~4), with the full set of Standard 
Model states predicted to be uncovered at a $\hG = \ee$ level as 
discussed for equation~\ref{lvto}.

 The gauge covariant derivative $D_{\mu}$ in equation~\ref{dlpvnbex} 
is associated with a gauge field $A_{\mu}(x)$, 
 in turn associated with a factor of the internal symmetry $G$ from 
equation~\ref{gbreak}. This might for example be identified as the 
electromagnetic gauge symmetry $\uo_Q$ and gauge field of 
equation~\ref{maxafree} or as the strong gauge group $\suth_c$ (as 
also listed in \cite{TimeE} figure~4). 
 In the case of a non-Abelian gauge group the gauge field can be 
written as $A_{\mu}(x) = \Aamu(x)E_{\alpha}$ with the $E_{\alpha}$ 
the corresponding Lie algebra generators in the appropriate 
representation acting on the $\psi$ and $\varphi$ components in 
equation~\ref{dlpvnbex} (this is closely analogous to the structure 
of the Lagrangian term of equation~\ref{lagdir} with reference to 
\cite{Unifi} equations 3.96--3.97).

  In general the constraint equations provide a collection of terms, 
and we then consider how each such term might transmit a field 
interaction on $M_4$ and contribute to equation~\ref{degplus} and 
hence equation~\ref{ppdd}, via corresponding strings of assignments 
of $R_{\mathrm{con}}(t_i) = 1$. For example we can consider the 
initial state of a solution for equation~\ref{gfromavt} to consist of 
propagating particle states associated with $\psi(x)$ and 
$\ol{\psi}(x)$ subcomponent contributions from $\bv_n(x)$, with each 
particle satisfying the Dirac equation of motion as considered 
towards the end of the previous subsection.    
 The impact on the external geometry through equation~\ref{gefv}, with the
 corresponding internal field oscillations expressed similarly as 
for equation~\ref{acosee}, can be written as:
\begin{eqnarray}
  G^{\mu\nu}(x) & = & f(\bv_n) \; = \; f(\psi,\ol{\psi})
    \label{gtwopsi} \\
   \psi(x) & = & \frac{1}{2} \psi(\bk_1)(e^{+ik_1\cdot x}
    + e^{-ik_1\cdot x}) \label{phiee} \\
	\ol{\psi}(x) & = & \frac{1}{2} \ol{\psi}(\bk_2)(e^{+ik_2\cdot x}
    + e^{-ik_2\cdot x}) \label{phieeb}
\end{eqnarray}
  While deriving from the same components of $\bv_n$ and the same 
fragment of equation~\ref{lpvn}, here $\psi$ and its conjugate 
$\ol{\psi}$ are associated with \textit{separate} particle states, 
with $k_1$ and $k_2$ independent, in composing \textit{two} factors 
in the first term of equation~\ref{dlpvnbex} and in making 
\textit{distinct} impressions in the spacetime geometry through 
equation~\ref{gtwopsi}. 
 While each propagates as an initial state particle according to the 
Dirac equation we can consider the possibility of their coming 
together in a local interaction as permitted through 
equation~\ref{dlpvnbex}.   

 As noted above, to determine a local degeneracy count for possible 
particle exchanges, rather than considering the full `real' fields 
  $\psi(x)$ and $\ol{\psi}(x)$ it is convenient to utilise the 
algebraic properties of exponentials and perform the analysis for the 
`complex' $e^{+ik\cdot x}$ and $e^{-ik\cdot x}$ modes (this 
complex substructure is independent of the octonion-valued origin of 
the $\psi$ components). The  $e^{\pm ik\cdot x}$ field 
components, with $+$ve and $-$ve signs in the exponents, can be embedded in the constraint of 
equation~\ref{dlpvnbex} which then determines the freedom in the  
choice of the fields making this contribution to the local geometric 
distortion of $G^{\mu\nu}(x)$ in equation~\ref{gfromavt} within the 
local degeneracy. This contribution may then be transmitted through a 
sequence of exchanges that leave equation~\ref{dlpvnbex}, as well as 
the local physical geometry, invariant.

 Here we describe a possible explicit means of transmitting 
 such field exchanges in $e^{\pm ik\cdot x}$ components
 through the constraint equations. We begin by writing 
equation~\ref{dlpvnbex} as:
%{\setlength{\baselineskip}{5.35\baselineskip}
\begin{eqnarray}
    \gamma^{\mu}\!\!\;\! \!\! & \!\!\! D_{\mu} \!\!\! & \!\!
   \Lsl_p(\bv_n)_{\mathrm{Lorentz}\times G}
     \; =  \nonumber \\
      &   & 	
	\ldots + \ol{\psi}(k_2)e^{+ik_2 \cdot x}e^{-ik_2 \cdot x}
	\,\gamma^{\mu}\! A_{\mu}\:\!
	  \psi(k_1)e^{+ik_1 \cdot x}e^{-ik_1 \cdot x} \, + \,              
            \ol{\varphi}\,\gamma^{\mu}\! A_{\mu}\:\! \varphi 
		  		+ \ldots  \,= \, 0 \qquad
     \label{ext0} 
\end{eqnarray}
%\par}
 This can be interpreted as a means of 
 embedding either the $+$ve or $-$ve exponential modes of 
equations~\ref{phiee} and \ref{phieeb} into the constraint 
equation~\ref{dlpvnbex}. Taking the case of the $+$ve modes, and 
recalling that in general $A_{\mu}(x) = \Aamu(x)E_{\alpha}$, 
considering only the two given terms the above expression can then be 
written identically as:
\vspace{-3pt}

{\setlength{\baselineskip}{1.6\baselineskip}
\vspace{-40pt}
{\setlength{\baselineskip}{1.35\baselineskip}
\begin{eqnarray}
    \gamma^{\mu}D_{\mu} 
   \Lsl_p(\bv_n)_{\mathrm{Lorentz}\times G}
     \; =  \qquad\qquad\qquad\qquad\qquad\qquad 
	\qquad\qquad\qquad\qquad\qquad & &  \nonumber \\
    \Aamu 
	 \left(\ol{\psi}(k_2)e^{+ik_2\cdot x}\right)
	\gamma^{\mu}E_{\alpha}
	 \left(\psi(k_1) e^{+ik_1\cdot x}\right)
	e^{-i(k_1+k_2)\cdot x} \; + \;              
    \Aamu\,
	   \ol{\varphi}\gamma^{\mu}E_{\alpha} \varphi 
		  	 \,    \,= \! & \! 0 & \quad  \label{ext1}  \\
\setlength{\unitlength}{25pt}
\begin{picture}(0.0,0.0)(0.0,0.0)
  {\Large
	 \put(2.7,1.1){\vector(0,-1){0.65}}
	 \put(5.9,1.07){\vector(-4,-1){2.433}}
   }
\end{picture}			 
		 =  \quad\:\!\!\;
    \left(\Aamu(k)e^{+ik\cdot x}\right)   
    (\ol{\psi}
	\gamma^{\mu}E_{\alpha}\psi)e^{-i(k_1+k_2)\cdot x}
	     \quad\;\:\,\,\, + \quad\;\,\,\,\,              
    \Aamu\,
	   \ol{\varphi}\gamma^{\mu}E_{\alpha} \varphi 
		  	\qquad\!  = \! & \! 0 &  \label{ext2} \\			
\setlength{\unitlength}{25pt}		
\begin{picture}(0.0,0.0)(0.0,0.0)
  {\Large
	 \put(3.87,1.09){\vector(+4,-1){2.6}}
   }
\end{picture}		
			 = \qquad\quad\!\!\!\!\:
		\Aamu\, \ol{\psi}\gamma^{\mu} 
	   E_{\alpha}	\psi \quad\;\:\,\,\, + \quad\;\,\,\,\,
		\left(\Aamu(k)e^{+ik\cdot x}\right)
		(\ol{\varphi}
	\gamma^{\mu}E_{\alpha}\varphi)e^{-ik\cdot x}  
	    	  \qquad\,\;\;	= \! & \! 0 &  \label{ext3} \\	
\setlength{\unitlength}{25pt}
\begin{picture}(0.0,0.0)(0.0,0.0)
  {\Large
	 \put(7.7,1.1){\vector(-1,-1){0.65}}
	 \put(8.3,1.1){\vector(+4,-1){2.433}}
   }
\end{picture}				
			 = \quad\!\!
		\Aamu\, \ol{\psi}\gamma^{\mu} 
	   E_{\alpha}	\psi \;\;\;\, + \;\;\;\,
    \Aamu
   \left(\ol{\varphi}(k^{'}_2)e^{+ik^{'}_2\cdot x}\right)
	\gamma^{\mu}E_{\alpha}
	 \left(\varphi(k^{'}_1)e^{+ik^{'}_1\cdot x}\right)
	  e^{-ik\cdot x}  
	    \,  	\,= \! & \! 0 &   \label{ext4}
\end{eqnarray}
\par}
\vspace{-33pt} 
\begin{equation}
 \mbox{provided} \quad  k_1 + k_2 \; = \; k \; = \; k^{'}_1 + k^{'}_2
  \qquad \qquad \qquad \qquad \qquad \qquad \qquad
  \qquad \qquad \; 
 \label{ksame}
\end{equation} 
\par}

 Between these equations the arrows track the components incorporated 
into the local structure of $G^{\mu\nu}(x) = f(A,\bv_n)$, 
generalising from equation~\ref{gtwopsi}.
  As seen in equations~\ref{ext1}--\ref{ext2} the $+$ve exponential 
parts of $\psi(x)$ and $\ol{\psi}(x)$ are free to \textit{slide over} 
to an $A_{\mu}(x)$ contribution of a similar form to a complex mode in 
equation~\ref{acosee} without impacting upon the other terms.
  Here the argument of $\Aamu(k)$ is the wave 4-vector $k$, rather 
than the 3-vector $\bk$, as these local intermediate exchanges will 
generally be `off-mass-shell', unlike for example the case of a 
propagating electromagnetic wave with $k^{\mu}k_{\mu} = 0$ 
 in equation~\ref{ecoeff} or for the `on-mass-shell' initial and 
final states more generally.
(Here we are strictly dealing with wave 4-vectors only, the connection with kinematic properties and `mass' will be considered in the following section).
 
  While we have initially considered the $+$ve  exponential modes
   since the exchanges involve different  ways of decomposing the 
\textit{same} terms of the constraints, as for 
equations~\ref{ext0}--\ref{ext4}, with the terms left invariant in 
all cases, the transfer of the $+$ve  and $-$ve  modes can be 
analysed  \textit{independently}. That is the $-$ve exponential 
components in equation~\ref{ext0} can also be propagated through 
terms in the constraints in a similar manner, contributing to the 
$D_-$ degeneracy count. This construction then makes use of both the 
multiplicative properties of the exponentials, with for example  
 $e^{+ik_1 \cdot x} \times e^{-ik_1 \cdot x} = 1$ substituted into equation~\ref{ext0} and
$e^{+ik_1 \cdot x} \times e^{+ik_2 \cdot x} =e^{+i(k_1+k_2) \cdot x} $ embedded identically in 
constraint terms such as in equations~\ref{ext1} and \ref{ext2}, as well as the 
additive property, with $e^{+ik_1 \cdot x} + e^{-ik_1 \cdot x} = 2\cos k_1\!\cdot\! x$ making a real field contribution to the external geometry 
through equations~\ref{gtwopsi} and \ref{phiee}.

 Exchanges of complex field components are then
 permitted provided that the modified hybrid field composition of 
equation~\ref{gfromavt} for $G^{\mu\nu}(x)$ describes the same real 
local geometry on $M_4$, taken to be a minimal distortion from the 
vacuum case as for processes studied in the HEP environment.
 The range of possible exchanges is also limited by the requirement of compatibility with the constraints through which they are transmitted. With 
for example no `selection rule' for the direct exchange between 
 $\ol{\psi}\psi \leftrightarrow \ol{\varphi}\varphi$ components the 
transfer can take place via the juxtaposition with $A_{\mu}$ 
components in the terms of equation~\ref{dlpvnbex}. A broader variety 
of underlying interactions will be allowed by the terms of the 
constraint equations~\ref{gchift}--\ref{emcon} more generally.

  The sequence of exchanges in equations~\ref{ext1}--\ref{ext4} 
contributes to equation~\ref{degplus} for $n=2$ with 
$R_{\mathrm{con}}(t_2)=1$ for the equation~\ref{ext1}$\to$\ref{ext2} 
field exchange and   
  $R_{\mathrm{con}}(t_1)=1$ for the 
equation~\ref{ext3}$\to$\ref{ext4} field exchange
 for an overall transition from an initial state described for 
equations~\ref{gtwopsi}--\ref{phieeb} to a final state associated 
with $\varphi(x)$ and $\ol{\varphi}(x)$ field components.
 With the temporal progression $t = 0 < t_2 < t_1 < T$ this evolution 
of the field contributions is pictured in figure~\ref{extofey}(a).
  While $\psibbar\!\!\!\!\:(x)$ and $\phibbar\!\!\!\!\:(x)$ may 
represent leptons or quarks, the gauge field $A_{\mu}(x)$ might 
represent an electromagnetic or weak gauge field, as exemplified in 
the related QFT diagram of figure~\ref{extofey}(b). 
  
%temporary
%\pagebreak  
  
\begin{figure}[htbp]  
\centering
\epsfxsize=14.4cm
\leavevmode
\epsffile[0 0 1639 516]{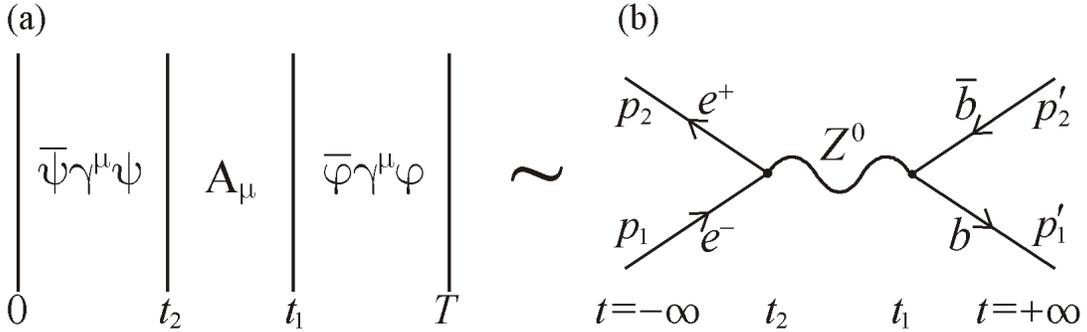}
\caption{\setb  (a) A depiction of the $+$ve exponential field 
component exchanges between time $t=0$ and $t=T$ 
 (equivalent to $t=-\infty$ and $t=+\infty$ in representing initial 
and final states respectively)
corresponding to the sequence of equations~\protect\ref{ext1}--\protect\ref{ext4}
  as
  contributing to the degeneracy count $D_+$ of 
equation~\protect\ref{degplus} for $n=2$ and (b) an associated Feynman 
diagram (with $t_2<t_1$ and
 $t_2>t_1$  equally implied by the mathematical structure of 
equation~\protect\ref{smatrix}) for the process $e^+e^- \to Z^0 \to
  b\bar{b}$ as identified in figure~\protect\ref{sldbb}.}
\label{extofey}
\end{figure}   
 
  In such a degenerate exchange of matter fields composing the local 
geometry $G^{\mu\nu} = f(A,\bv_n)$ as described above as well as the 
oscillation structure the normalisation for the \textit{amplitude} of 
the disturbance $G^{\mu\nu} =: -\kappa T^{\mu\nu}$ should be consistent, 
corresponding for example to a trade between 
 $\psi(k_1)/\ol{\psi}(k_2)$ and $\Aamu(k)$ components of the 
appropriate `magnitude' in the field exchange of equations~\ref{ext1}$\to$\ref{ext2}.
This compatibility must respect the fact that $G^{\mu\nu}(x)$ is 
everywhere a real tensor function, here composed of a consistent hybrid of complex $+$ve and $-$ve 
exponential field contributions  throughout such an 
interaction process.

 With terms such as $\ol{\psi}\gamma^{\mu}\! A_{\mu}\psi$ in 
equation~\ref{dlpvnbex} providing a `window of opportunity' for 
interactions
the likelihood of such a process should then in some sense be 
proportional to the relative `magnitude' of these terms.
 Relative `charge' factors associated with the terms in 
equation~\ref{dlpvnbex} will hence proportionately increase the 
probability of the corresponding field transitions
(see also the discussion in \cite{Unifi} following equations~11.33--11.35 and after 
figure~11.5). While the terms of equation~\ref{dlpvnbex} are closely 
analogous to Lagrangian terms in the form of equation~\ref{lagdir} 
further work is needed to connect the role of `charges' in the 
present theory with that in QFT calculations and to understand what 
determines the absolute scale of interaction rates.  
 Here the consistency in the amplitude of the geometric distortion 
described by the real function $G^{\mu\nu}(x)$ is then a further 
factor to be propagated through all terms in the constraints, such as 
for the sequence of equations~\ref{ext1}--\ref{ext4}.

 For the present theory propagating particles are associated with 
real field components, as for the full expressions of 
equations~\ref{acosee}, \ref{phiee} and \ref{phieeb} for example.
  In QFT initial and final state particles can be effectively 
represented by a single $e^{- ip\cdot x}$ or $e^{+ ip\cdot x}$ 
complex  component, with for example
 $\langle 0 \vert \hat{\phi}(x) \vert \bp\rangle = e^{- ip\cdot x}$ 
representing a one-particle wavefunction in spacetime for the state 
 $\vert \bp\rangle$ 
 (see also discussion here after equation~\ref{aacomr}, \cite{Unifi} equations~10.43 and 10.44, \cite{Pesk} 
equation~2.42). 
 More generally in the interaction picture, with quantum fields 
 such as $\hat{\phi}(x)$ in equation~\ref{kgosol2}
 composing terms in the interaction Hamiltonian $H_{\mathrm{int}}$
  of equation~\ref{htlagx},
  such as that of equation~\ref{lagfint},
 the action and algebra of creation $a^{\dag}(\bp)$
  and annihilation $a(\bp)$ operators 
 permit a series of non-zero contributions  to the 
 $S$-matrix in equation~\ref{smatrix} via  exchanges of $e^{\pm 
ip\cdot x}$ Fourier components between the fields. 
  The commutation relations of equation~\ref{aacomr},
  corresponding to the `quantisation' of the fields, lead to 
energy-momentum conservation in terms of preserving the total
 4-momentum, as identified directly with the
 wave 4-vector $p$, in each field interaction (as reviewed following 
equation~\ref{smatrix} with reference to \cite{Unifi} equations~10.45 
and 10.46 and figure~10.4).
 Hence the `quantisation rules' of equation~\ref{aacomr} as utilised 
in the expansion of equation~\ref{smatrix} effectively convert terms 
in the classical interaction Lagrangian into `selection rules' for 
contributions to the transition amplitude $\mcM_{fi}$ as a factor in 
determining a process probability  in proportion to  
$\vert \mcM_{fi} \vert^2$.

 As an intermediate stage in the construction of the $S$-matrix, 
before rearrangement for the time-ordering of operators in 
equation~\ref{smatrix}, the non-trivial part of the iterative 
solution for the unitary evolution operator in equation~\ref{uevolve} 
has the form (\cite{Unifi}~equation~10.31, as can be taken from an 
initial $t_0=-\infty$ to a final $t=+\infty$):
\begin{equation}
  U(t,t_0) - \mbox{\boldmath $1$} \; = \; \sum_{n=1}^{\infty} \;
     (-i)^n \! \int_{t_0}^t \! dt_1 \! 
	  \int_{t_0}^{t_1} \! dt_2 \ldots
	 \! \int_{t_0}^{t_{n-1}} \! dt_n  \,
  H_{\mathrm{int}}(t_1)  H_{\mathrm{int}}(t_2) \dots   
     H_{\mathrm{int}}(t_n) \,\,\,
    \label{uhiter}    	 				
\end{equation}
  Here we have subtracted the trivial $\mbox{\boldmath $1$}$ operator part which 
  does not contribute to $\mcM_{fi}$ (as noted before 
equation~\ref{sfi}).
This expression bears a close formal similarity to the structure of 
the degeneracy count $D_+$ in equation~\ref{degplus} described for the present theory.
 In equation~\ref{degplus} each factor $R_{\mathrm{con}}(t_i)$ collectively 
represents all terms of the constraint equations, while in 
equation~\ref{uhiter} each factor $H_{\mathrm{int}}(t_i)$ includes all terms of 
the interaction Hamiltonian, with non-zero contributions to both of 
these equations identified through consistent sequences of possible 
field exchanges.
 Given that for the present theory the total probability $P_{fi}$ in 
equation~\ref{ppdd} is proportional to $D_+D_-$ while in QFT the 
event cross-section is proportional to $\vert \mcM_{fi} \vert^2$, 
equations~\ref{degplus} and \ref{uhiter} in a sense each respectively 
correspond to a `square root' of the full calculation required. 
  The question is then to consider how these structures might be more 
closely related.

  In QFT the action of a creation operator $a^{\dag}(\bp)$ can be 
interpreted as `exciting' a positive energy quantum of the 
corresponding field in the form of a single $e^{\pm ip\cdot x}$ 
Fourier component while the annihilation operator $a(\bp)$ is 
associated with a corresponding \mbox{`de-excitation'} of the field 
(as discussed for \cite{Pesk} equation~2.47). For the present theory 
the act of `creation' or `excitation' for an $A(x),\bv_n(x)$ subcomponent simply 
means placing an $e^{+ ik\cdot x}$ or
 $e^{- ik\cdot x}$   field component \textit{into} the structure of the spacetime geometry
$G^{\mu\nu} = f(A,\bv_n) =: -\kappa T^{\mu\nu}$, hence making a corresponding 
contribution to the energy-momentum, while `annihilation' or `de-excitation' simply 
implies its \textit{removal}, as for example described for the 
sequence of wave 4-vector conserving $e^{+ ik\cdot x}$ exchanges of 
equations~\ref{ext1}--\ref{ksame}. (The connection between the 4-momentum 
and the wave 4-vector of equations~\ref{ehbaro} and \ref{phbark} for 
initial and final  physical particle states will be discussed in the 
following subsection).
 Such annihilation and creation sequences are seamlessly strung 
together through all terms of the constraint equations under
 the continuous enveloping geometric structure of 
 the local  external spacetime $M_4$.     

	 This is then analogous to quantum fields such as $\hat{\phi}(x)$ 
in equation~\ref{kgosol2} expressing the potential for field 
excitations in a flat Minkowski spacetime. 
 Further to this  analogy with the calculations of QFT, all the terms 
in the interaction Lagrangian for quantum fields are here replaced by 
terms in the constraint equations to act as selection rules for 
possible field exchanges contributing to event probabilities.  
   Rather than pragmatically constructing a Lagrangian 
$\lag_{\mathrm{int}}$, as for the model of equation~\ref{lagfint} or 
more specifically for the Standard Model itself including 
equations~\ref{lagsff}--\ref{lagdir}~\cite{Teub}, the possible field 
exchanges are here constrained by terms in expressions deriving 
directly from the general form of proper time in equation~\ref{lpvn} in the 
symmetry breaking projection onto $M_4$. 
The relation $\lpvng$ is \textit{not} considered as a function on a 
pre-existing spacetime manifold $M_4$, but rather provides the component 
structures out of which $M_4$, together with its geometric 
description, is \textit{itself} constructed through 
equation~\ref{gfromavt}.  
  From the symmetry breaking of $\lpvng$ implicit in this construction the resulting fragmented 
subcomponents of $A(x),\bv_n(x)$ are identified as potential matter field 
contributions to the spacetime geometry  $G^{\mu\nu} = f(A,\bv_n) =:  -\kappa T^{\mu\nu}$.

 That is,
for the present theory quantum phenomena and observable field 
interactions  arise through solutions for a specific local spacetime 
external geometry $G^{\mu\nu}(x)$ in equation~\ref{gfromavt} 
expressed over a local degeneracy of internal $A(x), \bv_n(x)$ field 
contributions, mutually linked by exchanges permitted within the 
constraint equations~\ref{gchift}--\ref{emcon}.
 While  $\psi(x)$ and  $\ol{\psi}(x)$ in equations~\ref{dlpvnbex} and \ref{ext0}--\ref{ext4}, as an application of equation~\ref{dlpvnb}, derive from  the same $\bv_n(x)$ fragment they make two individual 
 contributions to the geometry of $M_4$ in equation~\ref{gtwopsi},  as 
discussed after equations~\ref{phiee} and \ref{phieeb}.
 Similarly the 
cubic and quartic terms for \textit{non-Abelian} gauge fields 
underlying equation~\ref{gchift} imply the possibility of  locally
redistributing 
 factors of $e^{\pm ik\cdot x}$  between three or four contributions  
to $G^{\mu\nu}(x) = f(A)$ on $M_4$ from the same gauge field $A(x)$, such as 
associated with an internal $\suth_c$ strong gauge group, implying the possibility of `self-interactions' for non-Abelian gauge fields generally as alluded to after equation~\ref{gchift}.

  In QFT the likelihood of an event, as a transition between 
particular initial and final particle states, in terms of a 
cross-section $\sigma$ is determined by the non-trivial part of the  
unitary evolution operator in equation~\ref{uhiter} as arranged and 
expressed via a transition amplitude $\mcM_{fi}$ as described for 
equations~\ref{sfi}--\ref{smatrix} with 
  $\sigma \propto \vert \mcM_{fi}\vert^2$ (\cite{Unifi} 
equation~10.3).
 `Conservation of probability' is modelled in these calculations by 
the unitarity of the $S$-matrix, that is with $SS^{\dag}=\mbox{\boldmath $1$}$ 
or $\sum_f \vert S_{fi} \vert^2 =1$ as described for 
equation~\ref{fsumunit}, which also directly implies that a 
\textit{linear}  relationship between a cross-section and an 
amplitude can be identified.
   This is namely the `optical theorem' in which the total 
cross-section $\sigma_{\mathrm{tot}}$ for a process from a given 
initial state to any final state can be shown to be equal, up to a kinematic  
normalisation factor, to the imaginary part of the forward scattering  
transition amplitude $\mcM_{ii}$ via all possible intermediate states 
(\cite{Pesk} section~7.3 equation~7.50, \cite{Unifi} 
equations~10.87--10.96):
\begin{equation}
   \sigma_{\mathrm{tot}}(i \to \mbox{anything}) \; \propto \;    
\mbox{Im}({\mathcal M}_{ii}) 
    \label{optic}   
\end{equation}  

  Using a Feynman diagram analysis the optical theorem can be proven 
to all orders of perturbation theory, demonstrating a close link 
between the unitarity of the $S$-matrix and the Feynman rules. This 
involves employing `cutting rules' in which the propagators of a 
`cut' loop in a Feynman diagram are replaced with an on-mass-shell condition such that a 
sum over the contributions from each cut yields the imaginary part of 
the Feynman diagram (\cite{Pesk} figure~7.6 and equation~7.56, 
\cite{Unifi} figure~10.10 and equation~10.98, \cite{Cutk}, 
\cite{Velt} chapter~8). In breaking down the optical theorem to each 
perturbative order the cutting rules also demonstrate  how each 
individual possible final state contributes at each order to the 
total cross-section in equation~\ref{optic} (see discussion of 
\cite{Unifi} equations~10.99--10.101).
 
  For the present theory the calculation of an event probability 
$P_{fi}$ as constructed for equation~\ref{ppdd} in terms of the 
degeneracy count $D_+D_-$, in integrating over field exchanges for 
$e^{+i k \cdot x}$ and $e^{-i k \cdot x}$ parts, has significant 
features analogous to the determination of  $\mbox{Im}({\mathcal 
M}_{ii})$ in QFT described above.
Each contribution to $D_+D_-$ for a given final state can be 
correlated with a contribution to $\mbox{Im}({\mathcal M}_{ii})$ at 
the same order of perturbation.
 For both $D_+D_-$ and $\mbox{Im}({\mathcal M}_{ii})$ a real number 
is determined by integrating in time over a similar sequence of field 
interactions with a `fold' in the former case involving 
`on-mass-shell' components taking the place of the `cut' in the 
latter
(as described for \cite{Unifi} figures~11.7 and 11.8).
 In this manner equations~\ref{ppdd} are \ref{optic} can be 
provisionally linked, with the degeneracy count construction of an 
event probability for the present theory associated with standard 
cross-section $\sigma$ calculations in QFT (\cite{Unifi} 
equation~11.46) via:
\begin{equation}
      P_{fi} \, \sim \,  D_+ \, D_-  \, \sim \, 
	     \mbox{Im}({\mathcal M}_{ii}) \, \sim \, 
		      \vert \mcM_{fi}\vert^2    
    \label{pdddds}
\end{equation}

  As well as the care needed to match for example the time sequence 
structure and the normalisation of initial and final states, 
  a main caveat here is the need to disentangle the appropriate final 
state contributions for $\mbox{Im}({\mathcal M}_{ii})$ to link 
specifically with $\mcM_{fi}$ in these relations, in place of the 
inclusive composition for $\sigma_{\mathrm{tot}}$ in 
equation~\ref{optic}. The nature of this disentanglement might be 
initially explored at low orders of perturbation (as alluded to in 
the discussion in \cite{Unifi} after figure~11.8).
 With such caveats in mind,
  progression through equation~\ref{pdddds} from left to right 
provisionally represents the calculation for an event probability 
$P_{fi}$ in terms of the `real' measure degeneracy count
$D_+D_-$ being `complexified' through the introduction of an 
effective interaction Lagrangian and field `quantisation' structures 
with $\mbox{Im}({\mathcal M}_{ii})$ interpreted as modelling  
selection rules for allowed field transitions involving specific $i$ 
and $f$ states as ultimately expressed via the structure of the 
transition amplitude $\mcM_{fi}$. In this manner `unitarity' is 
employed to ensure probabilities are suitably normalised at each 
stage as ever higher orders of nested field exchanges are considered 
with all possible final state outcomes implicitly accounted for, addressing the comment after equation~\ref{ppdd}.   
  
   Looking at equation~\ref{pdddds} the other way the present theory 
is proposed to provide an underlying conceptual foundation, origin 
and explanation for calculations based on transition amplitudes in 
QFT.
  The unitary symmetry on the QFT side both models the conservation 
of probability and implies the optical theorem. From
 $\vert \mcM_{fi}\vert^2$ it is then through the optical theorem and structure of 
 $\mbox{Im}({\mathcal M}_{ii})$ that 
the link can be made with the count of the `number of ways' $D_+D_-$ a process 
can happen through
the local degeneracy in the spacetime composition for
 equation~\ref{gfromavt} as a means by which an event likelihood 
$P_{fi}$ can be determined.
  Here the relative probability for an observable process has a more 
\textit{classical} basis involving a direct \textit{summing}, or 
integrating, over temporal sequences of permitted field contributions 
underlying the construction of spacetime itself as described for 
equation~\ref{degplus}.

  The merging of the degeneracy count determination for the 
probability $P_{fi}$ with the transition amplitude $\mcM_{fi}$ 
calculation through terms of an interaction Lagrangian is discussed 
further in (\cite{Unifi} section~11.2). The link specifically with 
Feynman vertices and Feynman propagators, as initially intimated in
  figure~\ref{extofey}, as well as with the optical theorem is 
summarised in particular in (\cite{Unifi} section~11.2  points 
`1)--7)'). Feynman diagrams, composed of such vertex and propagator 
elements, are considered pragmatic tools for calculations, in 
particular associated with the restructuring of equation~\ref{uhiter} 
for equation~\ref{smatrix} on employing the time-ordering of operator 
products, and are \textit{not} interpreted as literally representing 
particle exchanges (as discussed for \cite{Unifi} figures~10.5 and 
10.6 and in the opening of section~10.5).
 Here the `virtual particles' corresponding to internal Feynman 
propagators, such as for the $Z^0$ in figure~\ref{extofey}(b), while correlated with particular field contributions  are 
considered a purely mathematical construction deriving from the 
reworking of the probability $P_{fi}$ calculation through 
equation~\ref{pdddds}.

  In contrast to QFT the present theory does provide an account of 
\textit{what is actually happening} in HEP events such as that of 
figure~\ref{sldbb} through an interplay of mathematical and physical 
elements. While all processes are enveloped under a continuous 
solution for the smooth \mbox{4-dimensional} spacetime geometry through 
equation~\ref{gfromavt} the likelihood of a range of outcomes from an 
`interaction point' or `decay vertex' in a particle process is 
determined by a direct \textit{mathematical} count of the degeneracy 
of possible matter field contributions, analysed in terms of complex $e^{\pm ik\cdot x}$ 
components, maintaining the equivalent external local geometry, as 
described in this subsection.
 On the other hand the macroscopically propagating \textit{physical} 
particles, corresponding to the initial and final states for the 
event of figure~\ref{sldbb} for example, are associated with extended 
elements of solutions for the spacetime geometry in 
equation~\ref{gfromavt} involving particular real matter field 
components and corresponding equations of motion, again as consistent with all the constraints of 
equations~\ref{gchift}--\ref{emcon}.
 These real physical particle states will include leptons as well as 
hadrons composed of quarks, such as the $B$-hadrons containing a 
$b$-quark component identified in the event of figure~\ref{sldbb}. 
 The nature of these particle states, as provisionally introduced in 
the previous subsection, will be considered further below and in 
particular in the following section.  

  While we described a plane wave solution  propagating in both the 
electromagnetic field, equation~\ref{acos}, and the gravitational field, 
equation~\ref{gconfu}, as noted in the discussion after 
equation~\ref{wavecom} a more localised wave-packet structure is 
anticipated to describe physical particle states such as photons and 
electrons. Similarly, while the QFT calculations described for 
equations~\ref{kgosol2}--\ref{smatrix} employ idealised plane wave 
expressions, initial and final states can be represented in QFT more 
realistically by localised wave-packets with the single complex mode 
in the 4-momentum $p$ replaced as (\cite{Pesk}~equation~7.39):
\begin{equation}
 e^{+ip\cdot x} = e^{ip^0 x^0} e^{-i\bsl{p} \cdot \bsl{x}} \to
   \int \frac{d^3 \bsl{p}}{(2\pi)^3}
   e^{ip^0 x^0} e^{-i\bsl{p} \cdot \bsl{x}} F(\bp)
  \label{fpack} 
\end{equation} 
   where $F(\bp)$ is the Fourier transform of the localised 
wave-packet shape in physical \mbox{3-dimensional} space. Provided 
$F(\bp)$ is a sufficiently narrow function about a central value in 
momentum space, the results of QFT calculations only depend upon that 
central momentum value and not the details of the shape (\cite{Pesk} 
section~4.5 opening of subsection on `The $S$-matrix').

   In position space the wave-packet might typically take the shape 
of a symmetric 3-dimensional Gaussian distribution centred on the 
particle trajectory which, classically and non-relativistically in 
terms of the wave 4-vector $k$, would be of the form: 
\begin{equation}
 e^{+ik\cdot x} = e^{ik^0 x^0} e^{-i\bsl{k} \cdot \bsl{x}} \to
      e^{ik^0 x^0} e^{-i\bsl{k} \cdot \bsl{x}}
	  \; A\, e^{\left[-(\vert\bsl{v}_g\vert x^0 - x_{\parallel})^2 - 
(x_{\perp})^2\right]/b^2}
  \label{wpack}
\end{equation}
  Here $A$ and $b$ are real parameters, $x_{\parallel}$ and 
$x_{\perp}$ are the spatial displacements parallel and transverse to 
the particle motion, and $\bv_g = dk^0/d\bk$ is the wave group 
velocity which is equivalent to the velocity of the particle 
described
 (unlike the wave phase velocity \mbox{$\bv_p = k^0/\bk$}).
For the non-relativistic case, with the kinematic relation 
$E=\bp^2/2m$ between energy, 3-momentum and mass, on adopting the 
identifications of equations~\ref{ehbaro} and \ref{phbark} with 
$\omega = k^0$ and $\hbar =1$ we have $k^0 = \bk^2/2m$ and hence
  $\bv_g = \bk/m = \bp/m = \vkin$ consistently.
   That is, the group velocity $\bv_g$ of the wave-packet is equal to 
the kinematic velocity $\vkin$ of the corresponding particle state.
 The same Gaussian amplitude in equation~\ref{wpack} can also of 
course be applied to an $e^{-ik\cdot x}$ mode and for a real oscillation  $\cos k \! \cdot \! x$ such as that  in equation~\ref{acosee}.

  For the present theory we are considering what might be termed 
`generalised wave-packets', describing a propagating localised mutual 
disturbance of a real field in the internal subcomponents of $A(x)$ or $\bv_n(x)$ 
\textit{as well as} in the external spacetime geometry 
$g_{\mu\nu}(x)$. These internal and external disturbances are related in providing a 
solution for equation~\ref{gfromavt}  
 consistently within the various constraint 
equations~\ref{gchift}--\ref{emcon} for the theory. It remains to 
determine an explicit shape for the generalised wave-packets, which 
may not be simple Gaussian functions in all cases, as well as the 
relation between the internal and external field wave-packet 
structure for different cases, including in relation to Maxwell's 
equation and the Dirac equation.
  The means described in this subsection for calculating the likelihood of a particle interaction process, concerning to the left-hand side of equation~\ref{pdddds}, might also be anticipated to carry over from the plane wave analysis to a more physically realistic wave-packet approach, as is the case for the associated  QFT calculations as linked on the right-hand side of  equation~\ref{pdddds} and as noted after equation~\ref{fpack}.

 Regardless of the specific wave-packet shape a group velocity $v_g = 
dk^0/d\bk$, as described for equation~\ref{wpack} and as also applies 
for the relativistic case, will be associated with the effective 
particle state for which the corresponding kinematic properties are 
 represented in the present theory by the energy-momentum tensor $T^{\mu\nu}(x)$ in 
equation~\ref{gfromavt}.
 The mathematical consistency between the structure of these wave and 
kinematic properties will be key to identifying discrete particle quanta as we 
 describe in the following subsection.
  Such `generalised wave-packets' are to be identified not only for 
direct particle trajectories, as seen in figure~\ref{sldbb}, but also 
more generally for single particle  solutions  
  such as that implied for the experiment of figure~\ref{dslit}
  with the state apparently `passing through both slits at the same 
time', as will  be discussed in subsections~\ref{qugr62} and in particular 
 \ref{qugr74}.

 While the generalised wave-packets may propagate over macroscopic 
distances, we have described in this subsection how the likelihood of 
an interaction between such particle states is determined in terms of 
local structure. 
The geometry associated with a local interaction between particle 
wave-packets, as described by a metric function $g_{\mu\nu}(x)$, will 
in general be consistent with solutions for equation~\ref{gfromavt} 
over a range of $A(x),\bv_n(x)$ field component contributions. This 
is what is meant by a `degeneracy of solutions for the 4-dimensional 
spacetime structure' as introduced in section~\ref{qugr4}. 
 The restriction on the range of degenerate field states is in fact 
somewhat looser in the sense that it is not actually a specific 
$G^{\mu\nu}(x)$ or $g_{\mu\nu}(x)$ \textit{function} that needs to be 
invariant but rather it is the \textit{intrinsic geometry} that is 
required to be preserved.

As alluded to in the discussion after equation~\ref{wavecom} in the 
previous subsection, the same physical geometry can be 
represented by a range of
 $g_{\mu\nu}(x)$ metric functions as related by freedom in the choice 
of coordinates on $M_4$, with 
same local geometry corresponding to local solutions for the metric 
$g_{\mu\nu}(x)$ that are equivalent within superfluous coordinate 
transformations. 
 As noted in the discussion of gravity waves in 
subsection~\ref{qugr22} such a `gauge' choice of coordinates may be 
arbitrary or \mbox{$k$-dependent} according to convenience. Hence in 
considering the transfer of matter field components underlying 
equation~\ref{gfromavt} for initial, intermediate, and final states, 
as described in equations~\ref{ext1}--\ref{ext4} for example, care 
will be needed to assess whether apparent changes in the metric 
function $g_{\mu\nu}(x)$ are equivalent to unphysical coordinate 
changes and are hence actually describing exactly the same intrinsic 
local geometry.

  These subtleties concerning the equivalent local geometry
 required for initial and final states to be connected through an 
interaction involving a sequence of degenerate field solutions 
	apply to some extent also for the description of individual 
propagating particles, even for the simplest case of a scalar field and 
scalar particle states. In practice real cases will involve both 
spin-1 gauge particles deriving from the internal symmetry $G$ in equation~\ref{gbreak}, such as associated with the electromagnetic 
field $A_{\mu}(x)$ of equation~\ref{ecoeff}, and spin-$\frac{1}{2}$ 
leptons and quarks (as components of hadronic states in the latter case) from the 
fragmentation of $\bv_n(x)$ in the symmetry breaking of 
equation~\ref{lpvn}. Such lepton and quark states were labelled $\psibbar\!\!\!\!\:(x)$ and 
$\phibbar\!\!\!\!\:(x)$ in equations~\ref{dlpvnbex}--\ref{ext4}. With spin 
associated with torsion the full wave-packet solutions for 
equation~\ref{gfromavt} for these states may require engaging 
elements of Einstein-Cartan theory as alluded to in subsection~\ref{qugr22} shortly after equation~\ref{einlamt}.
 In this case for non-trivial torsion both the metric $g_{\mu\nu}(x)$ 
and a linear connection
 $\Gamma^{\rho}_{\ph{\rho}\mu\nu}(x)$ are needed to describe the 
external geometry (see also \cite{Unifi} discussion in section~13.3 
pages 422--423). 
 
With a number of subtleties hence remaining 
  further rigorous mathematical development and investigation will be 
needed into all areas of the present theory, including the structure 
and links relating to equation~\ref{pdddds} for particle interactions 
described in this subsection. The aim has been to show how progress 
can be made in developing the mathematical structure of the theory in 
a manner capable of addressing significant questions that remain 
unanswered or untouched in most other approaches to quantum gravity.
 With this goal in mind in the following section we consider the 
origin of one of the most elementary properties of particle states.

%\pagebreak
\section{On the Underlying Origin of Particle Quanta}
\label{qugr6}

  One of the oldest and most elementary properties of quantum theory,
  and the eponymous protagonist of the theory itself,
   is the existence of discrete particle `quanta' as categorised by 
the de Broglie relations of equations~\ref{ehbaro} and \ref{phbark}.
 For the present theory we can consider how these relations and the 
nature of particle quanta at the most basic level of matter could 
originate ultimately from the simple foundation for all phenomena in 
a one-dimensional progression in time and the elementary structures 
involved, as we describe in this section.

\subsection{The Basic Argument from Constraints of the Theory}
\label{qugr61}

   Any real function $\Upsilon\in\rrr$ dependent upon time $s\in\rrr$ 
can be decomposed into Fourier modes, as elementary oscillations with  
angular frequency $\alpha\in\rrr$ and amplitude $B\in\rrr$, of the 
form:
\begin{equation}
  \label{fcosas}
     \Upsilon(s) = B\cos \alpha s 
\end{equation}
   In this theory with $s\in \rrr$ representing proper time the 
arithmetic composition of the real line allows the direct 
identification of a substructure for invariant intervals such as the 
form of equation~\ref{sfourd}:
\begin{equation}
  \label{sfourx}
  (\delta s)^2 \; = \; (\delta x^{0})^2 - (\delta x^{1})^2
                     - (\delta x^{2})^2 - (\delta x^{3})^2
\end{equation} 
  This introduces the Lorentz 4-vector of intervals
  $\delta x = (\delta x^{0},\delta x^{1},
           \delta x^{2},\delta x^{3}) \in \rrr^4$ upon which a 
Lorentz symmetry acts leaving $\delta s$ invariant.
  Through the translation symmetry over
   a full range of values for $x\in \rrr^4$  the 
   set of four intervals $\delta x \in \rrr^4$ in this structure 
provides the basis for an extended 4-dimensional manifold, with a 
flat Minkowski metric $\eta_{ab}$ and a global Lorentz symmetry 
(\cite{KKone} subsection~2.1).
  With the one-dimensional progression in time manifested in this 
4-dimensional spacetime form the elementary geometry of a 
3-dimensional spatial arena is incorporated, exhibiting Euclidean 
properties through the metric structure of the local coordinate 
intervals 
    $(\delta x^{1},\delta x^{2},\delta x^{3})$ in 
equation~\ref{sfourx} (see also discussion following 
equation~\ref{lpvn}). 

   Having identified an extended Minkowski manifold in this manner 
the simple oscillation of equation~\ref{fcosas} for the 
one-dimensional case under the generalisation of proper time to the 
form of equation~\ref{sfourx} can itself be represented
 as a function of the four parameters $x\in \rrr^4$ spanning 
  the 4-dimensional spacetime on applying Lorentz transformations such that: 
\begin{eqnarray}
   \Upsilon(s) = B\cos \alpha s \quad\! \to &  
                    \Upsilon(x^0) = B\cos k^0x^0
     &  \to \quad\!  \Upsilon(x) = B\cos k\! \cdot\! x 
	    \label{fcoskx} \\
    s \qquad  \to &   x^0  & \to  \;\;\; x=(x^0,x^1,x^2,x^3)  
	  \nonumber  \\
  \alpha \qquad  \to & 
             k^0 & \to \;\;\; k=(k^0,k^1,k^2,k^3)
		 \quad\;\;	  \nonumber  
\end{eqnarray}  		
  The wave 4-vector $k\in \rrr^4$, with
    $k\! \cdot\! x = \eta_{ab}k^ax^b$ and $(k)^2 = \eta_{ab}k^ak^b$, 
has the same Lorentz transformation properties as the coordinate 
4-vector $x\in \rrr^4$, as required for the argument 
\mbox{$\alpha s \to k^0x^0 \to k\!\cdot\! x$} to be a Lorentz 
invariant.
 That is $k\in \rrr^4$ is a generalisation from $\alpha \in \rrr$ via 
the assignment $\alpha \equiv k^0$ similarly as  $x\in \rrr^4$ is a 
generalisation from $s \in \rrr$ via the assignment $s\equiv x^0$, 
and as directly associated in equation~\ref{fcoskx}.

  The form of proper time in equations~\ref{sfourd} and \ref{sfourx} 
can be generalised further to equation~\ref{salpha} and written more 
conveniently in the form of equation~\ref{lpvn}. 
 From this augmented $n$-dimensional form, within this
  $L_p(\bv_n)_{\hG} = 1$ invariant structure, the projected
 4-vector  subcomponent $\bv_4 = \{v^a\} = \{\delta x^a/\delta s\} 
\in \TM_4$ in the local 4-dimensional tangent space can itself have 
varying magnitude $h(x) = \vert \bv_4(x) \vert = 
 (\eta_{ab}v^av^b)^{\frac{1}{2}}$, with $a,b = 0,1,2,3$.
   This variation is associated with conformal deformations from the 
otherwise flat spacetime with 
$g_{\mu\nu}(x) = \eta_{\mu\nu}/h^2(x)$, expressed in global general 
coordinates with indices $\mu,\nu = 0,1,2,3$, as described for 
equation~\ref{gwarph}.
 With the metric $g_{\mu\nu}(x)$ being the most elementary and 
intrinsic field on $M_4$, and also having the above direct means of 
variation in this theory, at the simplest level the oscillation 
$\Upsilon(x)$ in equation~\ref{fcoskx} can be associated with the 
metric function:
\begin{equation}
   g_{\mu\nu}(x) = \eta_{\mu\nu} (1 + B\cos k \!\cdot\! x) 
   \label{gbwave}
\end{equation}
 Here, with reference to the above discussion,  
 we have adopted an approximately flat Minkowski frame
 with a choice of appropriately scaled global coordinates   
corresponding to the transformation $h(x) \to (1+ B \cos k \!\cdot\! 
x)^{-\frac{1}{2}} \simeq (1 - \frac{B}{2} \cos k \!\cdot\! x)$ for $\vert B 
\vert\ll 1$.
  This tiny distortion from a flat metric propagating with wave 4-vector 
$k$ over values of $x \in M_4$ is of a very similar form to that in 
equation~\ref{gconfu} except for the factor of two in the cosine 
argument there and  with $(k)^2 = k^{\mu}k_{\mu} = \alpha^2 \neq 0$ 
here from equation~\ref{fcoskx} as originating in 
equation~\ref{fcosas}. 

  On assuming small perturbations from a flat spacetime the 
approximations of linearised general relativity can be adopted, with 
the Einstein tensor corresponding to the above metric for $\vert B 
\vert\ll 1$ obtained from the left-hand side of equation~\ref{Einlin} 
as:
\begin{equation}
  G^{\mu\nu}(x) = B\left( (k)^2 \eta^{\mu\nu}
      -k^{\mu}k^{\nu} \right) \cos k \! \cdot \! x
 \label{gkekk}	  
\end{equation} 
  Again, this is very similar to the second term in 
equation~\ref{gwave} as associated with equation~\ref{gconfu} for the 
original case with $(k)^2 = 0$.
  As for equation~\ref{gwave}, here equation~\ref{gkekk} is also a 
wave of purely Ricci curvature in being associated with the conformal 
metric perturbation of equation~\ref{gbwave}. The description in 
terms of the Einstein tensor $G^{\mu\nu}(x)$ is convenient owing to
 the contracted Bianchi identity of equation~\ref{Bian}, as satisfied 
here for the flat spacetime limit with:
\begin{equation}
  \pal_{\mu} G^{\mu\nu}(x) = -B\left( (k)^2 k^{\nu}
      -(k)^2 k^{\nu} \right) \sin k\! \cdot \! x  = 0
	\label{biankk}  
\end{equation} 

  In aiming to describe `particle phenomena' such disturbances can be 
combined to propagate as a localised wave-packet, as discussed for 
equations~\ref{fpack} and \ref{wpack}. 
 Here such a wave-packet is composed of a superposition of 
modes in the form of equation~\ref{gbwave},
 with each contribution considered to originate
 from the same 
oscillation in equation~\ref{fcosas} with frequency $\alpha$
  through an active
  Lorentz boost as described for  equation~\ref{fcoskx}  and 
hence with 
 $(k)^2 = \alpha^2 = (k^0)^2 - \bk^2$, in the collective form:
\begin{eqnarray}
   g_{\mu\nu}(x) & = & \eta_{\mu\nu} \left(1 + 
   \int d^3 \bk \,
   B(\bk) \cos k \!\cdot\! x \right) 
   \label{gbwavesup}  \\
    G^{\mu\nu}(x) & = &  \int d^3 \bk \,  B(\bk)
     \left( (k)^2 \eta^{\mu\nu}
      -k^{\mu}k^{\nu} \right) \cos k \! \cdot \! x
 \label{gkekksup}	
\end{eqnarray} 
     The second equation above follows from equations~\ref{gbwave}, \ref{gkekk} and \ref{gbwavesup} by the linearity of equation~\ref{Einlin}. This superposition is still of course consistent with the Bianchi identity 
    $\pal_{\mu} G^{\mu\nu}(x) = 0$ with equation~\ref{biankk} applying for each Fourier mode in equation~\ref{gkekksup}.
  
  With $k^0 = \sqrt{\bk^2 + \alpha^2}$ the group velocity for such a 
wave-packet is then:
\begin{equation}
  \bv_g = \frac{\pal k^0}{\pal \bk}
   = \frac{\bk}{\sqrt{\bk^2 + \alpha^2}}  = \frac{\bk}{k^0}
   \label{vgroup}
\end{equation}
  This expression $\bv_g = {\bk}/{k^0}$ has the same form as the 
kinematic relation for the \mbox{3-velocity} 
  $\vkin = {\bPP}/{P^0}$ for 
   a relativistic particle with 4-momentum $P^{\mu} = (P^0,\bPP)$. 
  However, unlike for the non-relativistic group velocity case 
discussed after equation~\ref{wpack}, here
 equation~\ref{vgroup} is determined as a purely \textit{wave} 
expression and we have \textit{not} yet introduced any notion of 
momentum  or assumed any connection between the wave \mbox{4-vector}~$k$ and a momentum \mbox{4-vector} $P$ associated with these 
wave-packet states or presupposed any kinematic relations.
 The \textit{implication} from equation~\ref{vgroup} is that, as a 
direct consequence of the required Lorentz transformation properties 
of  \mbox{$k = (k^0,k^1,k^2,k^3)$} in constructing  
equations~\ref{fcoskx} and \ref{gbwavesup}--\ref{vgroup} this wave 4-vector will 
necessarily be \textit{directly proportional to} the 4-momentum  
carried by the wave-packet, that is with a single real normalisation 
factor relating the 4-vectors $k$ and $P$. Further, as can be seen from the 
working for equation~\ref{vgroup}, the real parameter $\alpha$ will 
itself be proportional to the apparent invariant mass of the 
corresponding particle state. Hence from the elementary consistency 
requirement that $\vkin = \bv_g$ we surmise that for each propagating wave-packet 
state constructed in this manner:
\begin{equation}
         P^{\mu} \propto k^{\mu} \quad \mbox{with} \quad
		 m \propto \alpha
  \label{ppropk}		  
\end{equation}		 

  For the present theory energy-momentum $T^{\mu\nu}(x)$ is defined 
directly through identification with the Einstein tensor as described 
for equation~\ref{gfromavt} and as will apply here for the wave-packet of
equation~\ref{gkekksup}. Via equation~\ref{emcon} the energy-momentum 
for such oscillating structures will be conserved,  
with \mbox{$\pal_{\mu}T^{\mu\nu}=0$} as seen explicitly from 
equation~\ref{biankk} as discussed after equation~\ref{gkekksup}. However physical oscillations, as for all 
solutions for the spacetime geometry, will be obtained not as 
arbitrary perturbations from the vacuum but rather here the $\delta 
h(x)$ variations underlying equations~\ref{gbwave}--\ref{gkekksup}  will be moderated  through the constraints of 
equation~\ref{gfromavt} and equations~\ref{gchift}--\ref{emcon}. 
 In obtaining such a solution for 4-dimensional spacetime from 
generalised proper time through equation~\ref{gfromavt} oscillations 
in the external metric $g_{\mu\nu}(x)$ will be accompanied by oscillations in internal 
field subcomponents of $A(x),\bv_n(x)$ that are identified in the symmetry 
breaking of equation~\ref{lpvn} and determine the matter field associated with the wave-packet state. This notion of a `generalised 
wave-packet' was introduced in the discussion following 
equation~\ref{wpack}.

The condition of non-negative energy density, discussed in 
subsection~\ref{qugr22}, suggests that a physical solution identified 
in this manner may involve a further contribution to the geometry, 
similarly as seen in the first term    
 of equations~\ref{twave} and \ref{gwave} for the case of an 
electromagnetic wave,  incorporated into the full wave-packet 
structure. 
  This full wave-packet solution for equation~\ref{gfromavt} will still be associated with 
oscillatory components and the group velocity of 
equation~\ref{vgroup}.

  For such a wave-packet an integral can be performed of the 
\mbox{$T^{\mu 0}(x) := \frac{-1}{\kappa}G^{\mu 0}(x)$} components of the energy-momentum tensor over a 
3-dimensional spatial volume $V$ enclosing the wave-packet, with
  four time-independent quantities then defined as: 
\begin{equation}
   P^{\mu} = \int_{V} T^{\mu 0} d^3\bx
      \quad \implies \quad \pal_0 P^{\mu} = 0
 \label{pcon}	  
\end{equation} 
    The first expression implies the second given that
	$\pal_{\mu}T^{\mu\nu} = 0$ implies 
	$\pal_0T^{\mu 0} = -\pal_{i}T^{\mu i}$ (summing over $i=1,2,3$) 
together with an application Gauss's divergence theorem and the 
vanishing of $T^{\mu\nu}(x)$ on the boundary of $V$ for the localised 
wave-packet. (This then differs from the argument via Noether's 
theorem for energy-momentum deriving from a Lagrangian, discussed for 
\cite{Unifi} equation~3.102, as will be utilised for 
equation~\ref{pconcl}).
 The conserved Lorentz 4-vector $P^{\mu}$ in equation~\ref{pcon} is 
then identified with the actual \mbox{4-momentum} carried by the 
particle state, with mass $m=\sqrt{(P^0)^2-\bPP^2}$, associated with 
the generalised wave-packet propagation. 
 The conservation of 4-momentum in equation~\ref{pcon} will also 
apply for all localised energy-momentum, such as through a local 
interaction in a volume $V$ between initial and final particle 
states.

 The wave 4-vector $k$ is also conserved in all local particle 
interaction processes via the algebra of $e^{\pm k\cdot x}$ mode 
exchanges through factors of terms in the constraint equations~\ref{gchift}--\ref{dlpvnb}.
 This is seen explicitly
 for the example of equations~\ref{ext1}--\ref{ksame} through the 
constraint of the terms of equation~\ref{dlpvnbex}.
  This construction reflects the invariance of such terms through all intermediate internal field exchanges underlying the same local external physical spacetime geometry as permitted 
by the constraints in performing the internal field degeneracy count for the particle interaction.
 In particular for the initial $i$ and final $f$ states we have:
\begin{equation}
    \sum_i k^{\mu}_i \, = \, \sum_f k^{\mu}_f
\label{ksamif}	 
\end{equation}
 as exemplified by the left and right-hand sides of 
equation~\ref{ksame}, as a necessary condition for elementary 
interaction processes.

  Given the conservation of the total 4-momentum via 
equation~\ref{pcon} and of the total wave 4-vector in 
equation~\ref{ksamif}, the preservation of $\;\!\sum P^{\mu}$ and of 
$\;\!\sum k^{\mu}$ independently through any interaction implies that 
the proportionality between $P^{\mu}$ and $k^{\mu}$ for particle 
wave-packet states as described for equations~\ref{vgroup} and 
\ref{ppropk} must be the same for the initial and final states. 
 A universal relationship can hence be concluded for particle states 
propagating between such interactions as:
 \begin{equation}
         P^{\mu} = \hbar k^{\mu}
  \label{phbarku}		  
\end{equation}	
   This constancy in the proportionality, as denoted by $\hbar \in \rrr$, 
completes the kinematic interpretation of wave propagation 4-vector $k$. This 
fixed relationship is a consequence of the mutual constraints between 
the internal field structure, as described for 
equations~\ref{ext1}--\ref{ksame} and \ref{ksamif}, and the external 
geometric structure, with the properties of equations~\ref{emcon} 
and \ref{pcon}, in identifying minimal wave-like propagation and 
local interaction solutions for equation~\ref{gfromavt} in spacetime.   
  In this manner, leading to equation~\ref{phbarku}, we identify an 
argument to \textit{account} for the de Broglie relations of 
equations~\ref{ehbaro} and \ref{phbark}, and hence the origins and 
nature of discrete particle quanta, without needing to postulate this 
property.

 A key element of this argument is the necessary matching of the 
wave-packet group velocity $\bv_g$ of equation~\ref{vgroup}, deriving from the special 
relativistic transformation properties of the wave 4-vector $k$ in 
equation~\ref{fcoskx} as utilised in the geometric distortion of 
equations~\ref{gbwavesup} and \ref{gkekksup}, and the kinematic velocity $\vkin$ of the 
corresponding particle states, as implied for the 4-momentum $P$ identified via the geometric 
Bianchi identity of general relativity as described for 
equations~\ref{biankk} and \ref{pcon}.
 This is required in order to obtain a coherent solution for the 
geometric structure of 4-dimensional spacetime through 
equation~\ref{gfromavt}, hence explicitly incorporating the 
gravitational field in an irreducible manner.

  By contrast in a more standard approach to quantum particle 
phenomena gravity is completely neglected while the relation of 
equation~\ref{phbarku}, that is equations~\ref{ehbaro} and 
\ref{phbark}, is \textit{assumed} from the outset.
 For the non-relativistic case, with no intrinsic relation between the 
temporal $k^0$ and spatial $\bk$ components of a wave, it was 
necessary to assume both the de Broglie relations and the kinematic 
relation $E=\bp^2/2m$ in order to obtain the wave-packet group 
velocity $\bv_g = \vkin$ as discussed after equation~\ref{wpack}.
 In standard relativistic wave mechanics the fact that the group velocity of a 
wave-packet `turns out to equal the particle velocity' 
 (\cite{Wein} volume 1 noted after equation~1.1.1) \textit{allows} a physical 
picture in which `particles are thought of as wave packets'
 (\cite{Pea} section~6.1 opening) as a seemingly convenient 
observation.
 
 For the present theory, with gravity playing an essential role,
  it is through the mathematical gears of Riemannian geometry in 
4-dimensional spacetime that a time-dependent oscillation
with wave 4-vector $k$ in the metric geometry $g_{\mu\nu}(x)$
 can be expressed in the form of equation~\ref{gbwave} and wave-packet superposition of equation~\ref{gbwavesup} and then 
churned out through \mbox{$G^{\mu\nu}(x) =: -\kappa T^{\mu\nu}(x)$} into a  
conserved 4-momentum $P$ via  equations~\ref{gkekksup} and \ref{pcon}.
 The demands of consistency then \textit{result} in 
equation~\ref{phbarku} as described above.
  Since the determination of  $\bv_g = {\bk}/{k^0}$  in equation~\ref{vgroup} for the wave 4-vector $k$ and of   $\vkin = {\bPP}/{P^0}$ for the 4-momentum $P$ derive from a different basis, there is an open question  concerning to what  extent it is inevitable that they have the same functional form, or to what extent it `turns out' that way.
  In any case, with the \mbox{4-vectors}  $k$ and $P$ transforming in the same way under arbitrary Lorentz transformations the argument here based on the necessary equivalence of $\bv_g$  and $\vkin$ in leading to equation~\ref{phbarku} also holds in the non-relativistic limit.

  The wave 4-vector $k$ itself is obtained through the 4-dimensional 
generalisation of the angular frequency $k^0$, in turn originating 
from the oscillation rate $\alpha$ of equation~\ref{fcosas} as 
described for equation~\ref{fcoskx}, with practical units of 
[seconds]$^{-1}$.
 Through equation~\ref{phbarku} this wave component $k^0$ is directly 
associated with 
 the energy component $P^0$ as might be measured for example in 
[Joules].
  The empirically determined value of Planck's constant $\hbar \simeq
 10^{-34}\,$J$\,$s~\cite{PDG20}, being numerically very far from unity,  
reflects the historical origin of the everyday units employed. 
  For applications in the particle physics laboratory, where 
phenomena associated with equation~\ref{phbarku} are observed, it is 
typically more convenient to employ natural units with $\hbar =1$.
  In quantum field theory setting $\hbar=1$ corresponds to the 
 \textit{immediate} identification of a wave 4-vector with a particle 
4-momentum, as labelled by $p$ and noted for equation~\ref{kgosol2}, 
 explicitly incorporating the \textit{postulate} of
  equations~\ref{ehbaro} and \ref{phbark}.    
  Here, by contrast, the wave 4-vector $k$ and particle 4-momentum 
$P$ are introduced as \textit{independent} constructions that are then found 
to be intimately related in equation~\ref{phbarku} through the 
constraints involved in identifying interacting particle states.

  The role of $\hbar$ in equation~\ref{phbarku} is analogous to that 
of a constant `speed of light' $c$ that could be introduced into 
equation~\ref{sfourx}, with
  $(\delta s)^2  = (\delta x^{0})^2 -
     c^{-2} \vert\delta \bx \vert^2$ as the basis for 4-dimensional 
spacetime, and where in natural units $c=1$ can also be chosen.
 With the empirical units of $c$ being [metres][seconds]$^{-1}$ and  
of $\hbar$ being [Joules][seconds] these constants respectively 
reflect close connections of space and energy with time, which ultimately derive from a 
fundamental basis in generalised proper time for the present theory.

 The third fundamental constant of nature may be considered to be the  
	gravitational constant $\kappa = \frac{8\pi \GN}{c^4}$ in the 
Einstein equation~\ref{Eineq}, with $\GN$ the Newtonian constant of 
gravitation. Introduced in the present theory through 
equation~\ref{gfromavt} this constant here represents the direct 
identification of the energy-momentum $T^{\mu\nu}(x)$ with the 
Einstein tensor $G^{\mu\nu}(x)$ in 4-dimensional spacetime, which can 
be constructed within the constraints implied in obtaining solutions 
through the general form of proper time.
 In this case it is generally less convenient to set $\kappa = 1$ 
since although they are equated
  $G^{\mu\nu}(x)$ and $T^{\mu\nu}(x)$ not only have quite distinct 
meanings but for all practical purposes in the laboratory the former 
is unobservably small and only the latter can be measured. Setting 
$\kappa = 1$ might be more appropriate for theoretical studies regarding the properties of black holes or extreme spacetime curvature more generally.

  In standard physics $c$ appears in the equations of both quantum 
theory and general relativity, while typically $\hbar$ only appears 
in the former and $\kappa$ only in the latter.
 With $\hbar$ in equation~\ref{ehbaro} and $\kappa$ in 
equation~\ref{Eineq} both relating to energy, and with 
equation~\ref{gfromavt} employed in the present theory in an 
intrinsic manner in the process of deducing equation~\ref{phbarku} 
some relation between these constants is implied in this unified 
theory. In particular the amplitude ($B$ in equation~\ref{gkekksup}) 
and extent ($V$ in equation~\ref{pcon}) of physical wave-packets as 
distortions in the spacetime geometry in explicit solutions for 
equations~\ref{gfromavt} exemplifying equation~\ref{phbarku} will 
relate to both $\hbar$ and $\kappa$. We explore some of these 
features of the wave-packets in the following subsection.

%\pagebreak
    
\subsection{The Source of Mass for Elementary Particle Quanta}  
\label{qugr62}
  
  The empirical fact that at the most elementary level of matter the 
energy of a particle quantum \textit{only} depends upon the evolution 
in time of the state
 rather than any spatial properties or specific matter content, as 
signified by equation~\ref{ehbaro}, is 
itself evidence for a primary role for \textit{time} at this 
fundamental microscopic level. This is explicitly the case for the 
present theory in which indeed 4-dimensional spacetime and all the 
matter it contains are manifestations of the general form of proper 
time. In beginning with time alone at the most elementary level, and 
analysing its intrinsic arithmetic substructure, space (via 
equations~\ref{sfourd} and \ref{sfourx}), matter (via 
equations~\ref{salpha}--\ref{gbreak}) and energy-momentum (via 
equation~\ref{gfromavt}) are \textit{all purely derived entities}. 

 Here we do not begin by positing forms of matter and energy content 
in a pre-existing arena of space and time. Rather these entities are 
all intimately connected in deriving from time alone, down to the 
most elementary level of the `quantisation' \mbox{$E=\hbar \omega$} 
of particle states.
  It is proposed then that the consequences of the constraints in 
constructing the physical world through such a simple origin 
\textit{include} the characteristic quantum relations of 
equations~\ref{ehbaro} and \ref{phbark} as we have argued for 
equation~\ref{phbarku}. 

 From a classical physics perspective given a background of space and 
time the mass and energy of a particle associated with a localised 
disturbance in field values would be expected to vary continuously 
with the spatial extent and amplitude of the disturbance  as well as 
with the field type, with a large degree of arbitrariness permitted 
in these parameters. The hypothesis of the particle quanta relations 
of equations~\ref{ehbaro} and \ref{phbark}, as found to be generally 
applicable for a broad range of phenomena and particle states, was 
hence a particularly puzzling aspect of the old quantum theory in the 
early 20$^{\mathrm{th}}$ century. It remains as a \textit{postulated} 
feature of quantum theory into the early 21$^{\mathrm{st}}$ century,
 as we shall consider again in the following subsection.

  From basic special relativistic kinematics in the rest frame of a 
massive particle the relation $E^2 = \bp^2c^2 + m^2c^4$ reduces to 
the iconic equation
 `$E = mc^2$'. In this frame, while both sides of 
equation~\ref{phbark} vanish, equation~\ref{ehbaro} becomes
 $m = \hbar \omega/c^2$.
  Not only the energy but also 
  the invariant mass of the particle is then 
intrinsically associated with an angular frequency, further highlighting the significance of dependence on time. Here we began with such a basic oscillation rate $\alpha$ in 
equation~\ref{fcosas} which we hence expect to be directly related to 
particle mass with:
\begin{equation}
  m = \frac{\hbar\:\! \alpha}{c^2}  
  \label{mhac}
\end{equation}
  Given  that we can adopt natural units with $\hbar=1$ and $c=1$ we 
can then ask how such a direct identification of the angular 
frequency $\alpha$ with the mass $m$ of quanta for a spectrum of 
different particle states, as initially suggested by 
equation~\ref{vgroup} as noted in equation~\ref{ppropk}, might arise 
in the present theory. 

 As described for equation~\ref{gbwave} and after equation~\ref{ppropk} the oscillations are 
associated with variations $\delta h(x)$ in the magnitude of the 
 4-vector $\bh(x) \equiv \bv_4(x) \in \TM_4$ projected out of the full 
form for proper time in equation~\ref{lpvn} through 
equations~\ref{gwarph} and \ref{gmnconh}. This full 
\mbox{$n$-dimensional} expression for proper time is fragmented in 
the construction of the \mbox{4-dimensional} spacetime manifold 
$M_4$, 
 with field subcomponents of $\bv_n(x)$ such as $\psi(x)$ in 
equations~\ref{gtwopsi} and \ref{phiee}  associated with a particular particle state 
impinging upon $\bh(x)$ and hence the external geometry through the 
terms of equation~\ref{lpvnb}. These terms have the form of `Yukawa 
couplings' in a manner analogous to Higgs mass terms of the Standard 
Model Lagrangian as noted for equation~\ref{lpvnb}
 by comparison with equation~\ref{lagmas}.

  While the oscillatory structure of equations~\ref{gwave} and 
\ref{gconfu} for the geometry associated with the electromagnetic 
wave $A_{\mu}(x)$ of equations~\ref{acos} and \ref{ecoeff} was obtained as a solution 
for equation~\ref{tmnem}, itself identified as a Kaluza-Klein type relation  
between the Einstein tensor and the gauge field strength~\cite{KKone}, here for 
the fragmented subcomponents of $\bv_n(x)$ impinging upon $\bh(x)$ we 
are considering a direct perturbation of the metric geometry of 
$M_4$. However, in both cases at the most elementary level the 
corresponding particle states are associated with \textit{smooth} 
undulations in the metric, via variations in the value of $h(x)$ in 
the case of equations~\ref{gbwave}--\ref{gkekksup}, as an element of an extended 
solution for the 4-dimensional spacetime structure.

 Such minimal but non-trivial solutions for the linearised form of
  the differential equation~\ref{gfromavt}
   would be expected to be wave-like at the simplest level, whether 
or not considered as composed of Fourier modes as suggested in the opening of 
subsection~\ref{qugr61}.
   The assumption of a flat spacetime limit allows the employment of oscillations described for example by the real function $\cos k\! \cdot\! x$ to a good approximation, with a continuum of such oscillation modes underlying the construction of localised wave-packets, as described for equations~\ref{gbwavesup} and \ref{gkekksup}, representing a physical particle state. With reference to the discussion following equations~\ref{fpack} and \ref{wpack} the properties and interactions of such a state may be represented by the properties of a central plane wave mode, as will be the case for the discussion and equation~\ref{yhpsi} below.

  The smooth continuous exchanges between for example the 
subcomponent $\psi(x)$ of $\bv_n(x)$  in equations~\ref{gtwopsi} and 
\ref{phiee}  and the field $\bh(x)$ through terms of 
equation~\ref{lpvnb}  associated with a propagating particle as a 
physical deformity in spacetime contrast with the abrupt exchanges 
between  $\psi(x)$  and  the gauge field $A_{\mu}(x)$ through 
equation~\ref{dlpvnb} as exemplified in 
equations~\ref{ext1}--\ref{ext4} underlying a local field degeneracy 
in an interaction between particle states.  The latter local 
degeneracy in the field solution for equation~\ref{gfromavt} involves 
matching externally indistinguishable internal field contributions to the local 
spacetime geometry through complex $e^{\pm i k\cdot x}$ functional 
modes consistent with terms in the constraint of 
equation~\ref{dlpvnb} in the form of equations~\ref{dlpvnbex} and \ref{ext0}. 
  While relative `charges' in these terms will proportionally 
determine the particle interaction rate, as discussed after 
figure~\ref{extofey}, the relative `Yukawa couplings' identified in 
the terms of an explicit form for equation~\ref{lpvnb} are expected to similarly determine 
the oscillation rate for the perturbations in $\delta h(x)$ and the corresponding particle mass.

   For example, while emphasising the provisional nature of this construction, a term in the constraint equation~\ref{lpvnb} 
including three factors of the form $Y\,\ccdot\, h\,\ccdot\,\psi$, with $Y$ 
identified as the effective Yukawa coupling, could accommodate the complex  
$e^{+ik_1\cdot x}$ mode of $\psi(x)$ in equation~\ref{phiee} in an 
invariant manner via the substitution:
\begin{equation}
 \label{yhpsi}
 Y\,\ccdot\, h\,\ccdot\,\psi
   \quad \Rrightarrow \quad
 Y\,\ccdot\, h(\bk_1)\, e^{-ik_1\cdot x}\,\ccdot\,\psi(\bk_1)
                         e^{+ik_1\cdot x}
\end{equation} 
  with a similar embedding possible for the complementary complex exponential mode in equation~\ref{phiee}. These perturbations can be
 interpreted as contributions to a real oscillation in $h(x)$ and 
hence to the geometric structure of spacetime via equations~\ref{gwarph} and \ref{gmnconh} as associated with equations~\ref{gtwopsi} and \ref{gbwave}--\ref{gkekksup}.
 The above coupling factor 
$Y$ will then relate to the amplitude $B$ and spatial extent $V$ of the generalised wave-packet, as alluded to 
at the end of subsection~\ref{qugr61}, as well as the central wave-vector $k_1$ in constructing a consistent solution under equation~\ref{gfromavt}.
  In this manner such effective
  Yukawa couplings might then determine the permitted angular frequency 
$\alpha$ in equation~\ref{fcosas}, as manifested in a 4-dimensional 
form for $G^{\mu\nu}(x)$ via equations~\ref{fcoskx}--\ref{gkekksup}, and 
ultimately act as the source of the particle mass as consistent with equation~\ref{mhac}.

  These distinct undulating impressions in the metric geometry 
$g_{\mu\nu}(x)$ of $M_4$ associated with propagating particles are 
 punctuated by mediating local particle interactions 
associated with local field degeneracies, collectively as elements of 
a solution for equation~\ref{gfromavt}. From the constraints in this 
construction one criterion for particle quanta propagating between 
interactions, such as the
initial and final particle states connected through the local 
degeneracy of such an interaction, will be consistency with the 
universal relation of equation~\ref{phbarku}.
 As we described in  subsection~\ref{qugr61} it is the mutual 
constraints from the internal field and external geometry aspects of 
both the propagating and interacting particle elements of solutions 
for  equation~\ref{gfromavt}, hence 
  involving the broad scope of the present theory based on 
generalised proper time, that is needed to connect with one of the 
earliest and most elementary properties of quantum theory in 
equations~\ref{ehbaro} and \ref{phbark}.

  As we have noted in the opening of subsection~\ref{qugr23} and 
towards the end of subsection~\ref{qugr51} in many approaches to 
`quantum gravity' it is proposed that this quantum postulate should 
also apply for `gravitons' as quanta of the gravitational field. From 
the point of view of the present theory gravitational waves, as 
oscillations of Weyl curvature as reviewed in subsection~\ref{qugr22} 
culminating in equations~\ref{wavepol} and \ref{waveee}, are 
perturbations \textit{purely} in the \textit{external} gravitational 
field, as a kind of classical `surface run-off' from disturbances in 
the geometry of spacetime, with no constraints imposing any `quantum' 
conditions on these phenomena. On the other hand the form of external 
gravitational undulations in Ricci curvature associated with states 
of matter are \textit{mutually} constrained through compatibility 
with perturbations in the corresponding \textit{internal} field 
components and their potential for interactions through the 
constraint equations~\ref{gchift}--\ref{emcon} as exemplified  in equation~\ref{yhpsi}.
 The spatial extent and field amplitude for
 the generalised wave-packets of
 matter particles propagating as solutions for 
equations~\ref{gfromavt} and \ref{gavt} are then determined 
\textit{in conformity with} these underlying constraints,  necessarily incorporating 
the relations of equations~\ref{phbarku} and \ref{mhac}.

 The construction of a localised particle in terms of a localised 
wave-packet requires a sum of component plane waves over 
a continuous range of wave 4-vectors $k$ as described for 
equations~\ref{gbwavesup} and \ref{gkekksup}. While we have focussed upon the collective 
group velocity of equation~\ref{vgroup}, the phase velocity $\bv_p = 
k^0/\bk$ (alluded to after equation~\ref{wpack}) has a 
$k$-dependent value. This may raise the question of stability and 
suggests that such a wave-packet might be expected to become 
distorted and spread out, and hence less `particle-like', as the 
state propagates. However, we are not considering a 
disturbance created at one time to propagate freely into the future 
or taking place in a pre-existing material medium as for a wave 
across water. Rather here the generalised wave-packet incorporating a 
disturbance in the gravitational field is embedded within a full 
\mbox{4-dimensional} solution for the spacetime geometry itself.
 Unlike a freely propagating wave-packet prone to dispersion, the 
construction here of generalised wave-packets incorporates internal 
field interactions, including the $\delta h(x)$ perturbations, as an 
intrinsic feature in obtaining such coherent propagating elements in 
solutions for equation~\ref{gfromavt}.

The primary requirement for any particle state is to be 
incorporated within a consistent solution for the 4-dimensional 
spacetime geometry as we have described for equation~\ref{gfromavt}.
 Within the constraints, including the implication of 
equation~\ref{phbarku}, between interactions such a particle state 
described by a generalised wave-packet disturbance essentially 
propagates as a `gulp' in the spacetime fabric, while in principle 
taking a range of possible extended spatial forms. 
 As noted above it is important to emphasise that a particle state is \textit{not} 
here conceived of as a field structure that once generated will 
\textit{then} propagate in a stable manner (as would typically be the 
case for a classical particle or wave). Rather such states are 
minimal but non-trivial basic elements of a \textit{full} 
4-dimensional spacetime solution for a metric geometry 
$g_{\mu\nu}(x)$ under equation~\ref{gfromavt}, as linked by local 
interactions associated with local internal field degeneracies in 
 the composition of the spacetime geometry.

 Particle solutions in this theory can take the form of the 
quintessentially particle-like trajectories seen in the `tracks' of 
laboratory events such as that of figure~\ref{sldbb} as will be 
described for figure~\ref{sldbbv} in subsection~\ref{qugr72}.
 However, while rooted as an elementary oscillation in time, the 
solution for a single particle state can be spatially more diffused 
and less localised as might be constrained by laboratory apparatus 
such as a diffraction grating or the double-slit arrangement of 
figure~\ref{dslit} as will be described for figure~\ref{dslitg} in 
subsection~\ref{qugr74}.
 In such experiments the notion of a strict `particle' correlate of 
these phenomena becomes less appropriate, except at the points of 
local interactions such as at $S$ and $I$ in figures~\ref{dslit} 
and~\ref{dslitg}.
 The kinematic properties key to the impression of local 
particle-like interactions are a demonstration of the energy-momentum 
conservation with $\pal_{\mu}T^{\mu\nu}(x)=0$ which, via 
equation~\ref{emcon}, applies everywhere in this flat spacetime 
limit, as we shall also elaborate in the following subsection.

 A formalism of `quantisation' is not then required to identify particle-like behaviour here, rather the 
underlying source of these properties can be accounted for through a 
theory based on generalised proper time. In this section  we have 
focussed on particles essentially as `energy quanta' with a distinct 
invariant mass, however there are of course further characteristic 
discrete features associated with the quantum phenomena of particle 
states.
 In particular, in addition to equations~\ref{ehbaro} and 
\ref{phbark} the quantisation of intrinsic spin, as discussed near 
the end of subsection~\ref{qugr52} in relation to torsion, as well as 
of structures with orbital angular momentum in relation to the 
properties of the energy-momentum tensor defined in 
equation~\ref{gfromavt}, will need to be more fully described. While 
a particle 4-momentum $(E,\bp)$ is related to \mbox{4-dimensional} 
spacetime translations, the spin and orbital quantum properties 
relate to representations of the 6-parameter Lorentz group.
 Hence more generally collectively under the 10-parameter 
Poincar\'{e} group in the flat external spacetime limit the full 
nature of particle classification by mass and spin, rather than mass 
alone, will need to be accounted for (as alluded to in \cite{Unifi} 
section~15.2).

 Multi-particle systems can also exhibit the less straightforward 
properties of entanglement and EPR-type correlations as will be 
discussed in subsection~\ref{qugr73}, again here to be identified within a single solution 
for equation~\ref{gfromavt}.
 In general full solutions for the configuration of generalised 
wave-packets, as described by an external metric $g_{\mu\nu}(x)$ and 
internal $A(x),\bv_n(x)$ field structures, within a given set of 
boundary conditions, may not be easy to determine. Indeed exact 
solutions are difficult to find in general relativity alone for the 
general case as we discussed in subsection~\ref{qugr22}.
  The nature of such solutions will be further complicated if for 
example intrinsic spin and torsion components are also to be 
included as alluded to above. However as a robust feature of all particle solutions the 
elementary relations for 4-momentum in equation~\ref{phbarku} and 
mass in equation~\ref{mhac} for interacting particle quanta are 
anticipated to also apply for these more general cases.

%\pagebreak

\subsection{Relation to Particle States in Quantum Theory}
\label{qugr63}

   The manner in which elementary particle quanta in the present 
theory can be distributed and distorted in space is to some degree 
analogous to the properties of a wavefunction
 $\Psi(\bx,t)$ in quantum mechanics.  
  However here wave-like functions in $A(x),\bv_n(x)$ matter field 
components are \textit{enveloped} within and mutually related to a 
geometric form for the external spacetime $G^{\mu\nu}(x)$ through 
equation~\ref{gfromavt}, with constraints on the possible solutions
 for the metric $g_{\mu\nu}(x)$ limiting the range of observable 
events in 4-dimensional spacetime.
 On the other hand while
 a wavefunction $\Psi(\bx,t)$ has some properties similar to those of 
the internal fields $A(x),\bv_n(x)$, in propagating in a flat 
3-dimensional space in the non-relativistic case or in a flat 
4-dimensional spacetime background in QFT, there is no non-trivial 
metric geometry $g_{\mu\nu}(x)$ associated with the wavefunction in 
quantum theory.

 Stripped of this external geometric constraint and describing 
internal fields alone the wavefunction of quantum theory in some 
sense represents `naked quanta', as an incomplete account of the 
actual elementary physical structures of the world.
 Much of the mystery shrouding quantum theory arises through our 
inability to directly empirically observe the minute but non-trivial 
metric $g_{\mu\nu}(x)$ geometry on the scale of microscopic  
laboratory  events and its corresponding neglect in the theoretical 
constructions
 of the quantum formalism.
  This formalism then requires a series of \textit{postulated} 
features, including the particle quanta relations of 
equations~\ref{ehbaro} and \ref{phbark}, as well as an 
interpretation, in particular in relation to `wavefunction reduction' 
as also reviewed in subsection~\ref{qugr21}, in order to make 
practical calculations and extract results for comparison with 
experiments.

 As an explicit expression of the comment at the end of 
subsection~\ref{qugr22}, concerning general relativity being 
\textit{external} to quantum theory, the structure of the present 
theory incorporates the spacetime geometry itself one layer 
\textit{outside} the standard approach to quantum theory, with 
particle states that are no longer `naked' and implying constraints 
on their properties.
  Here there is no attempt to retain any conception of an elementary 
particle as a `building block' introduced \textit{on top of} a 
spacetime background, which might then be assigned properties to 
model quantum behaviour. Rather particle states are inferred as 
elements of specific minimal disturbances identified in the 
construction of the spacetime geometry itself, as a consistent 
solution for equation~\ref{gfromavt}, from which the quantum 
mechanical properties might be \textit{derived} in the appropriate 
limit.
 This geometric structure, with a non-trivial metric $g_{\mu\nu}(x)$, 
in solutions for equation~\ref{gfromavt} is considered to offer a 
physically real and conceptually complete description of particle 
states, unlike the pragmatically introduced `naked quanta' exhibiting 
properties associated with a wavefunction $\Psi(\bx,t)$ in the 
quantum mechanical account.

  In `semi-classical quantum gravity' an attempt is made to connect 
quantum matter with a non-trivial spacetime geometry. This is 
achieved by promoting the energy-momentum on the right-hand side of 
the Einstein equation~\ref{Eineq} to a quantum operator for which the 
expectation value is then coupled to the classical Einstein tensor on 
the left-hand side with (\cite{Hugg} section~4, \cite{Carl4} 
section~1):
\begin{equation}  
  G^{\mu\nu} = -\kappa \langle \Psi \vert
    \hat{T}^{\mu\nu} \vert \Psi \rangle
  \label{Einqm}	 
\end{equation}
  There are several problematic features for this to be considered 
 a fundamental theory, including the fact that abrupt changes in the 
right-hand side under wavefunction collapse are inconsistent with 
equation~\ref{Bian} and the conservation of the left-hand side. Such 
a model then highlights the difficulty of associating a probability 
distribution, such as that of $\vert\Psi(\bx,t)\vert^2$ in figure~\ref{dslit},  with a gravitational field.

 Similarly the wavefunction in quantum mechanics, such as 
$\Psi(\bx,t)$ in figure~\ref{dslit}, cannot itself be considered to 
carry energy, otherwise there would be an abrupt discontinuous jump 
in the distribution of that energy in the collapse event at $I$.
 The wavefunction can only represent the \textit{likelihood} for 
where a quantum of energy might be observed. 
 The unitary evolution {\bf U} of the wavefunction channels the range 
of possible outcomes, with a summed probability of unity maintained, 
until punctuated by reduction {\bf R} to the state of the observable 
such as energy or location measured.  
 This is here considered an incomplete description in which, together 
with the conceptual issues reviewed in subsection~\ref{qugr21}, 
energy-momentum conservation essentially only applies intermittently 
at such points of measurement.

 An essential feature of this incompleteness is that quantum theory 
does not recognise the primary role played by the non-trivial 
geometry of the 4-dimensional spacetime for these systems. 
 For the present theory this geometry \textit{defines} the 
energy-momentum, through equation~\ref{gfromavt}, implying a 
continuous distribution in $T^{\mu\nu}(x)$ for all systems and all 
processes, as enveloped under the smooth external geometry, and with 
energy-momentum \textit{conserved} throughout via 
equation~\ref{emcon}. 
  This conservation of energy-momentum then applies throughout 
spacetime both for all events that we can see (as might be observed 
in the laboratory at points of apparent `wavefunction collapse' in 
the quantum theory description) as well as for everything we cannot.
 This theory then `tightens up' the above intermittent 
energy-momentum conservation of quantum theory to apply everywhere in 
spacetime as blanketed under a continuous metric $g_{\mu\nu}(x)$ 
geometry.

 We noted in the previous subsection how the elementary quanta 
relations discussed for equations~\ref{ehbaro} and \ref{mhac}
 suggest a fundamental role for time at the most elementary level of 
matter, as is directly employed for the present theory through a basis in generalised 
proper time.  	
 In quantum theory a fundamental basis in time \textit{is} reflected 
in the Schr\"{o}dinger equation~\ref{Schro} with the Hamiltonian  
$\hat{H}$ acting as the generator of infinitesimal transformations of 
the quantum state $\Psi$ in time, with time an independent background 
parameter. This distinguished role for time in quantum theory was 
also noted for the relativistic case after equations~\ref{Schroq} and \ref{uevolve}.

 The construction of the Hamiltonian operator, as for the 
wavefunction, lacks any input from any non-flat external spacetime 
geometry.
 Hence from the perspective of the present theory the propagation of 
the likelihood of the range of outcomes forwards in time through 
equation~\ref{Schro} or \ref{Schroq} carries only partial information about the 
system with a limited set of restrictions. This unitary evolution of 
the quantum system will be accompanied by a parallel, and more 
complete, evolution as described in the present theory incorporating 
a full metric $g_{\mu\nu}(x)$ account of the system as obtained from 
complete 4-dimensional spacetime solutions for 
equation~\ref{gfromavt}.

 In quantum theory the Hamiltonian operator represents the total 
energy of the system, which through a correspondence between the 
dynamical elements can be obtained from the classical Hamiltonian 
(see for example \cite{Rae} section~4.2). 
 In classical field theory in a flat spacetime an energy-momentum 
tensor  $t^{\mu\nu}(\lag)$ can be defined by applying Noether's 
theorem for the spacetime translation invariance of the field
 Lagrangian $\lag$ (as alluded to after equation~\ref{pcon} with 
reference to \cite{Unifi} equation~3.102, see also \cite{Pesk} 
equation~2.17). The four conserved `charges' correspond to a 
4-momentum $p^{\mu}$ defined similarly as for equation~\ref{pcon}, 
with the Hamiltonian $H$ identified with the first component (\cite{Pesk} 
equation~2.18):
\begin{equation}
   H \equiv p^{0} = \int t^{00} d^3\bx
\label{pconcl}	
\end{equation}
  
  However in general the energy-momentum $t^{\mu\nu}(x)$ derived from 
a Lagrangian may be neither symmetric nor gauge invariant, and with a 
degree of ambiguity can be augmented by additional terms to address 
these drawbacks (although this does not change above the 4-momentum $p^{\mu}$ as noted for \cite{Pesk} equations~19.144 and 19.145, see also \cite{Unifi} 
equation~3.104). 
  For the present theory the energy-momentum $T^{\mu\nu}(x)$ is 
defined in a less ambiguous and more directly symmetric and 
gauge-invariant manner through the definition in terms of the 
Einstein tensor in equation~\ref{gfromavt}. While $t^{\mu\nu}(x)$ 
represents an energy-momentum for matter fields \textit{in} a flat 
spacetime, $T^{\mu\nu}(x)$ in the present theory essentially 
represents the energy-momentum \textit{of} a curved spacetime as 
warped through the mutual relation with its internal matter field 
composition. 

  In making a connection between the two approaches we note that 
while in this theory \textit{all} energy-momentum is defined through 
 $T^{\mu\nu}(x) := \frac{-1}{\kappa}G^{\mu\nu}(x)$ via 
equation~\ref{gfromavt}, it will be the measurable and 
macroscopically discernible elements of this energy-momentum that 
provide the input for $t^{\mu\nu}(x)$ and determine the Hamiltonian 
in classical and quantum theory. This Hamiltonian can then be 
interpreted as indirectly modelling the imprint of some elements of 
the spacetime geometry $G^{\mu\nu}(x)$, such as described by the 
configuration of experimental apparatus, that channel and guide the 
evolution of the state of the system, with the unknown elements  and underlying source of indeterminacy
effectively modelled by a wavefunction and an associated propagation of 
probabilities.    

  The relation between the $P^0$ component deriving from 
$T^{\mu\nu}(x) := \frac{-1}{\kappa}G^{\mu\nu}(x)$ in 
equation~\ref{pcon}, and connected with a temporal oscillation rate for particle quanta
with $P^0 = \hbar k^0$ in equation~\ref{phbarku},  and $H \equiv p^0$ deriving 
from $t^{\mu\nu}(\lag)$ in equation~\ref{pconcl}, as in turn 
connected with the Hamiltonian operator $\hat{H}$ in quantum theory 
as the generator of evolution in time and employed in equations~\ref{hpehw} and \ref{hathqft} below, may prove pivotal in relating 
the present theory to the quantum formalism.
 This applies both for QFT in the limit of flat spacetime and the 
quantum mechanical case in a Newtonian background of space and time, 
for which the focus on the $T^{00}(x)$ and $t^{00}(x)$ elements may be more 
natural in this non-relativistic limit. 

  In generating the probability-conserving unitary transformations of 
the quantum state the Hamiltonian is itself an Hermitian operator
 for which eigenstates can be determined with corresponding 
eigenvalues
  representing the energy that can be observed in a measurement. For 
a simple complex wavefunction of the plane wave form 
  $\Psi(\bx,t) = e^{(i\bsl{k\cdot \bsl{x}} - i\omega t)}$, 
representing such an eigenstate for either a single free non-relativistic particle or an 
initial or final state particle in a QFT interaction process, the 
energy eigenvalue $E$ can be extracted as:
\begin{equation}
   \hat{H}\Psi \, = \, i\hbar\frac{\pal}{\pal t}\Psi \, = \, E\Psi
              \quad \mbox{with} \quad E = \hbar \omega
	\label{hpehw}		  
\end{equation} 

   The formalism of quantum mechanics in terms of operators and 
wavefunctions was largely formulated explicitly to provide  this 
link,
 given the experimental evidence that had accumulated by the 1920s 
for the direct universal association in equation~\ref{ehbaro} between 
wave-like frequency $\omega$ and particle-like energy $E$ for 
elementary phenomena  (\cite{Pen} sections~21.3--21.5).
  While energy is hence related to a time derivative in 
equation~\ref{hpehw} the 3-momentum is similarly related to an 
operator
  $\hat{\bp} =-i\hbar\nabla$ in the form of spatial derivatives
   and also applied to wavefunctions such as $\Psi(\bx,t) = 
e^{(i\bsl{k\cdot \bsl{x}} - i\omega t)}$ in an eigenvalue equation as 
consistent with equation~\ref{phbark}. 
 More generally an Hermitian operator can be identified for any 
observable characterising the quantum state as alluded to in subsection~\ref{qugr21}.

  In QFT the Hamiltonian operator representing the energy of a 
particular state can be expressed in terms of creation and 
annihilation operators. For example for states associated with the 
free scalar quantum field of equation~\ref{kgosol2} the Hamiltonian is:
\begin{equation}
   \hat{H}= \int \frac{d^3\bsl{p}}{(2\pi)^3}\, \omega_{\bsl{p}}
      \left(a^{\dag}(\bp)a(\bp) +\frac{1}{2}
	  \left[a(\bp),a^{\dag}(\bp)\right] 
	  \right)
 \label{hathqft}
\end{equation}  
  which is consistent with equation~\ref{pconcl} for a classical Klein-Gordon field (see \cite{Pesk} 
equations~2.8, 2.18 and 2.31). The first term under the integral acts 
on multi-particle states, utilising equation~\ref{aacomr}, such that 
  $\hat{H}\vert \bp_1,\bp_2,\bp_3,\ldots \rangle = 
(\omega_{\bsl{p}_1}+\omega_{\bsl{p}_2}+\omega_{\bsl{p}_3}+\ldots)
  \vert \bp_1,\bp_2,\bp_3,\ldots \rangle$  in the appropriate manner
  with $\omega_{\bsl{p}_i}$ the energy eigenvalue of the state
   $\vert \bp_i \rangle$.
   However the second term yields an infinity which has to be 
subtracted or ignored, as a further unsatisfactory feature of this 
means of defining the energy of a system via a Lagrangian approach.
 As well as introducing a formalism that is contrived to some degree, 
as noted for equations~\ref{kgosol2} and \ref{aacomr} the properties 
of $a^{\dag}(\bp)$ and $a(\bp)$ are \textit{defined} in such a way as 
to create and annihilate particle quanta under the 
\textit{assumption} of equations~\ref{ehbaro} and \ref{phbark}, with 
the 4-momentum $p$ already identified with the wave 4-vector of the associated oscillation modes.

  In subsection~\ref{qugr61} we have described how these de Broglie 
relations can be \textit{derived} in equation~\ref{phbarku} by 
defining energy-momentum in terms of the external spacetime geometry 
for physical particle quanta constructed explicitly as generalised wave-packets in the external spacetime metric structure as well as in internal matter field components through a basis  
in generalised proper time.
For the present theory rather than applying  operators
 similar to those  in equation~\ref{hpehw} or \ref{hathqft}
 to a complex component of a field such as that in equation~\ref{acosee} 
(as considered for \cite{Unifi} equation~11.52)
 or to an abstract Fock space of states, here
 the \mbox{4-momentum} $P^{\mu}$ of interacting wave-packet particle 
states is \textit{explicitly} identified as described for equation~\ref{pcon} and found to be directly 
connected to the corresponding wave 4-vector $k^{\mu}$ through the constraints of 
the theory as we have described for equation~\ref{phbarku}. This construction hence employs 
a non-flat metric geometry $g_{\mu\nu}(x)$ in an essential way.
 
 This latter element is the crucial ingredient missing in quantum 
theory, with the mathematical constructions behind 
equations~\ref{hpehw} or \ref{hathqft} and the corresponding quantum states here considered as providing a 
description of `naked quanta' in a manner that \textit{models} the 
corresponding evident empirical phenomena.  
 Here we then propose to provide a more complete account of particle 
quanta as \textit{deriving} from the elementary principles and 
constraints of the theory based on generalised proper time. 
 In addition to particle `quanta' the nature of `eigenstates' for observables in 
general, as restrictions on what \textit{can} be measured, is here 
also proposed to originate from the constraints imposed both in 
finding a consistent solution for external and internal fields in 
equation~\ref{gfromavt}, as described for 
equations~\ref{gchift}--\ref{emcon}, as well as from the boundary 
conditions under which the observations are made.

   The ability of the present theory to account for the fundamental 
particle quanta relations of equation~\ref{ehbaro} and \ref{phbark} 
as we have shown in this section for equation~\ref{phbarku} further 
expands on the mathematical connections with QFT developed in the 
previous section, while potentially going somewhat beyond the 
explanatory power of quantum theory.   
As outlined in section~\ref{qugr4} quantum mechanical characteristics 
such as indeterminacy, non-locality and wavefunction collapse can 
also be accounted for consistently with general relativity within 
this theory founded upon generalised proper time.
The framework within which equation~\ref{phbarku} has been derived
 and the origin of elementary quanta elucidated 
 already presents then the possibility of addressing major conceptual 
issues surrounding quantum theory itself as well as general relativity and 
  the coherent union of these two theoretical structures.
These questions were reviewed in section~\ref{qugr2}.
 In the following section we return to consider the extent to which 
these conceptual difficulties might be resolved in the present 
theory of quantum gravity, which would hence provide a firm and robust basis for the 
further technical development of the calculational side described 
here and in the previous section.

%\pagebreak
\section{Conceptual Framework for Quantum Gravity}
\label{qugr7}

\subsection{For a General Curved Spacetime}
\label{qugr71}

  In this section we elaborate upon the conceptual significance of 
this framework for quantum gravity, ranging from the full general 
relativistic environment through the flat spacetime limit to the 
non-relativistic case and the elementary laboratory setting, with the 
implications for quantum theory assessed at all stages.
 In this first subsection
 we consider the conceptual issues raised on attempting to combine 
quantum theory and general relativity for the general case of 
arbitrarily curved spacetime as introduced in 
subsection~\ref{qugr23}. This situation is characterised in 
particular by the need to incorporate the full scope of general 
relativity, without employing any approximations such as linearised 
gravity as considered in the previous two sections. With no such 
approximations being made the mathematical structure for these more 
general systems is likely to be harder to express than that required 
for making connections with the QFT limit and the HEP laboratory 
environment, with that limit considered further in the following 
subsection. Given the difficulty of performing experiments involving 
large gravitational fields it will also prove more challenging to 
empirically test any theory for a highly curved spacetime, but there 
remains the question of establishing a theoretically consistent and 
conceptually coherent theory of quantum gravity for the general case.

  In many approaches to quantum gravity one such conceptual issue 
that arises is the `problem of time'. The identification of an 
unambiguous time parameter $t$ is essential to describe the evolution 
of the wavefunction via the Schr\"{o}dinger equation~\ref{Schro} or 
of the quantum state for QFT in equation~\ref{Schroq}, while the 
nature of such a temporal parameter is ill-defined for a background 
independent theory like general relativity, particularly if a 
superposition of 3-dimensional spatial geometries is considered, as 
reviewed in subsection~\ref{qugr23}.
 The problem of time, arising from this very different treatment of 
time in quantum theory and general relativity, is particularly acute 
for attempts at a quantum gravity theory that \textit{assume} gravity 
should itself be quantised.
 In subsection~\ref{qugr23} we alluded to various strategies for 
addressing the problem of time, ranging from defining time through a 
space-filling reference fluid such as a material dust~\cite{Brown} to 
a fundamentally timeless conception of quantum gravity~\cite{Rove1}. 

 For the present theory we not only take `time' to be present before 
any quantisation, but \textit{begin} with the concept of time 
\textit{alone}, about which the entire theory, including all quantum 
theoretic and general relativistic aspects, is constructed.
 Here time is \textit{the} fundamental entity with the general 
arithmetic form for proper time in equations~\ref{salpha} and 
\ref{lpvn} conceptually preceding the construction of a single 
extended \mbox{4-dimensional} spacetime manifold $M_4$ through 
equation~\ref{gfromavt} with the underlying \mbox{one-dimensional} causal progression 
in time parametrising the local probabilistic matter field 
composition of the geometry and in turn underlying the significance of time 
evolution in quantum theory.
 This picture for merging quantum and gravitational phenomena was 
introduced in subsection~\ref{qugr41}.

 For this theory gravity is \textit{not} quantised, rather a complete 
and continuous \mbox{4-dimensional} spacetime is incorporated 
similarly as for general relativity.
Here this single continuous geometric spacetime is identified as a 
solution for equation~\ref{gfromavt} consistent with the constraint 
equations~\ref{gchift}--\ref{emcon} for the matter fields.
 The breaking of the symmetry of the full generalised form of proper 
time in equation~\ref{lpvn} in identifying 4-dimensional spacetime 
generates all of the matter field content as originally noted for equation~\ref{gbreak}, without needing to 
postulate any material substratum independently of time itself.
 As described for equation~\ref{emcon} the constraints include the 
basis for energy-momentum conservation throughout spacetime. These 
equations constrain the structure of matter in terms of the possible 
observables and the intermediate evolution of physical systems 
between observations collectively in forming solutions for 
equation~\ref{gfromavt}. 

In the quantum mechanical limit it will be required to account for an 
apparent parallel description in terms of the unitary evolution of a 
pragmatically constructed wavefunction as governed by a Hamiltonian 
operator via the Schr\"{o}dinger equation~\ref{Schro}, as considered 
in subsection~\ref{qugr63}. As alluded to there the nature of 
energy-momentum conservation is anticipated to play a key role in 
making this connection between the present theory and quantum theory.
  The familiar properties of quantum theory are to be attained in the 
appropriate spacetime limits in which an unambiguous time parameter 
can be identified, such as in the laboratory setting. This applies 
both for the case of QFT and the construction of the $S$-matrix as a solution for 
equation~\ref{uevolve} as well as for the non-relativistic limit and 
the Schr\"{o}dinger equation~\ref{Schro}.

  In this manner the focus on one-dimensional temporal evolution in 
quantum theory and the significance of full 4-dimensional spacetime 
structures in general relativity might be consistently convolved and 
unified, through generalised proper time as the underlying 
fundamental entity.
  This approach is essentially a `theory of time'
 and hence in principle avoids any `problem of time'. In the context 
of this theory the problem of time arises in other approaches largely 
from assuming the formalism  of quantum theory to be fundamental and 
hence taking it too literally and too far in the combination with 
gravity.

  While the realm of quantum theory is more limited in the present 
theory and  originates from a different conceptual basis to the 
standard approach, with gravity taking priority in the combination as 
initially motivated in subsection~\ref{qugr24}, it is nevertheless 
the case that there are also some differences from the standard 
conception of general relativity. The typical interpretation 
presented for the Einstein field equation~\ref{Eineq} is that the 
introduction of matter into spacetime warps the geometry (sometimes 
pictured by analogy with a `heavy weight' such as a bowling ball being 
placed on a large rubber sheet that is correspondingly distorted,  
although with an \mbox{element of circularity}).
More generally the nature of the mutual relationships between space, 
time and matter is a key characteristic both for theories of the 
gravitational field and approaches to quantum gravity~\cite{Kief2}.

   For the present theory it is \textit{not} the case of 
\textit{having} a flat spacetime that is \textit{then} warped by the 
presence of matter nor a question of describing a curvature 
propagating through a \textit{pre-existing} spacetime. Rather here we 
begin with proper time alone and \textit{then} construct spacetime 
with an intrinsic curvature inextricably \textit{linked together} 
with an apparent matter content through equation~\ref{gfromavt}. 
Through this equation energy-momentum is \textit{defined}, in a 
manner consistent and compatible with the Einstein field 
equation~\ref{Eineq} but with a somewhat different conceptual basis 
in generalised proper time.
 There is however an historical connection with unified field 
theories as a generalisation of general relativity as noted in 
section~\ref{qugr3} (citing \cite{Gener} subsections~1.2 and 5.1, \cite{Ufield}).

  The most extreme spacetime curvature is encountered in the 
singularity of a black hole or of the Big Bang at the origin of 
cosmic time, where classical general relativity apparently breaks 
down as reviewed in subsection~\ref{qugr22}, representing the 
greatest conceptual difficulty for general relativity alone. This 
provides a significant testing arena for the internal theoretical 
structure in the full expression of any quantum gravity theory, in 
aiming to describe such an environment in a consistent and finite 
mathematical manner. Here the goal will then be to show how this 
conceptual blind spot for general relativity in itself might be 
addressed through the coherent incorporation of apparent quantum 
phenomena for this extreme case via the corresponding solutions for 
equation~\ref{gfromavt} in the present theory.

 While compared with the immediate vicinity of a singularity the 
spacetime curvature is less extreme at the event horizon of a black 
hole, a complete quantum gravity theory is still desired to determine 
the precise nature of black hole thermodynamics. 
A significant feature of the present theory is a perspective of placing 
the priority more on the side of gravity over quantum theory, rather 
than the other way around.
Since here QFT is considered a pragmatic approximation applicable 
in the flat spacetime limit, it is not clear the extent to which QFT 
tools adapted for curved spacetime conditions will provide a reliable 
means for determining the properties of black hole evaporation for 
example. Indeed the environment of a black hole is very different to 
that of the laboratory for which the formalism  of quantum theory was 
initially motivated and where its implications have been profusely 
tested with great success.

     In particular the apparent `information paradox' could in part 
be a symptom of pushing the formalism of quantum mechanics too far 
and too faithfully beyond its familiar domain of application. 
  A central element of the paradox is the conflict that would be 
implied between loss of information and the requirements of unitarity 
in quantum theory. For the present theory however unitarity is 
\textit{not} part of the fundamental basis, but rather is introduced 
through establishing the link with the practical calculations of 
quantum theory via equation~\ref{pdddds}.  
Before the full development of this theory a range of 
possibilities are open, as noted towards the end of 
subsection~\ref{qugr23}, from the case of  information escaping the 
black hole and being fully conserved to the case of information being 
lost in the evaporation as an intrinsically irreversible feature of 
black hole phenomenology.
   Indeed, as for other new approaches to quantum gravity, even the 
question of whether black holes actually evaporate at all might need 
to be reconsidered. 
 To address the information paradox it may also be necessary to first 
understand the quantum gravity nature of the very centre of a black 
hole, to assess whether information could be lost in the 
`singularity' itself, or stored there to be retained permanently or eventually evaporated.  

 As noted above such technical questions, for both black hole and 
initial singularities, provide a challenging theoretical testing 
ground for any theory of quantum gravity. For the present theory the 
nature of the spacelike initial singularity and general vicinity of 
the Big Bang for the full 4-dimensional cosmological solution for 
equation~\ref{gfromavt} will describe the very early universe 
(potentially relating to topics such as inflationary theory and the 
horizon problem), the evolution from which may in principle have 
observable empirical consequences for present day cosmology. 
  As for other theories of quantum gravity these connections with the 
early universe and the evolution of cosmological structure might then 
provide empirical tests (as alluded to in subsection~\ref{qugr24} 
with reference to~\cite{Carl1} section~VI, \cite{Ansel}).

   Theories of quantum gravity can also have implications for
 understanding the nature of the vacuum, as will be the case for the 
full `vacuum solution' for equation~\ref{gfromavt} for the present 
theory.
 This would involve a very different definition and calculation 
compared with that based on the expectation value for the vacuum 
energy $\langle 0 \vert \hat{H} \vert 0 \rangle$ in quantum theory -- 
 which is dependent on the cut-off scale introduced to deal with the infinite contribution from the second term in equation~\ref{hathqft} discussed in the previous subsection
 (\cite{Pesk} section~22.2, \cite{Carr} discussion of equation~4.75).
 The degree of deviation from a flat spacetime in terms of a `vacuum 
energy' for the present theory might then be assessed in the context 
of the `cosmological constant problem' reviewed in 
subsection~\ref{qugr23} with reference to equation~\ref{einlamt}. 
This in turn could have observable empirical consequences for the 
large scale evolution of the universe in relation to the apparent 
properties of `dark energy' (as for other approaches such as the 
models in~\cite{Berg} cited  in subsection~\ref{qugr24}).

  Besides further potential astronomical means of testing models of 
quantum gravity \cite{Calc,Jacob} in subsection~\ref{qugr24} we also 
noted that
  on the opposite distance scale experiments have been proposed for a 
laboratory setting which may be able to test in general whether 
gravity is quantised or not \cite{Haine,Faure,Howl,Kamp}.
  Given the progress made in recent years in improving the 
feasibility of  a variety of possible approaches the technology could 
be in reach to realise at least one of these terrestrial experiments 
in the foreseeable future. 
 With gravity itself being unquantised in the present theory, unlike 
the case for many approaches to quantum gravity, this aspect might 
hence then be testable in such an experiment.

However it is anticipated that the main area for testing this theory 
could be found in the QFT limit of traditional HEP experiments, 
augmenting the range of predictions presently possible with QFT for 
the Standard Model of particle physics in such an environment. The 
approximations that can be employed in approaching the flat spacetime 
case suggest that the corresponding mathematical structure for the 
present theory may also be more accessible than for the more general 
case, as noted in the opening of this subsection. While the 
development of the corresponding mathematical formalism was described 
in the two previous sections, in the following subsection we consider 
conceptual questions pertaining to this special relativistic limit.

%\vspace{-2pt}
%\pagebreak
\subsection{Quantum Field Theory Limit}
\label{qugr72}

   In this subsection we describe how the conceptual picture for the 
theory presented in sections~\ref{qugr3}--\ref{qugr6} can be 
compatible with, provide an explanation for, and address conceptual 
issues regarding the calculations of QFT in the flat spacetime limit. 
 Perhaps the main technical concern, and
a prominent feature of QFT, is the prevalence of infinities that 
arise on the calculational side and the need for renormalisation, as 
we noted towards the end of subsection~\ref{qugr21}. 
 An inability to renormalise generally proves fatal for a QFT, as has 
been the case for direct attempts to quantise the gravitational field 
as noted in the opening of subsection~\ref{qugr23}.
 However even if renormalisation is possible, as it is for the 
Standard Model of particle physics, the occurrence of infinities 
suggests that the description of processes in terms of a transition 
amplitude $\mcM_{fi}$, as introduced for equation~\ref{sfi}, may be 
incomplete with a fuller picture  desired, as also considered at the 
end of subsection~\ref{qugr21}.   

  The present theory purports to provide such a complete description 
as identified through equation~\ref{pdddds}, with event likelihoods 
determined by the local degeneracy of matter field solutions for the 
local spacetime geometry in equation~\ref{gfromavt}.
 Here `infinities' naturally arise in the degeneracy `counts' $D_+$ 
and $D_-$ of equation~\ref{ppdd} since there are an infinite `number 
of ways' to divide up a finite interval of the continuum of time in 
describing field exchanges underlying the local spacetime 
composition. This kind of infinity is dealt with simply by replacing 
discrete sums with continuous integrals such as 
equation~\ref{degplus}  (as described for \cite{Unifi} 
equations~11.40 and 11.41). 

  For intermediate states involving multiple fields there is a 
further infinity that arises through the degeneracy in the 
individually unbounded wave 4-vector $k_1$ and $k_2$ contributions in  
factors of  the Fourier components $e^{ik\cdot x} = 
e^{i(k_1+k_2)\cdot x}$ 
 (as would be the case if the $\ol{\psi}\gamma^{\mu}\psi$ stage in 
equation~\ref{ext1} and figure~\ref{extofey}(a) were an internal 
transitional link, as discussed for \cite{Unifi} figure~11.6 and equation~11.44),
closely analogous to the divergent 4-momentum loop integrals that are 
encountered in QFT  (\cite{Unifi} figure~10.9 and equation~10.86). However in all cases the process probability 
$P_{fi}$ in equation~\ref{ppdd} corresponds to a \textit{relative} 
frequency and will be normalised by compensating infinities 
associated with the degeneracy count for other possible processes, as 
noted after equation~\ref{ppdd} (and discussed for \cite{Unifi} figure~11.11 and  
equation~11.48). This trivial normalisation with $\sum_f P_{fi} =1$ 
over all potential outcomes is the same as that 
 for \textit{any} classical probability, without any evident need for 
\textit{re}normalisation (see also the discussion of a classical `dartboard' 
analogy shortly after \cite{Unifi} equation~11.48).

    If the complexification of this calculation through to the 
right-hand side of equation~\ref{pdddds} is indeed able to connect 
with a standard formalism of QFT, then the manner in which QFT 
infinities arise despite unitarity and the apparent constraint of 
equation~\ref{fsumunit} might be further analysed.
 However the direct statistical normalisation might then be hidden 
and need to be replaced by a conventional program of QFT renormalisation that 
involves calibrating against empirical observations in a manner 
insensitive to very high-energy phenomena (\cite{Unifi} 
section~11.3).
  The related empirical phenomenon of `running coupling' 
 (\cite{Unifi} equation~11.47 and figure~11.10) should, on making the 
connections from right to left through equation~\ref{pdddds}, also be 
a property identified for the present theory, which in addition 
should directly apply for an arbitrarily high energy scale.

   In demanding the consistency of the present theory with QFT one 
implication for any relativistic interacting particle theory that 
might be of concern is the Coleman-Mandula theorem~\cite{ColMan}.
 This states that the most general Lie algebra describing symmetries 
of the $S$-matrix consists of a set of internal symmetry generators 
that commute with the set of Lorentz transformation and spacetime 
translation generators. That is,
 in order for the $S$-matrix of equation~\ref{smatrix} to be 
non-trivial, the external Lorentz symmetry and internal gauge 
symmetry $G$ must be unrelated and utilised in the form of a trivial 
direct product $\mbox{Lorentz} \times G$, rather than associated as  
subgroups of a larger Lie group symmetry $\widetilde{G}$ for the 
theory with:
\begin{equation}
 \mbox{Lorentz} \times G \, \subset \, \widetilde{G}
 \label{gbreakc}
\end{equation}

  On the other hand the existence of such an inclusive group 
$\widetilde{G}$ would in some sense be more consistent with the ideal 
of \textit{unification}, in implying a \textit{common} source for 
both the external Lorentz and internal $G$ symmetries. For the 
present theory we indeed \textit{begin} with such a unifying full 
symmetry $\hG$ as the symmetry of the full generalised 
\textit{mathematical} form for proper time in equation~\ref{lpvn}. 
However compatibility with the Coleman-Mandula theorem for the QFT 
limit is also ensured by the \textit{absolute} nature of the symmetry 
breaking over 4-dimensional spacetime down to the subgroup in 
equation~\ref{gbreak}, with $\hG$ broken \textit{necessarily} in 
order to identify 4-dimensional spacetime itself.
 The \textit{physics}, including all connections with QFT, then 
\textit{begins} with \textit{only} 
 the direct product symmetry $\mbox{Lorentz} \times G$ 
  surviving in spacetime.

 It is not the case that we are \textit{starting} with 4-dimensional 
spacetime and a local Lorentz symmetry that is \textit{then} 
augmented to a larger symmetry $\widetilde{G}$ to encompass also the 
internal symmetry $G$ in the form of equation~\ref{gbreakc}. Rather 
we \textit{start} with the `one dimension' of proper time which can 
be directly arithmetically expressed in the \mbox{$n$-dimensional}  
generalised form of equation~\ref{lpvn} with full symmetry $\hG$.
 The necessary identification of the extended 4-dimensional spacetime 
manifold itself, through which we perceive all events in the world, 
\textit{then} breaks the symmetry $\hG$ absolutely down to 
 $\mbox{Lorentz} \times G$ as the starting point for all physics.

   This fundamental symmetry breaking, through which the physical 
world itself is identified from the original mathematical substructure of time, 
is not itself the consequence of a `Higgs mechanism', which typically 
concerns a model constructed \textit{within} the \mbox{4-dimensional} 
physical world. On the other hand the subset of field components
 \mbox{$\bh(x) \equiv \bv_4(x) \in \TM_4$} projected onto the local external 
spacetime, out of $\bv_n$ in equation~\ref{lpvn} and as a central 
element of the symmetry breaking here,  \textit{is} associated with 
the origin of Higgs phenomenology at the electroweak scale in the 
Standard Model, as discussed in section~\ref{qugr3} and for 
equations~\ref{gwarph}--\ref{lpvnb}. Here the Standard Model gauge group \mbox{$\SML$} is 
predicted to be identified directly as the internal component $G$ in 
equation~\ref{gbreak} for the full $\hG = \ee$ case
  of equation~\ref{lvto}, with no further symmetry breaking mechanism 
required (\cite{TimeE} equation~69).

  Since there is only one underlying source of symmetry breaking, and 
no Higgs mechanism required for example at the `Grand Unification' 
scale, any potential `hierarchy problem' is in principle avoided.
 That is, there is no reason for the GUT scale at which the strong, 
weak and electromagnetic running couplings approximately converge, 
and as not here associated with any Higgs-like field, to be closely 
related to the electroweak scale and parameters relating to the $\bh(x)  
\equiv \bv_4(x) \in \TM_4$ component projection, as associated with the 
Standard Model Higgs field, since these are two different aspects of the 
overall symmetry breaking structure
 (as discussed in \cite{KKone} towards the end of subsection~5.3).

  In QFT it is also required that the internal symmetry $G$ should 
consist of \textit{compact} groups, or factors of $\uo$, to ensure 
positive kinetic energy for the Lagrangian terms in the 
gauge field strength $F^{\alpha}_{\ph{\alpha}\mu\nu}(x)$ of 
equation~\ref{fofa}
 described for equation~\ref{lagsff}; which more specifically are then 
of the form:
\begin{equation}
 \label{lagkff}
 \lag  = - \frac{1}{4}K_{\alpha\beta} 
  F^{\alpha}_{\ph{\alpha}\mu\nu} F^{\beta \:\!\mu\nu}
\end{equation}
 with a corresponding negative definite Killing form $K$ (as 
discussed for \cite{TimeE} equation~88, see also \cite{Witt} 
section~1 discussion of equation~1.1).
 While here we are considering the 4-dimensional spacetime physics 
for the present theory, this has implications for the general form of 
proper time itself in equation~\ref{lpvn}, in order that the 
resulting internal $G$ of equation~\ref{gbreak} should be composed of 
factors of the appropriate consistent compact form.
   This is the case for the specific real forms $\hG = \esig$ and 
$\hG = \eseg$ utilised for the particular forms of proper time 
described in section~\ref{qugr3} and suggests that for the predicted 
further extension to $\hG= \ee$ the real form $\eeg$ of this largest 
exceptional Lie group might be sought (as explained for \cite{TimeE} 
equation~89).

  Given the internal symmetry $\SML$ non-trivial conditions are also 
placed on the particle content by the demand that anomalies should 
cancel.
 In QFT `anomalies', divergences that wreck the renormalisability of 
a theory, occur if for example a local gauge symmetry of a classical 
Lagrangian is broken through the interaction terms under 
quantisation. Hence for a consistent QFT all such anomalies must 
cancel to zero (\cite{Pesk} section~20.2 on `Anomaly Cancellation').
  For the chiral gauge group $\SML$ this introduces constraining 
relations between the number of colours and flavours for each 
generation.
 These constraints are both very restrictive and also consistent with 
the Standard Model multiplet structure (\cite{Unifi} equation~7.36). 
For a consistent QFT limit for the present theory such a multiplet 
structure should also arise from the breaking of the predicted full 
form of proper time $\lvtfep$ in equation~\ref{lvto}, and might even 
help guide the identification of this structure.

  If the universe was to be considered a fundamentally quantum world 
then all possible couplings between different fields consistent with 
the symmetries of the theory, such as gauge invariance, would be 
expected to play a role. For a quantised gravity theory that would 
also include the gravitational field (as discussed for \cite{Carr} 
equation~4.32). Even for the Standard Model alone for QCD in addition 
to equation~\ref{lagkff} the Lagrangian term (\cite{Unifi} 
equation~11.39, \cite{Teub} section~3.3, \cite{Pecc} equation~17):
\begin{equation}
  \lag_{\theta} = \frac{\alpha_s}{4\pi} \, \theta \, 
F_{\alpha}^{\ph{\alpha}\mu\nu}
    \past{F}^{\alpha}_{\ph{\alpha}\mu\nu}
	   \label{stocp}
\end{equation}
 with $\alpha_s$ the strong coupling and 
$\past{F}^{\alpha}_{\ph{\alpha}\mu\nu}(x)
	     = \fh\varepsilon_{\mu\nu\rho\sigma}
	   F^{\alpha\,\rho\sigma}(x)$ the dual field strength tensor, 
necessarily arises from the complicated nature of the QCD vacuum.
 This term implies \textit{CP} violation in the strong sector, 
contrary to observation, unless the $\theta$-parameter  takes an 
unnaturally very small value, with an experimental limit of
   \mbox{$\vert\theta\vert \lesssim 10^{-10}$} 
    (\cite{PDG20} section~9.1). 
 In the present theory, with gravity taking priority and QFT obtained 
only as a limit, this `strong \textit{CP} problem' is potentially 
avoided since, while $FF$ terms of the form of equation~\ref{lagkff} 
are identified through the \textit{geometric} construction described 
for equation~\ref{gchift}, there are no $F\past{F}$ contributions to associate with equation~\ref{stocp}. 
This approach might then come under the category of `unconventional 
dynamics' (\cite{Pecc} section~2), which  here arise through a new 
theoretical framework and the resulting constraint 
equations~\ref{gchift}--\ref{emcon} which prescribe all possible 
field couplings. 
  
   As a further property of QFT `spin-1' particles, associated with the 
gauge fields of the internal symmetry, behave 
statistically as boson states, while `spin-$\frac{1}{2}$' particles, 
corresponding to leptons and quarks, behave as fermions
 (see for example \cite{Pais2} section~20(c).3, \cite {Pesk} 
section~3.5 and references therein).
  For the present theory the spin-1 gauge fields $A(x)$ derive directly from 
the internal symmetry $G$ component of $\hG$ in 
equation~\ref{gbreak}, while under the external Lorentz symmetry the 
spinor states of leptons and quarks are identified in the fragmented 
components of $\bv_n(x)$ of equation~\ref{lpvn} in the symmetry 
breaking over $M_4$ to equation~\ref{lpvnb} (see \cite{TimeE} figure~4 for the 
$\hG = \ese$ case). The relation between `spin' and `statistics' in 
QFT  will then also need to apply for the present theory in order to 
account for the well-established empirical properties of 
\mbox{spin-1} boson and spin-$\frac{1}{2}$ fermion particle states.

  That is the fermionic properties of spinor fields such as $\psi(x)$ 
in equation~\ref{phiee} as identified in equations~\ref{dlpvnbex} and \ref{ext0}
 will need to be accounted for similarly as these properties apply 
for the spinor field in the standard Dirac equation (\cite{Pesk} 
equation~3.31) under quantisation. In the standard approach for the 
latter case the need to consistently define a positive energy for a 
state through a Hamiltonian operator similar to 
equation~\ref{hathqft} 
 requires that equation~\ref{aacomr} must be replaced by 
\textit{anti}commutation relations for the creation and annihilation 
operators (\cite{Pesk} equations~3.90--3.97) implying in turn that 
particles of the Dirac field obey Fermi-Dirac statistics.
 Analogous consistency requirements may be sought for the present 
theory to identify fermionic properties for the spinor states.
The spacetime geometry of particle wave-packet states and the finite 
torsion associated with spinors, discussed towards the end of both
subsections~\ref{qugr52} and \ref{qugr62}, 
might also play a role here in the relation between spin and 
statistics.

    While the quantitative calculational side for the present theory 
is still at a preliminary stage, as described for the constructions 
relating to equation~\ref{pdddds} in section~\ref{qugr5}, a 
significant contribution that the theory does in principle provide is 
a seamless and consistent conceptual picture concerning \textit{what is 
happening} in the processes observed in HEP experiments. Here we 
again consider in particular the particle interaction processes at an 
electron-positron collider of the type $e^+ e^- \to Z^0 \to b\bar{b}$
  of equation~\ref{eezbb}.
  Such events, with $e^+$ and $e^-$ lepton states colliding to 
produce  a $Z^0$ gauge boson which decays into a pair of $b$ and 
$\bar{b}$ quarks, are identified through the vertex structure of the 
subsequent decays of the heavy $B$-hadrons, as pictured in the event 
of figures~\ref{sldbb}(a) and \ref{sldbb}(b) in 
subsection~\ref{qugr42} and again on a yet smaller scale in 
figure~\ref{sldbbv}(a) below.  

%temporary
%\pagebreak

\begin{figure}[htb]  
\centering
% \hspace*{-48pt}
%\hspace*{-28pt}
\hspace*{-13pt}
%\hspace*{-14pt}
\epsfxsize=15.2cm
\leavevmode
\epsffile[0 0 3235 1553]{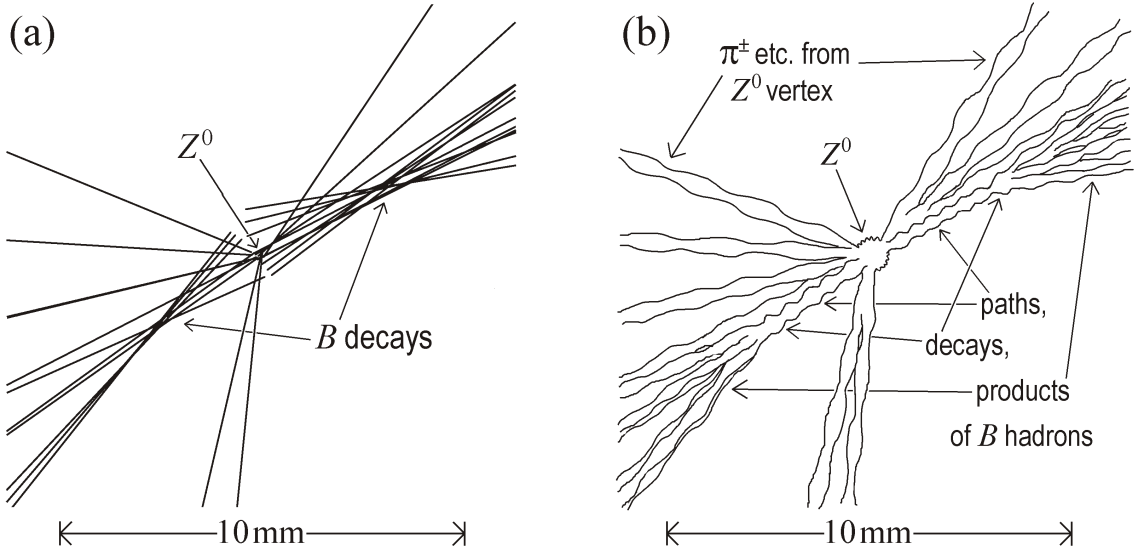}
%\hspace*{-48pt}
\caption{\setb  (a) The $b$-quark/$B$-hadron production event of 
 figure~\protect\ref{sldbb}, here zoomed in to show the secondary 
vertex structure more clearly~\protect\cite{SLDweb}. (b) Artistic 
impression of the \mbox{4-dimensional} spacetime metric geometry 
$g_{\mu\nu}(x)$ enveloping such a process as a solution for 
equation~\protect\ref{gfromavt} in the present theory.
  Further geometric distortions, not shown, will be associated with 
the incoming $e^+$ and $e^-$ beams perpendicular to the diagram 
plane.}
\label{sldbbv}
\end{figure}

  In physically generating the initial $e^+e^-$ state in 
equation~\ref{eezbb} the final focussing magnets are situated well 
over a metre from the central interaction point perpendicular to the 
plane of the event pictured, while for observing the final state 
 the innermost tracking measurements are provided by the vertex 
detector, as shown in figure~\ref{sldbb}(b), with an inner layer 
situated at an average radius of 28$\,$mm just outside a thin beam-pipe 
 (\cite{VXD3} figures~3--5 and table~2).
 Hence  the cross-sections and lifetimes, for the particle 
interaction and decay processes occurring fully \textit{inside} the 
detector volume such as observed in figure~\ref{sldbbv}(a), are 
computed purely in terms of the calculational tools of QFT in an 
otherwise vacuum region with no macroscopic matter present.

   At the same time in reconstructing these events the extrapolation 
of the tracking data from the detector elements, located between a 
few centimetres to a few metres from the interaction point, to the 
region within the beam-pipe, as pictured in figures~\ref{sldbb}(b) and 
\ref{sldbbv}(a), treats particles as essentially classical objects 
following classical trajectories. This is seen for example in vertex 
reconstruction algorithms that utilise the measurement uncertainties 
in the track trajectories, as determined by the detector, as purely 
classical statistical errors (see for example~\cite{ZVTOP} equation~1 
and figure~2).

  In the Copenhagen interpretation, described in 
subsection~\ref{qugr21}, a line is drawn between the quantum system 
and the classical measuring apparatus which, while the explicit 
nature of the line is not clearly specified, permits no overlap 
between the two realms. However whatever is `happening' 
\textit{within} the detector volume in figure~\ref{sldbbv}(a) appears 
to be neither a purely continuous evolution of a quantum state 
\textit{nor} an interaction with the experimental apparatus, unless 
the interaction with the vertex detector elements can `act backwards 
in time' as well as non-locally on an apparent wavefunction. 
The inter-particle interactions described by QFT could be interpreted 
as stimulating the `wavefunction collapse', similarly as for an 
interaction with a detector element; however the calculations of QFT 
are of an entirely quantum nature -- determining an array of 
probabilities for possible measurements, as for the propagation of 
the wavefunction in figure~\ref{dslit} for example,
 and give no account of the physical decay vertices of the 
  specific outcome observed.

  The intrinsically quantum mechanical nature of these systems can be 
seen for example in the observation of
  $B^0_d \leftrightarrow \overline{B^0_d}$ meson mixing  which takes 
place between the corresponding $B$ production via a $Z^0$ decay at 
the central interaction point and the \mbox{$B$ decay} at the 
secondary vertex, well within the radius of the innermost detector 
elements (see for example \cite{SLDmix} figure~7). Hence the quantum 
mechanical wavefunction, describing an evolving superposition of 
$B^0_d$ and $\overline{B^0_d}$ states, appears to collapse to a 
specific decay outcome \textit{within} the quantum system and 
manifestly \textit{before} any interaction with any classical 
measurement apparatus.

 Hence it appears that something \textit{decisive} must indeed 
\textit{happen} at the decay vertex at the level of purely 
microscopic particle or field interactions effectively acting to 
`collapse the wavefunction' and generate the macroscopic physical 
form of the observed vertex structure. These observations suggest 
pushing some of the characteristic features associated with the 
`classical realm' into that of an otherwise distinctly quantum 
environment. This is essentially the reverse of attempts to extend 
the `quantum realm' into that of macroscopic classical objects, as 
explored for example in the `Schr\"{o}dinger cat' thought experiment 
and leading to considerations such as the `many worlds' 
interpretation of quantum theory as also reviewed in 
 subsection~\ref{qugr21}.

  Here we conclude that the ability to reconstruct a structural 
pattern of secondary and even tertiary decay vertices (see 
\cite{ZVTOP} figures~1 and 8 for a simulation description) 
\textit{within} the territory of a quantum system at a macroscopic 
distance from the nearest detector elements suggests that something 
of a tangible `semi-classical' nature is apparently happening in such 
processes. While QFT can be utilised to determine interaction rates, 
average decay lengths and mixing probabilities it offers no explicit 
description of what is physically transpiring in the extended 
spacetime volume of decay chains such as empirically identified in 
figures~\ref{sldbb}(b) and \ref{sldbbv}(a) 
(beyond pragmatically representing particle states by complex 
wave-packet functions through which the likelihood of observable 
outcomes can be determined).

 However the present theory directly incorporates a conception of the 
structure of spacetime itself through solutions for a generally 
non-flat metric geometry $g_{\mu\nu}(x)$ with corresponding Einstein 
tensor $G^{\mu\nu}(x)$ and a locally degenerate composition in terms 
of  $A(x),\bv_n(x)$ field components.
 The local degeneracies underlie a causal \mbox{one-dimensional} 
accumulation of probabilistic events embedded in the 4-dimensional  
spacetime solution as alluded to in subsection~\ref{qugr41} (and 
discussed further in \cite{Unifi} section~11.3 pages 328--330).  
 These solutions derive from the breaking of the symmetry of the full 
form of proper time of equation~\ref{lpvn} as we have described for 
equation~\ref{gfromavt}. This picture consistently and seamlessly 
subsumes both microscopic `quantum' processes, such as the 
interactions, decays and any mixing underlying the vertices in 
figures~\ref{sldbb}(b) and \ref{sldbbv}(a), as well as the 
macroscopic `classical' detector elements recording such phenomena as 
shown in
 figures~\ref{sldbb}(a) and \ref{sldbb}(b). Here there is no 
conceptual boundary or `line' to be drawn between the `quantum' and 
`classical' realms as for some interpretations, such as the 
Copenhagen analysis, of quantum theory.    

   Rather than acting as a \textit{source} of spacetime curvature 
here all matter is defined through solutions for 
equation~\ref{gfromavt} for constructing spacetime itself. An 
elementary particle state is considered as a \textit{minimal  
disturbance} in this geometry over the vacuum state as permitted by 
the constraint equations~\ref{gchift}--\ref{emcon}. Such a particle 
will propagate \textit{literally} in the form of a `generalised 
wave-packet' in a real field in subcomponents of $A(x)$ or $\bv_n(x)$ 
and \textit{also} in the spacetime geometry $g_{\mu\nu}(x)$ itself, 
consistent with a global extended solution for 
equation~\ref{gfromavt} as described in sections~\ref{qugr5} and 
\ref{qugr6}.
 While the `tracks' in figure~\ref{sldbbv}(a) represent the 
empirically reconstructed time-integrated trajectories of particle 
states, the corresponding particle wave-packets as propagating  
distortions modulated by sinusoidal undulations in the smooth 
geometry of spacetime for this theory are loosely illustrated in 
figure~\ref{sldbbv}(b) as also integrated over time. 
 In figure~\ref{sldbbv}(b) the transverse width of particle states, 
suggested by the single contour in the $g_{\mu\nu}(x)$ geometry, is 
very much not to scale; indeed the tracking resolution of the vertex 
detector is itself of order tens of microns, much more narrowly 
focussed than the paths depicted.

  The specific form for equation~\ref{lpvn} and the resulting 
symmetry breaking structure will determine the available components 
of the fields $A(x)$ and $\bv_n(x)$ and the set of possibilities for 
discrete exchanges between them via the constraint 
equations~\ref{gchift}--\ref{emcon}. As reviewed in 
section~\ref{qugr3} the natural progression through $\hG = \esi$ and 
$\hG = \ese$ forms for equation~\ref{lpvn} leads to a close 
connection with features of the Standard Model of elementary particle 
physics  (\cite{TimeE} figure~4).
 The nature of hadrons (as composed of quarks, including the $B$ 
states, pions, kaons and so on observed in experiments such as that 
in figures~\ref{sldbb} and \ref{sldbbv}), atoms, and other 
non-elementary particle states with quantum properties are also 
anticipated to be determined by these underlying constraints, as will 
be discussed further in section~\ref{qugr8}.

  The constraints on interacting generalised wave-packet states have 
already been shown to account for the elementary relation between the 
associated 4-momentum $P$ and wave 4-vector $k$ 
  for particle quanta as argued for equation~\ref{phbarku} in 
subsection~\ref{qugr61}.
 As an element of this construction 
  through the definition of the energy-momentum tensor in 
equation~\ref{gfromavt} the wave-like frequency component of 
solutions for particle quanta 
 is proposed to determine the corresponding invariant mass spectrum 
and kinematic properties of particle states as described for equation~\ref{yhpsi}  in 
subsection~\ref{qugr62}. More specifically the 
constraints, including interactions of the matter fields with the 
`Higgs' components of $\bh(x) \equiv \bv_4(x) \in \TM_4$ as discussed for 
equations~\ref{gwarph}--\ref{lpvnb} and utilised in equation~\ref{yhpsi}, are anticipated to restrain the 
wave 4-vector $k$ in the undulating $\cos{k \!\cdot \! x}$ components 
of the geometric wave-packet structure of equations~\ref{gbwavesup} and \ref{gkekksup}
 such that the particle mass $m$, with 
  $m^2 = P^2 = k^2$, is directly linked with a rate of change in time 
at the most elementary level as described for equation~\ref{mhac}, and as we have to some extent attempted to 
represent in figure~\ref{sldbbv}(b).

 Given the range of Standard Model particle states with known masses 
this in principle may be a significant testable and predictive 
feature for the theory. 
 As well as leptonic and hadronic states this construction should 
also apply for the massive $Z^0$ and $W^{\pm}$ gauge bosons to be 
identified in accounting for the full nature of electroweak symmetry 
breaking in the present theory. 
  This particle quanta description will also need to include the case 
of photons associated with the electromagnetic wave solutions with 
$m^2 = P^2 = k^2 = 0$ as alluded to towards the end of 
subsection~\ref{qugr51}, for which a rest frame description is not 
possible, but with the argument leading from equation~\ref{fcoskx} to 
equation~\ref{phbarku} holding in the limit $k^2 = \alpha^2 \to 0$ 
and equations~\ref{phbarku} and \ref{mhac}  
 applying uniformly in all cases.

  Such wave-packet solutions as the slightest ripples of Ricci 
curvature  in the spacetime geometry, associated with particular 
$A(x),\bv_n(x)$ component matter field compositions, as an element of 
extended global solutions for $g_{\mu\nu}(x)$ through 
equation~\ref{gfromavt}, will continuously blend into the 
4-dimensional geometric structure of macroscopic entities such as 
particle detectors, as will be exemplified in figure~\ref{dslitg} for 
the case of the double-slit experiment. Here it is the extension of 
solutions for the spacetime geometry inwards, down to arbitrarily 
microscopic scales of matter, in a unifying theory of `quantum 
gravity' that is proposed to fully account for the nature of both 
non-relativistic and HEP particle processes, such as depicted in 
figures~\ref{dslit} and \ref{sldbb} respectively. Rather than neglecting gravity, 
which is typically the case in the laboratory environment and for QFT 
calculations, here gravity is very much an irreducible part of 
understanding the full nature of quantum and particle phenomena, as 
described for the example of figure~\ref{sldbbv} above.

  While in principle providing a complete conceptual picture it 
remains to explicitly reproduce the calculational successes of QFT, 
let alone fulfil the potential of going further in computing particle 
masses and other empirical parameters of the Standard Model, with 
provisional progress in this direction described in 
sections~\ref{qugr5} and \ref{qugr6}. In general however, as noted 
above, solutions to equation~\ref{gfromavt} for microscopic systems 
studied in the laboratory will inevitably and inseparably involve the 
extended environment of the experimental apparatus. Such an 
inseparability has also been a central feature of attempts, such as 
the Copenhagen interpretation, to understand the application of 
quantum mechanics in itself. From the point of view of the present 
theory the conceptual problems with quantum mechanics arise precisely 
since the role of gravity, while utterly immeasurable  in itself in 
such an environment, cannot be neglected. We continue to describe the 
conceptual picture of the present theory, and the manner in which 
long-standing issues in non-relativistic quantum mechanics might be 
resolved, in the remainder of this section.

%\pagebreak
\subsection{Non-Relativistic Limit}
\label{qugr73}

  In this subsection we elaborate on how the approach to quantum 
gravity in this paper can address a series of prominent conceptual 
issues regarding quantum mechanics itself, as originally formulated 
in the 1920s and reviewed in subsection~\ref{qugr21}. In the unified 
theory presented here gravity takes precedence over quantum theory. 
Typically for the classical theory of general relativity, while exact 
solutions for the Einstein equation~\ref{Eineq} may be hard to find, 
the spacetime geometry can in principle be uniquely identified in a 
deterministic manner for given boundary conditions, as reviewed in 
subsection~\ref{qugr22}.
 Similarly free bodies follow deterministic geodesic trajectories 
through spacetime as a further postulate of general relativity. At 
first sight the inherent randomness of quantum phenomena might appear 
to be at odds with this standard conception of general relativity.

  However the field equation~\ref{Eineq} itself does \textit{not} 
prohibit an element of indeterminacy from entering the physical 
world, no more than it prohibits for example classical 
electromagnetic phenomena. The structures of Riemannian geometry 
imply the Bianchi identity, which via the contracted form
 $\nabla_{\:\!\!\!\mu}G^{\mu\nu}=0$  of equation~\ref{Bian} implies 
energy-momentum conservation as reviewed for equation~\ref{emcon}. 
Nevertheless for example the decay of a $Z^0$ boson into two jets of 
particles containing mutually recoiling $B$-hadrons pictured in 
figures~\ref{sldbb} and \ref{sldbbv}(a) can still occur by pure 
chance in \textit{any} direction over the full $4\pi$ solid angle, 
with further randomness in the decay length and decay products of the 
$B$-hadron states, all being completely compatible with 
energy-momentum conservation. In turn such events are also consistent 
with the continuous enveloping geometry $G^{\mu\nu}(x)$ of 
equation~\ref{gfromavt}, with the metric solution $g_{\mu\nu}(x)$
 represented in figure~\ref{sldbbv}(b), and the corresponding 
properties of Riemannian geometry.
 Here it is the `slippery ambiguity' of the underlying local 
degeneracy in the $A(x),\bv_n(x)$ field composition  of 
$g_{\mu\nu}(x)$ through equation~\ref{gfromavt}, describing the invariant 
local geometric structure for example at the $Z^0$ and $B$ decay vertices in 
 figure~\ref{sldbbv}(b), that provides the \textit{source} of quantum 
uncertainty while maintaining compatibility with the Einstein field 
equation~\ref{Eineq}, as initially described in subsection~\ref{qugr41}. 

  As noted in subsection~\ref{qugr21} indeterminism in quantum 
mechanics is not necessarily a conceptual \textit{problem}, it can 
simply and consistently be posited as a `brute fact' about the world. 
However the existence of this randomness has been intuitively 
puzzling since the early history of quantum mechanics, dating back to 
the debate over whether `God plays dice' as alluded to 
subsection~\ref{qugr21}. The origin of the indeterminacy is 
\textit{explained} in the present theory which begins with the one 
dimension of time alone, as expressed directly through the general 
multi-dimensional arithmetic form of equation~\ref{lpvn} with $\hG$ 
the full symmetry. This symmetry is broken absolutely in 
equation~\ref{gbreak} through the necessity of identifying 
4-dimensional spacetime with a local $\mbox{Lorentz} \subset \hG$ 
symmetry, leading to an \textit{intrinsic} degeneracy in the 
construction of the spacetime geometry. In principle there are many 
possible solutions for such a geometry through equation~\ref{gfromavt}, 
each described by $g_{\mu\nu}(x)$ up to a non-physical choice of 
coordinates, all exhibiting local degeneracies in the $A(x),\bv_n(x)$ 
composition and with all bar one of which \textit{not} realised as 
\textit{our} universe. This potential for `many solutions', 
offsetting the asymmetry implied in assuming the possible 
construction of the specific features of one universe alone, is an 
\textit{intrinsic} feature of the theory, rather than an 
\textit{interpretation} as for the `many worlds' view of quantum 
mechanics.  

  As perhaps an even more characteristic property of quantum theory 
the \textit{discreteness} of particle phenomena, as encapsulated in 
the de Broglie relations of equations~\ref{ehbaro} and \ref{phbark}, 
is also entirely compatible with the \textit{continuity} of the general 
relativistic spacetime geometry in this theory, as described in 
subsection~\ref{qugr61} culminating in equation~\ref{phbarku}.
 This relation for interacting particle wave-packets is a direct 
consequence of the constraints implied in constructing 4-dimensional 
spacetime together with its geometric structure and mutually related 
matter field content from generalised proper time.
 It is indeed these apparent particle quanta that display the 
probabilistic nature of elementary phenomena.

  The likelihood of detecting a particle at any given position in 
space is determined in quantum mechanics by a wavefunction 
$\Psi(\bx,t)$ which possesses a distinctly non-local character, as 
also discussed in subsection~\ref{qugr21} and as seen for example in 
the interference effects that can be observed through accumulating 
single electron events in the double-slit experiment of 
figure~\ref{dslit}. For the present theory such elements of 
non-locality are proposed to derive from the fact that we do not 
\textit{begin} with an external 4-dimensional spacetime arena within 
which to describe events, but rather the spacetime manifold $M_4$ is 
\textit{itself} built out of substructures of the general form of 
proper time in equation~\ref{lpvn}.

  While we begin with generalised proper time there is an \textit{a 
priori} necessity to perceive the world in space as well as through 
time. It is in the nature of extended \mbox{3-dimensional} space of 
\textit{itself}, by definition, to be a non-local entity. This 
property of space and its geometric structure can be described in 
simple mathematical terms which in turn can be extracted from 
substructures of the general form for proper time (\cite{Struct}, 
\cite{KKone} section~2).
The four projected components of $\bv_4 \in \TM_4$ are mathematically a 
\textit{subset} within the $\bv_n \in \rrr^n$ components of 
equation~\ref{lpvn}, however, in exhibiting through equation~\ref{sfourd} the local metric structure 
of a 4-dimensional Lorentzian spacetime, physically they provide the 
basis for the extended external manifold $M_4$ that 
\textit{accommodates} the residual subcomponents of 
equation~\ref{lpvn} from the symmetry breaking of equation~\ref{gbreak} as apparent matter 
fields.
 Here it is not a case of space being \textit{given}, with a 
geometric structure and a matter content \textit{then} being 
introduced. Rather extended 4-dimensional spacetime, of necessity
 incorporating the \textit{holistic} property of 3-dimensional space,
 is fabricated together with a geometric structure and corresponding 
matter field composition collectively as 
a solution  for equation~\ref{gfromavt}.

  While the geometry of the 4-dimensional manifold $M_4$ determines a 
causal light cone structure \textit{within} spacetime, the 
construction of the spacetime solution itself is necessarily 
identified \textit{outside} this causal structure in an intrinsically  
and irreducibly non-local manner, as also discussed in 
subsection~\ref{qugr41}. 
 That is, regions of the spacetime manifold $M_4$ that are causally 
unconnected, with respect to the light cone structure, still 
necessarily mesh together in a smooth and continuous manner both in 
terms of an extended space and a coherent geometric structure, as 
described by $g_{\mu\nu}(x)$ through equation~\ref{gfromavt}, 
everywhere. This includes across all spacelike hypersurfaces, through 
the very notion of space as an extended non-local entity.

 In general relativity all solutions for the Einstein 
equation~\ref{Eineq} correspond to a complete 4-dimensional spacetime 
structure, such as for the Schwarzschild solution or for cosmological 
models as reviewed in subsection~\ref{qugr22}, and are generally 
interpreted as solutions for the geometry of a pre-existing 
spacetime, in particular with the `fabric of space' warped by the 
presence of matter. The image of space as a malleable rubber sheet 
bending under the weight of massive objects that are subsequently  
introduced, as alluded to in subsection~\ref{qugr71}, reinforces this 
interpretation of general relativity.
 That is, general relativity can be interpreted as a theory of the 
gravitational field \textit{in} spacetime rather than as a theory 
\textit{of} spacetime itself -- as consistent with a logical order of 
first positing a flat spacetime, then introducing matter, and finally 
assessing the resulting curvature of the given spacetime.

 Here however through equation~\ref{gfromavt} spacetime itself is 
being constructed \textit{together with} its geometric structure \textit{and} 
implicit matter content.
  With reference to the above analogy, here the `bowling ball', the 
`rubber sheet' and its warping are all \textit{inseparably} 
aspects of the same solution for equation~\ref{gfromavt}.  
 While consistent with the Einstein equation~\ref{Eineq} and the 
notion of full 4-dimensional spacetime solutions this subtle 
difference from the meaning of general relativity was discussed in 
subsection~\ref{qugr71}. For the present theory the non-local 
character implicit in solutions for equation~\ref{gfromavt} is then 
imported into quantum phenomena. Hence the non-locality in the 
apparent evolution of the wavefunction $\Psi(\bx,t)$ in quantum 
mechanics is effectively subsumed under that of gravity
 as identified with the geometric construction of spacetime itself.

  The non-local nature \textit{of} `space', as external to all 
events, is in a sense  opposite to the conception of a `particle' as 
localised \textit{in} space and identified through interaction 
events. For the present theory particle entities are \textit{not} 
introduced \textit{into} spacetime as the building blocks of matter, 
as noted near the opening of subsection~\ref{qugr63}. Rather the 
observable empirical effects of particle states and their 
interactions are inferred from  equation~\ref{gfromavt} in solutions 
\textit{for} spacetime emerging through exchanges of $A(x), \bv_n(x)$ 
field components providing a locally degenerate composition of 
$g_{\mu\nu}(x)$ as allowed by the constraint 
\mbox{equations~\ref{gchift}--\ref{emcon}}, which also determine the equations of motion for particle states propagating between interactions. Quantum particle 
properties are then made apparent through an interplay of these 
non-local and local aspects to the construction of spacetime.  

 As well as the interference effects observed for single particle 
states, such as in the double-slit experiment, this construction
applies also for arbitrary systems of apparent multi-particle states
 (see for example~\cite{Pen} chapter~23). These include the 
statistical properties of identical particles such as bosons and 
fermions. In the latter case the connection between spin and 
statistics alluded to in the previous subsection will relate to the 
`Pauli exclusion principle' (\cite{Pais2} section~13(b)) -- with no 
two electrons able to occupy the same state in an atom, as essential 
to understanding atomic structure in the non-relativistic limit.
  The consideration of multi-particle systems  will also incorporate 
 the case of entangled particles, such as in EPR-type experiments or 
more generally, again all enveloped under single solutions for 
equation~\ref{gfromavt}.
 Given the holistic construction of spacetime and a matter 
distribution in the present theory an inherent element of 
entanglement is somewhat less surprising than for a framework in 
which a posited spacetime is then furnished with a matter content 
exhibiting an apparent `spooky action at a distance'.

 All particle interactions, quantum uncertainty and apparent 
non-local correlations are sewn into the solution for the spacetime 
structure, albeit with the transmission of signals faster than light 
prohibited by the light cone geometry as associated with the local 
Lorentz metric of equation~\ref{sfourd} as extracted from 
equation~\ref{lpvn}.
 Here causality is rooted in the original ordered one-dimensional 
progression in time that is manifested in an extended 4-dimensional 
spacetime form (see discussion on causality in \cite{Unifi} 
section~5.3 opening, chapter~11 and section~13.3 for example).
 This underlying causal progression in time is also anticipated to be 
key in establishing a quantum theory limit connecting with a standard 
formalism that emphasises the evolution of quantum states in time, as 
noted for the structure of equation~\ref{pdddds} and discussed for example in subsection~\ref{qugr63}.

  In addition to the apparent non-locality in the evolution of the 
wavefunction, a further non-locality in quantum mechanics concerns 
the apparent reduction or collapse of the wavefunction. This 
non-locality was also discussed in subsection~\ref{qugr21}, as 
exemplified by the detection of an electron hit $I$ in the 
double-slit experiment of figure~\ref{dslit}. The nature of 
wavefunction reduction and the corresponding `measurement problem' is 
perhaps the most significant internal conceptual difficulty with 
quantum mechanics in itself. In this case it cannot simply be 
dismissed as a `brute fact' about the world, as for indeterminacy and 
the non-locality of the wavefunction evolution, since it is not at 
all clear what \textit{it is} that wavefunction collapse actually 
says about the world.

 The Copenhagen interpretation allows for predictions from quantum 
theory to be made and tested against empirical observations in 
practice, but does not provide a definition for the distinction and 
boundary between the quantum system and classical measuring 
apparatus.
 Nor is there any empirical evidence for what constitutes the point 
at which, or means by which, the apparently classical mechanical 
properties of laboratory equipment take over in the reduction of the 
apparently
 quantum mechanical wavefunction on registering a measurement.

   For the present theory the partitioning of a system into an 
indefinite state and a specific outcome can be interpreted in terms 
of the local form of equation~\ref{gfromavt}. The uncertainty in the 
future evolution from a particular local form for 
$G^{\mu\nu}=f(A,\bv_n)$ corresponds to a \textit{degeneracy} in 
$A(x),\bv_n(x)$ field  solutions describing the \textit{same} local  
\mbox{4-dimensional} spacetime metric $g_{\mu\nu}(x)$ geometry. In 
evolving in time away from this state a \textit{specific} matter 
field composition, in terms of subcomponents of $A(x)$ or $\bv_n(x)$, 
will result in a distinct  
 propagation \textit{in} the spacetime geometry, consistent with the 
equation of motion for that matter field as permitted within the 
constraint equations~\ref{gchift}--\ref{emcon}, all enveloped in the 
full extended spacetime solution for equation~\ref{gfromavt}.  
 (This enveloping geometry $g_{\mu\nu}(x)$ is sketched in 
figure~\ref{sldbbv}(b) for the HEP event of figures~\ref{sldbb} and 
\ref{sldbbv}(a) for example).
 The `reduction' effectively takes place at the point at which 
$g_{\mu\nu}(x)$ describes a distinguished external geometry, such
 as for example specifically in terms of the electromagnetic field 
$A(x)$ as described for equations~\ref{tmnem}--\ref{wavecom} or in 
terms of specific subcomponents of $\bv_n(x)$ such as discussed for
 equations~\ref{gtwopsi}--\ref{phieeb} and as might represent 
leptonic states for example.
  This \textit{geometric} structure
  will generally be \textit{far} too small to be directly observable 
in terms of the minute gravitational fields involved
 and hence this `reduction' will not be immediately apparent.

   Since this `indelible mark' in the spacetime geometry is 
unobservably small an experimenter will need to `wait' until a 
\textit{non-gravitational interaction} associated with further 
$A(x),\bv_n(x)$ field exchanges takes place involving an element of 
detector apparatus through which a measurement can be made.
 (A single interaction may be sufficient, as described for  $I$ in figure~\ref{dslit}, while  
 in practice for the HEP environment a large number of such 
interactions may be needed, as for the event in figure~\ref{sldbb}).
 These interactions with detector elements, typically separated by 
macroscopic distances, are generally attributed to apparent particle 
effects.
 Such effects are elements of solutions to equation~\ref{gfromavt} 
which are determined as consistent full 4-dimensional spacetime 
geometries, as for solutions to equation~\ref{Eineq} in general 
relativity, with a particle state propagating between interactions 
effectively as 
 a minimal `gulp' in the geometry of spacetime as discussed in 
subsection~\ref{qugr62}.
 The contention here then is that the likelihood of such an 
observable particle interaction can be \textit{modelled} in the 
non-relativistic limit by a wavefunction $\Psi(\bx,t)$ and the 
methods of quantum mechanics, as considered in 
subsection~\ref{qugr63} and further below.

  For the present theory such phenomena, involving inanimate 
experimental apparatus, are objective features of the world, not 
explicitly requiring a human observer. This excludes the notion that 
`consciousness' might collapse the wavefunction and approaches such 
as the `many minds' interpretation of quantum mechanics as proposed within the class of
   subjective theories alluded to in subsection~\ref{qugr21}.
  We also note that at the macroscopic level of the well-known 
hypothetical experiment Schr\"{o}dinger's cat is here very much 
either in an alive state or in a deceased state, and definitely not 
in a superposition, somewhat before the `measurement' is observed by the 
experimenter. 

   With the indelible mark, or decisive impact, in the actual 
spacetime geometry $G^{\mu\nu}(x)$ of equation~\ref{gfromavt} 
generally made \textit{before} we can observe the consequences; the 
wavefunction, while not in itself a real physical entity, represents 
our best knowledge of the state of the system and the corresponding 
range of possible outcomes. This knowledge `catches up' with the 
actual state of the system when the observation is made, that is at 
the point described by quantum mechanics as `wavefunction reduction'. 
While our knowledge, as for the state of the wavefunction, in relying 
upon observable $A(x),\bv_n(x)$ field interactions, is somewhat 
disjointed, both the 4-dimensional spacetime $M_4$ and its geometry 
$g_{\mu\nu}(x)$ are perfectly smooth and continuous, subsuming an 
ordered temporal evolution in indeterministic exchanges of the local 
$A(x),\bv_n(x)$ matter field component contributions in 
equation~\ref{gfromavt}.

 The enigmatic character of quantum mechanics largely stems from the 
lack of our ability to detect the underlying cause of the apparent 
`wavefunction collapse'. On the other hand it is not necessarily 
surprising that there exist significant physical structures, such as 
the local geometry $g_{\mu\nu}(x)$, that are effectively impossible 
to detect directly. Nevertheless this theory, which takes this local 
physical geometry seriously, can still be highly predictive, as we 
noted towards the end of the previous subsection for example.
It then remains to develop the calculational aspects of the present 
theory, guided by and building upon the progress made in 
sections~\ref{qugr5} and \ref{qugr6}, to determine such predictions 
while also assessing the quantitative consistency with the successes 
of quantum theory.

The underlying conceptual picture of this unified theory based on the 
construction of spacetime through generalised proper time might then
 be represented pragmatically in a limiting approximation by 
the formalism of quantum mechanics, incorporating wavefunction 
evolution {\bf U} and reduction {\bf R} processes. As noted in 
subsection~\ref{qugr21} it is precisely this alternation between the 
very different properties of the continuous deterministic evolution 
{\bf U} and the discontinuous probabilistic reduction {\bf R} that 
suggests  quantum mechanics in itself represents an incomplete 
description of the physical world. 
 This incompleteness provides part of the motivation for seeking a 
more unified theory, for which the procedures of quantum theory might 
be justified as a limiting case. 
The incompleteness is here characterised in particular by the absence 
of any non-trivial  metric $g_{\mu\nu}(x)$ geometry, with particle 
states in quantum theory considered `naked quanta' from the 
perspective of the new theory, as described in 
subsection~\ref{qugr63}. 
  
  For the limiting case of the present theory to converge with 
quantum mechanics 
   the distinctive probability postulate, with the likelihood of an 
outcome being proportional to the squared magnitude $\vert \Psi 
\vert^2$ of the system wavefunction $\Psi$ must be accounted for. 
Here this relationship is proposed to derive from the connection 
between an event probability $P_{fi}$, associated with a degeneracy 
count in the present theory, with the squared magnitude $\vert 
\mcM_{fi} \vert^2$ of the transition amplitude $\mcM_{fi}$ in QFT 
calculations with $P_{fi}\propto \vert \mcM_{fi} \vert^2$ as 
described for equation~\ref{pdddds}. When taken to the 
non-relativistic limit the aim will be to identify the probability relation 
$P\propto \vert \Psi \vert^2$  consistent with quantum mechanics.

 The recovery of the non-relativistic quantum mechanical limit from 
the present theory might then be analysed in parallel with the 
corresponding limit for standard QFT; and for which for example the 
properties of electrons in atomic physics should be seamlessly 
related to the properties of electrons in HEP experiments. Further 
while this full unified theory coherently accommodates both general 
relativity and high energy particle physics the non-relativistic 
limit should consistently convolve their two historical precedents in 
classical and quantum mechanics respectively; shedding light on the 
historical origins of quantum mechanics and the debate over the 
Copenhagen and other interpretations.

 For the present theory
  the evolution of the $A(x),\bv_n(x)$ field components and their 
exchanges must be consistent with the constraint 
equations~\ref{gchift}--\ref{emcon}  in constructing the spacetime 
geometry through equation~\ref{gfromavt}. These constraints will 
limit the possible structures of the 4-dimensional spacetime and 
imply a limited set of possible detector responses and observables in 
the corresponding experiments. The discrete set of possible 
\textit{eigenstates} for a measurement in quantum mechanics, as for 
the discrete set of possible elementary particle types, should be 
largely determined by these constraints, as also considered in 
subsection~\ref{qugr63}. These derive in particular from the fact 
that the construction of 4-dimensional spacetime itself implies the 
absolute symmetry breaking in equation~\ref{gbreak} for the full 
symmetry $\hG$ of the general form of proper time in 
equation~\ref{lpvn}, with the constraints including the need for a 
coherent extended smooth geometric structure -- as discussed for 
equation~\ref{emcon} in connection with the implication of 
energy-momentum conservation continuously throughout spacetime.   

  The setup of any experiment consists of a specific distribution of 
energy-momentum $T^{\mu\nu}(x)$ in the form of the matter of the 
apparatus and hence, as a known part of the extended solution for 
equation~\ref{gfromavt}, acts as a significant `boundary condition' 
for determining the overall solution for the system being studied.  
These boundary conditions also include the potential for permitting 
and recording $A(x),\bv_n(x)$ field interactions via the macroscopic 
detector elements. The experimental apparatus itself then 
\textit{necessarily} shapes the possible forms of the overall 
extended solution for equation~\ref{gfromavt} and the possible 
observations that can be made, 
 incorporating for example the structures depicted in figures~\ref{sldbb} and
 \ref{sldbbv}.

 The `reduction' in the structure of a solution for  $G^{\mu\nu} = 
f(A,\bv_n)$  from a locally degenerate state in evolving locally to a 
resolved specific matter field composition, ultimately associated 
with an observable `eigenstate', is again not of course something 
that happens when somebody `decides' to make a measurement, but 
rather is a profusely and microscopically finely grained property of 
the physical world. The \mbox{$B$-hadrons} in figures~\ref{sldbb} and 
\ref{sldbbv} for example would `decay' whether we choose to observe 
them or not. However the geometric solution represented in 
figure~\ref{sldbbv}(b) only exists as part of a \textit{complete} 
4-dimensional spacetime solution incorporating also the macroscopic 
environment comprising the detector elements of figure~\ref{sldbb} 
and corresponding potential for interactions as an indivisible whole.

 This is consistent with the Copenhagen interpretation of quantum 
mechanics for which mutually complementary observables, and their 
corresponding range of possible measurement outcomes, can be 
determined only by complementary arrangements of detector apparatus.
In all cases the observer does not `make' any event happen, but 
rather observes a possible course of events given the intrinsic 
indeterminacy in local field states and as irreducibly channelled by 
the measurement apparatus and macroscopic environment more generally 
as an extended 4-dimensional spacetime solution for 
equation~\ref{gfromavt}.
 On the other hand the experimenter can \textit{choose} the 
arrangement of the apparatus itself and hence the nature of the 
permitted observations.
 We describe this conception of such laboratory phenomena further in 
the following subsection for a specific and archetypical 
non-relativistic example.

%\vspace{-9pt}
%\pagebreak
\subsection{Analysis of Double-Slit Experiment}
\label{qugr74}

In section~\ref{qugr2} we discussed how not only the issue of 
consistently combining quantum theory and general relativity but also 
the long-standing internal conceptual questions regarding quantum 
theory itself might ideally be addressed by a theory of quantum 
gravity. These questions include in particular the nature of 
indeterminacy, non-locality and wavefunction reduction,
 as well as of particle quanta themselves, 
 as reviewed in subsection~\ref{qugr21}. In the previous subsection 
we have described how the approach to quantum gravity based on generalised proper time presented in 
this paper incorporates and in principle \textit{explains} these 
well-known features of quantum mechanics in the non-relativistic 
limit, as we summarise below.

   \textit{Indeterminacy} arises from a local degeneracy in field 
solutions for equation~\ref{gfromavt}, permitting interactions 
between $A(x), \bv_n(x)$ matter field components of a probabilistic 
nature within the constraint equations~\ref{gchift}--\ref{emcon}, with the initial conditions framing each indeterministic event dependent upon the outcome of prior states as ordered through a causal progression in time.
 An element of \textit{non-locality} is implicit in the necessity of 
identifying an extended 4-dimensional spacetime manifold 
itself together with a global Riemannian geometric structure 
as a solution for equation~\ref{gfromavt} as deriving from the 
original one-dimensional progression in time expressed  
through the general multi-dimensional form of equation~\ref{lpvn}.
 The demands of consistency between the internal field $A(x), 
\bv_n(x)$ exchanges and external metric $g_{\mu\nu}(x)$ constructions 
lead to the identification of particle states propagating between 
interactions as generalised wave-packet \textit{quanta} subject to 
equation~\ref{phbarku}, with the invariant mass of the states directly related to an underlying rate of oscillation in time as described for equation~\ref{mhac}.

  The apparent \textit{reduction} of the wavefunction incorporates a 
combination of the above properties of indeterminism and 
non-locality, and is associated with a reduction in the degeneracy in 
the composition of equation~\ref{gfromavt} in evolving away from a 
locally more degenerate structure for the spacetime geometry. With 
the structure of the geometry, described by $G^{\mu\nu}(x)$ as a 
function of $g_{\mu\nu}(x)$, and its evolution far too weak to detect 
gravitationally, the physical implications are not observed until a 
non-gravitational interaction of the underlying $A(x),\bv_n(x)$ field 
components transfers a detectable form of energy-momentum 
$T^{\mu\nu}(x)$, defined through equation~\ref{gfromavt} and in 
particular equation~\ref{gavt}, to the recording apparatus, updating 
our knowledge of the state of the system under investigation. 

  It is not the observation of this transition in the structure of  
$T^{\mu\nu}(x)$  in the macroscopic measurement apparatus that 
\textit{forces} the `collapse of the wavefunction', rather the 
outcome is objectively resolved, within the intrinsic element of 
uncertainty associated with the local degeneracies, by the demands of 
a consistent extended $G^{\mu\nu}(x)$ spacetime geometry. This global 
solution necessarily subsumes both the microscopic system under study 
and the experimental apparatus, which acts as a boundary condition 
and hence itself constrains the possible measurements that can be 
observed. In the absence of any possibility to observe the external 
spacetime metric geometry $g_{\mu\nu}(x)$ of the system a 
wavefunction $\Psi$ representing particle states and associated with 
internal fields alone can be pragmatically constructed to describe 
the limit of our knowledge, albeit with the `measurement problem' in 
that no explicit explanation for wavefunction reduction is presented 
in the conventional formalism of quantum mechanics.

 All of the above quintessentially quantum mechanical features are 
exhibited in the detection of a single electron state having 
traversed the double-slit apparatus as described for 
figure~\ref{dslit}. In this case the rigidly mounted apparatus 
composed of the source $S$, double-slit screen $D$ and measurement 
screen $M$, together with their physical properties including the 
potential for $A(x),\bv_n(x)$ field interactions, constitute a fixed 
element of $T^{\mu\nu}(A,\bv_n)$ in equation~\ref{gavt}. For the 
present theory this known physical structure acts as a boundary 
condition for the full solution subsuming the emission and detection 
of an electron state, as collectively depicted through a single 
contour in the gravitational field in figure~\ref{dslitg}. 
%\vspace{-8pt}
%moved figure from here

 For such a weak gravitational field $g_{\mu\nu}(x)$ the electron 
state may be described in terms of small geometric distortions 
$d_{\mu\nu}(x)$ away from a flat metric $\eta_{\mu\nu}$ in a 
linearised approximation. While equations~\ref{gconfu} and 
\ref{wavecom} describe the undulating gravitational disturbance $d_{\mu\nu}(x)$ associated with  
a plane electromagnetic wave, subsequently following equations~\ref{fpack} 
and \ref{wpack} and in section~\ref{qugr6} we discussed the notion of 
`generalised wave-packets' to be associated with particle states, 
such as those depicted in figure~\ref{sldbbv}(b), for the present 
theory. Here the form of a particle wave-packet solution is 
generalised further for the case in figure~\ref{dslitg} of a 
\textit{single} electron state spatially separated into \textit{two} 
components. This structure is consistent with both the geometry of 
the double-slit $D$ and the potential for the coherent emission and 
detection of such an electron state at $S$ and $I$ respectively. 
  For this emission and detection of a particle quantum subject to 
the relation $P^{\mu} = \hbar k^{\mu}$, as demonstrated for 
equation~\ref{phbarku}, this is a case in which the intermediate 
energy-momentum $T^{\mu\nu}(x)$ for the single state is hence 
distributed \textit{non-locally} in 3-dimensional space.

  Figure~\ref{dslitg}, similarly as for figure~\ref{sldbbv}(b), represents a 
time-integrated depiction of the gravitational element of the passage 
of the generalised wave-packet, here propagating from left to right 
with the electron state literally split between the two slits of the 
intermediate screen $D$. It may seem unreasonable that the two 
wave-packet fragments emerging from $D$ apparently `know where they 
are going' in order to recombine at $I$. That would indeed be 
difficult to comprehend for ordinary wave-packets in components of 
 $A(x),\bv_n(x)$ alone  propagating through an independent spacetime 
$M_4$ background. However here the \textit{generalised} wave-packets 
incorporate also a related distortion in the spacetime geometry 
itself, as sketched in figure~\ref{dslitg}, as part of a full 
coherent \mbox{\textit{4-dimensional}} solution for the smooth 
spacetime manifold $M_4$  itself through equation~\ref{gfromavt}, 
which hence uniformly accommodates the whole time span of the system.

%figure moved down two paras
%\pagebreak % ********** temporary ***********
\begin{figure}[htbp]  
\centering
\epsfxsize=11.0cm
\leavevmode
\epsffile[0 0 1097 807]{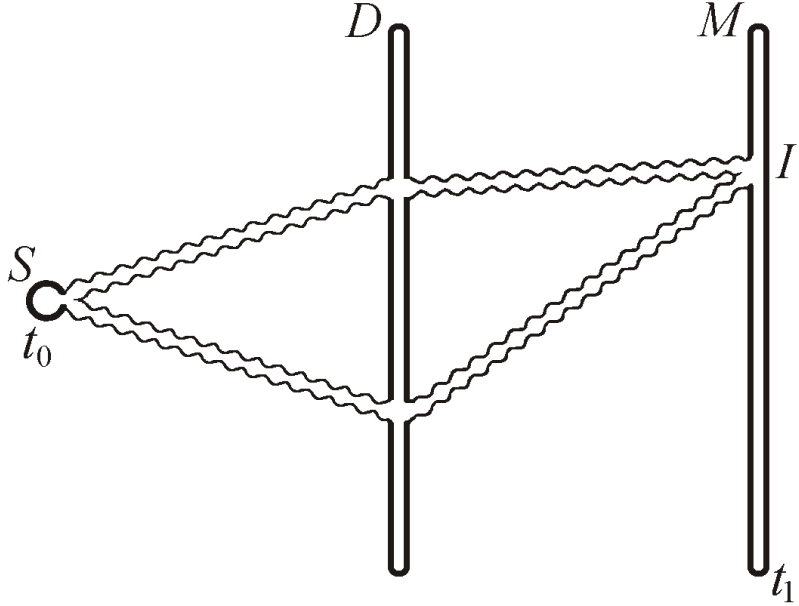}
\vspace{-7pt}
\caption{\setb Artistic depiction of the enveloping gravitational 
field $g_{\mu\nu}(x)$ associated with the transmission of a single 
electron generalised wave-packet state, illustrated here as 
integrated over time, through the double-slit apparatus described for 
figure~\protect\ref{dslit}.
The undulating propagation in the spacetime geometry $g_{\mu\nu}(x)$ 
corresponds to a solution for equation~\ref{gfromavt} involving a 
function in particular of the  subcomponents of $\bv_n(x)$ identified 
with an electron state.
(Note that compared with practical apparatus the vertical scale and 
distance between the slits is greatly exaggerated here and in 
figure~\protect\ref{dslit}).
}
\label{dslitg}
\end{figure}

  This full 4-dimensional solution has the character of events that
   from \textit{our} perspective
    have taken place in the \textit{past}, as a now immutable 
sequence of happenings whether of a classical or quantum nature.
 Hence taking a full 4-dimensional outlook seems more reasonable when 
considering for example the outcome in figure~\ref{dslitg} 
\textit{after} the experiment has been performed, with the events 
immersed within the 4-dimensional spacetime inside our past light 
cone. On the other hand events in our \textit{future}, particularly 
where indeterministic elements are involved such as in the double-slit experiment, seem to have a different, more open, nature.
 It may then seem less reasonable to adopt a full 4-dimensional 
worldview when describing potential events as they may unfold in the 
future, which are uncertain from our present perspective. 
 However it would be inconsistent for the fundamental laws of physics 
to be different for events in the past compared with those in the 
future, in particular given that all future experiments and events 
will inevitably from some point in time become subsumed `in the 
past'.
 A single universe obtained as a full 4-dimensional solution for 
equation~\ref{gfromavt}, as one of a vast array of possibilities 
given the local degeneracies involved in finding solutions, provides 
a coherent view of the world, independent of our `current' subjective 
view of events.

   While from our subjective perspective the evolving state of the 
world points to an open range of possibilities all events are 
necessarily part of a coherent 4-dimensional whole as sewn into the 
solution for the extended spacetime geometry. The apparent 
past/future divide says something about ourselves and \textit{our} 
place in the structure of the universe rather than about that 
physical structure itself, which should not depend upon our 
perspective on the past, the present and the future. As we propagate 
forwards in time there are changes in \textit{our} knowledge of the 
world and not in the full 4-dimensional solution itself. On the other 
hand with temporal progression being such a primitive element of our 
experience of the world an intimate connection \textit{is} 
anticipated between this subjective aspect of time and the objective 
basis of time as \textit{the} fundamental entity of the present 
theory -- with all matter and 4-dimensional spacetime itself a 
manifestation of time through substructures of its generalised form 
in equation~\ref{lpvn}. (A speculative attempt to expand upon this  
idea is explored in \cite{Unifi} chapter~14).

   Returning to figure~\ref{dslitg}, a local degeneracy in the 
underlying field composition of the external geometry at $S$ results 
in an indeterminacy in both the specific time $t_0$ and the 
directional aspect of the emission of an electron state.
 The final screen $M$ itself presents a uniform potential for field 
interactions corresponding to the detection of an electron. 
 The global extended 4-dimensional geometric solution $g_{\mu\nu}(x)$ 
for equation~\ref{gfromavt} naturally encompasses the entire 
apparatus, which is made more complicated by the presence of the 
double-slit screen $D$, 
 resulting in a range of possible solutions described by the 
probability distribution for actually detecting the electron on  
screen $M$ indicated in figure~\ref{dslit}.
 Our inability to observe and map out the distortions in the 
spacetime geometry, represented by the wavy  contours in 
figure~\ref{dslitg}, implies that we necessarily remain ignorant of 
the state until a macroscopically measurable interaction via the form 
of $T^{\mu\nu}(x)$ is signalled at $I$ at time $t_1$.

  This interaction can be attributed to an apparent collapse of a 
wavefunction $\Psi(\bx,t)$ as constructed to represent our best 
knowledge of the state in quantum mechanics as described for 
figure~\ref{dslit}.
Given the form of a typical solution for the present theory as 
sketched in figure~\ref{dslitg} such an event might indeed be 
interpreted as `the electron passing through both slits at the same 
time' as a non-local feature analogous to the description of the 
process in quantum mechanics.
 Here though an explicit account of the physical structure of the 
generalised wave-packet electron state as split between the two slits 
at the same time is provided.

  The classical gravitational field associated with the fixed 
apparatus $S,D,M$ in figure~\ref{dslitg} is of course in itself 
immeasurably small by a very large margin. The gravitational field of 
the single electron traversing the apparatus is in turn enormously 
smaller than even that, as represented by the thinner contour line  
(while illustrated by a single contour in figure~\ref{dslitg} in all 
cases the actual gravitational field of course extends out smoothly 
in $M_4$).
  In quantum mechanics the question of the gravitational field for a 
quantum system such as that in figure~\ref{dslit} is either neglected 
or not addressed satisfactorily, as noted in the discussion of 
equation~\ref{Einqm}.   
 Associated with this and the measurement problem in quantum 
mechanics there is also the significant question concerning the 
location and conservation of energy-momentum between $t_0$ and $t_1$ 
in figure~\ref{dslit} which may also be unanswerable as also 
discussed in subsection~\ref{qugr63} (see also \cite{Unifi}~figure~11.13(b) discussion). For the present theory although 
the continuous geometric structure $G^{\mu\nu}(x)$ for 
figure~\ref{dslitg} is too small to observe, through 
equations~\ref{gfromavt} and \ref{emcon} it does directly account for 
full energy-momentum conservation throughout all processes. In terms 
of $T^{\mu\nu}(x)$ this will imply for example a mechanical recoil of 
the source $S$ at time $t_0$ which will also be immeasurably small in 
practice.  

   The wave 4-vector $k^{\mu}$, describing the undulations of the electron state pictured 
in figure~\ref{dslitg}, directly relates to the 4-momentum $P^{\mu}$, 
as felt in the kinematic recoil at $S$ and impact at $I$, through 
equation~\ref{phbarku}.
 The `on-mass-shell' condition with $m^2 = k^2$ for particle mass $m$, alluded to after equation~\ref{ksame} 
and associated with equation~\ref{mhac}, for wave-packet particle 
states also applies in these interactions.
 The propagation of the 4-momentum $P^{\mu}$, and more generally the 
energy-momentum $T^{\mu\nu}(x)$ from which it is defined via 
equations~\ref{gfromavt} and \ref{pcon}, will be split between the 
two slits in $D$. The 4-momentum and `apparent mass' of the 
propagating state passing through each slit will then be around half 
of the total, while the undulation frequency will be preserved 
throughout, and hence the relation  $P^{\mu} = \hbar k^{\mu}$ of 
equation~\ref{phbarku} and the `on-mass-shell' condition does \textit{not} hold in the vicinity of each 
slit of $D$, where there is no interaction and passage of the state 
is not detected. However the propagation 3-velocity will still be 
consistent with equation~\ref{vgroup}
  with \mbox{$\bv_g = {\bk}/{k^0} = \bPP/P^0 = \vkin$}, since $\bPP$ and 
$P^0$ are each scaled by the same fraction for the propagation 
through a given slit in figure~\ref{dslitg} (as would also apply for 
the different paths through a diffraction grating or as opened up by 
other apparatus).

  The degree of positive interference in the overlap between the two 
wave-packet components from the two slits as they converge at 
locations such as $I$ on  detection screen $M$ should determine the 
relative likelihood to identify a coherent local electron quantum 
with 
 $P^{\mu} = \hbar k^{\mu}$ as consistent with the constraints 
underlying equation~\ref{phbarku} for such an interaction.
 This likelihood, as always proportional to the `number of ways' a 
process can happen in terms of underlying field compositions, may 
also here relate to the longitudinal extent of the wave-packets 
compared with the wavelength of the plane wave components in 
equations~\ref{gbwavesup} and \ref{gkekksup}. 
   This measure of the probability for an event to take place based 
on a relative local degeneracy count for the possible local 
interactions identified in solutions for equation~\ref{gfromavt} in 
regards to an \textit{overlap} of wave structure is analogous to the 
role of `charges' in relation to wave \textit{amplitudes} discussed 
after figure~\ref{extofey}.
 Here the full expression for the present theory should be expected 
to reproduce the  quantum mechanical 
  $\vert \Psi (\bx,t) \vert^2$ probability distribution of 
figure~\ref{dslit}, and indeed converge with the latter calculation 
in the quantum mechanical limit of equation~\ref{pdddds} as discussed 
in the previous subsection.
  
  The conceptual picture described here as exemplified in 
figure~\ref{dslitg} hence incorporates both `wave-like' and 
`particle-like' properties, with particle states represented by 
generalised wave-packets in the spacetime geometry $g_{\mu\nu}(x)$ as 
well as in $A(x)$ or $\bv_n(x)$ field subcomponents, encapsulating 
the quintessential quantum mechanical trait of `wave-particle 
duality', as associated with the Heisenberg uncertainty principle for 
example. Such properties are counter-intuitive if attributed to an 
independent \mbox{`particle'} entity, which can only represent a 
  convenient but  
 rather limited approximation to the full physical picture under 
equation~\ref{gfromavt} as illustrated  in figure~\ref{dslitg}.

  The interference pattern observed for the experiment in figures~\ref{dslit} and 
\ref{dslitg} reflects the possible range of global solutions for 
equation~\ref{gfromavt} for the boundary conditions of the given 
fixed double-slit apparatus together with the potential for  electron 
state transitions as permitted by possible $A(x),\bv_n(x)$ field 
interactions at $S$ and $M$.
 If a modified apparatus is set up to observe `which of the two slits 
the electron passes through' that would imply a \textit{different} 
known macroscopic $T^{\mu\nu}(x)$ with corresponding possible field 
interactions also in the vicinity of $D$.
 In turn that would involve
  a \textit{different} boundary condition structure for 
equation~\ref{gfromavt} and hence a \textit{different} range of 
possible 4-dimensional geometric solutions would result, involving in particular an interaction with an `on-mass-shell' electron state consistent with equation~\ref{phbarku} at one of the slits. This would 
be reflected in a very different probability distribution for 
detecting the electron on the final screen $M$, such as with a direct 
`shadow' cast by the double-slit screen $D$ corresponding to a peak 
for each slit and no interference pattern as reviewed in 
subsection~\ref{qugr21}.

 By way of a metaphor the delicate web spun by a spider in 
3-dimensional space as strung out between the irregular but stable 
surfaces of the environment  (bushes, walls etc.) provides an analogy 
for the minute gravitational field 
 associated with particle states in 4-dimensional spacetime as strung 
out between macroscopic objects of both natural and man-made origin  
(e.g. for cosmic rays and HEP events) as exemplified in 
figures~\ref{sldbbv}(b) and \ref{dslitg}. 
 In both cases there is a malleability in these distinctive 
structures that can to some degree adapt to changes in the 
environment, while large changes will require wholly different solutions. 
While there are internal rules, encapsulated in the spider's 
instincts, for the manner in which the web can be constructed it is 
never a free-standing entity that can be separated from the stable 
supporting environment, which acts as a `boundary condition' through 
which the web is located and shaped. Similarly a solution  
$g_{\mu\nu}(x)$ such as pictured in figures~\ref{sldbbv}(b) and 
\ref{dslitg},
  while constructed within the constraints of 
equations~\ref{gchift}--\ref{emcon}, is in no respect a stand-alone 
geometry but is located and shaped by the macroscopic environment to 
which it is irreducibly attached as part of a full 4-dimensional 
solution for  equation~\ref{gfromavt}.
  Under typical lighting conditions the fine web may be invisible to 
the eye, with its presence only betrayed by the clear visibility of 
the spider suspended upon it. On the other hand the tiny geometric distortion of 
$g_{\mu\nu}(x)$ in figures~\ref{sldbbv}(b) and \ref{dslitg}  is 
always beyond our perception, with its presence only indirectly 
illuminated by the detection of the associated particle interactions.

  While the known elements of $T^{\mu\nu}(x)$ for laboratory 
equipment in practice act as a boundary condition for finding 
solutions to equation~\ref{gfromavt}, conceptually it is the 
construction and composition of the 4-dimensional geometry 
$G^{\mu\nu}(x)$, as a function of $g_{\mu\nu}(x)$, that 
\textit{precedes} and defines the energy-momentum of the apparent 
matter content in spacetime, including the properties of apparatus 
and the setup of experiments.
 Even given such boundary conditions the determination of solutions 
for equation~\ref{gfromavt} may not be technically straightforward, 
as noted at the end of subsection~\ref{qugr62}.
 Indeed even for classical general relativity alone exact solutions 
are hard to find unless for example a high degree of symmetry can be 
assumed, with approximate iterative techniques 
 used for cases such as the 3-body problem, as reviewed in 
subsection~\ref{qugr22}.
 Also for non-gravitational interactions in a flat spacetime 
approximate mathematical methods, beyond that of perturbation theory 
in QFT, may be needed as for the case of `strong QCD' with the 
continuum of spacetime modelled instead by a discrete lattice of 
points (\cite{Pesk} section~22.1).

  For the present theory in finding solutions for 
equation~\ref{gfromavt} for the double-slit experiment there is both 
the non-trivial boundary condition structure of $T^{\mu\nu}(x)$ for 
the apparatus in figure~\ref{dslit} and also the underlying potential 
for $A(x),\bv_n(x)$ field interactions to take into account. However 
on the gravitational side, since the distortions away from a flat 
spacetime are so minutely small, employing the methods of linearised 
general relativity should prove to be a comfortably sufficient 
simplifying approximation for the laboratory environment as recalled 
before figure~\ref{dslitg}. 
More generally new mathematical techniques may be needed
 in order to make full calculations and determine  solutions for the 
present theory.

  For example in figure~\ref{dslitg} we have provisionally depicted 
the oscillatory propagation of the electron state by a contour in 
$g_{\mu\nu}(x)$ in a nominal direct manner connecting $S$, the two 
slits in $D$, and $I$. However the actual solution might involve a 
more diffuse spread in $g_{\mu\nu}(x)$, and the associated 
energy-momentum $T^{\mu\nu}(x)$, between the elements of the 
apparatus, as might be determined through a more complete and 
rigorous expression of this theory. 
 In being primarily determined as a structure of Riemannian geometry 
as described on the left-hand side a solution for 
equation~\ref{gfromavt} 
 need not take a shape characteristic of a particle trajectory 
 but rather 
may be counter-intuitive when interpreted as an apparent flow of 
matter through space as associated with the distribution of 
energy-momentum as defined on the right-hand side.
 While matter is everywhere constrained by the Bianchi identity 
through equation~\ref{emcon} it is also intermittently channelled 
  through more particle-like local interactions, such as at $S$ and 
$I$ in figure~\ref{dslitg}, as elements of a solution for 
equation~\ref{gfromavt} associated with particle phenomena. 
As we considered in subsection~\ref{qugr71}
 the approximations for the present theory utilised for such 
calculations  are likely to be further stretched in the environment 
of a highly curved spacetime, where in particular a more exact 
mathematical expression will be needed to describe more extreme forms 
of matter in a manner that may provide a significant theoretical test 
in itself.

  Regardless of the potential technical difficulties, conceptually 
exact solutions \textit{can} in principle exist (as for classical 
general relativity and strong QCD for example) and here in the 
context of a new coherent theory of quantum gravity that is capable 
of addressing the main conceptual issues concerning quantum theory 
itself. 
Even with such a complete theory the tools and methods of QFT in the 
flat spacetime limit and quantum mechanics in the non-relativistic 
limit would continue to be precise enough, and perhaps more readily 
applied, for most practical purposes, albeit with care needed over 
the interpretation. 
  This need to rely on pragmatic rules of application and a desire for
 a more rigorous quantitative explanation and understanding of the  
quantum mechanical conception of the world 
 as we reviewed in subsection~\ref{qugr21}, 
 not only for the non-relativistic case considered here and in the 
previous subsection but also for  particle interactions in HEP 
experiments, provides a significant element of the motivation for the work presented 
in this paper.

 From the perspective of the present theory the conceptual and 
technical problems encountered in applying quantum theory largely 
arise from taking the mathematical formalism too literally and its 
scope of application too far, whether into the realm of macroscopic 
matter or that of gravity and spacetime itself. Rather here we have 
described a coherent unified framework in which the construction of a 
continuous spacetime and smooth gravitational field encompasses 
macroscopic and microscopic systems on all scales in a seamless 
manner. As described for the phenomena in figures~\ref{dslit} and 
\ref{dslitg} as well as figures~\ref{sldbb} and \ref{sldbbv} the 
gravitational field associated with the apparatus extends into and 
immerses the realm of quantum systems, for which quantum theory 
itself survives only as a limiting approximation. 
  Some of the possible practical implications of prescribing limits for the applicability of non-relativistic quantum mechanics will be considered in the following section.

 As discussed for figure~\ref{sldbbv} in subsection~\ref{qugr72} and 
figure~\ref{dslitg} above,
 while the gravitational field in these laboratory experiments is far 
too small to detect, the theory could still be tested by accounting 
for the measurements we \textit{can} make.
  These ambitions for the present theory then include not only
 reproducing 
 the spectrum of Standard Model particle states and their masses but 
also in principle making predictions for new observable phenomena 
\cite{Gener,Ufield}. 
  In this manner this theory based on the simple idea of generalised 
proper time provides a unifying framework for establishing a 
consistent theory of quantum gravity that is also accessible to 
empirical investigation.

\vspace{-7pt}
%\pagebreak
\section{Discussion and Conclusions}
\label{qugr8}

  In this paper we have described a unified theoretical framework for 
quantum gravity in which the essential features of both quantum 
theory and general relativity are retained and combined in a mutually 
compatible manner. This is achieved through the composition of the 
geometric structure of 4-dimensional spacetime as constructed through 
a basis in generalised proper time. As a theory of quantum gravity 
here the approach has been to \textit{begin} with an intuitively  
appealing conceptual picture, that might in principle for example 
address conceptual questions in quantum theory itself, before 
developing a rigorous quantitative mathematical expression for the 
theory. This contrasts with the direct technical mathematical 
approaches, of string theory and loop quantum gravity for example, 
that adapt the tools of quantum theory in attempting to quantise  
gravity or the structure of spacetime, for which technical issues 
still remain unresolved (see for example~\cite{Kief,Carl1}) and a 
number of conceptual problems, including the question of why 
\textit{anything} should be quantised, are left largely untouched.

 Here in stepping back from both quantum theory and general 
relativity to consider a broader framework, based upon generalised 
proper time as the progenitor of spacetime and all the matter it 
contains,
    the focus has been not so much on quantising gravity but rather 
we have been led to emphasise the role and priority of gravity in 
accounting for quantum phenomenology itself. The resulting conceptual 
picture incorporates the appropriate characteristic features of both 
quantum theory and general relativity as we summarise here:

\begin{itemize}

\item[A.]{With analysis of time itself central to the theory, through 
the generalised form of equation~\ref{lpvn}, the underlying 
one-dimensional flow of time parametrises the evolution of all 
entities, including the apparent quantum state according to the 
Schr\"{odinger} equation~\ref{Schro} in the quantum mechanical 
limit.}

\item[B.]{The necessity of perceiving the world in space as well as 
through time is represented by a single, smooth and continuous, 
global 4-dimensional spacetime solution for equation~\ref{gfromavt},
 identified from substructures of generalised proper time,
 incorporating the Einstein equation~\ref{Eineq} of general 
relativity on all scales.}

\item[C.]{The intrinsic uncertainty of quantum events derives from 
degeneracies in the local $A(x),\bv_n(x)$ field subcomponent 
composition of the spacetime metric geometry $g_{\mu\nu}(x)$ as a 
solution for equation~\ref{gfromavt}, which is akin to a classical 
probabilistic construction here in terms of the `number of ways' for 
building spacetime itself.}

\item[D.]{In addition to indeterminacy the principal conceptual issues 
of quantum theory regarding non-locality and wavefunction reduction, 
underlying the `measurement problem', are addressed here through the 
properties of the enveloping gravitational field and spacetime 
structure as described for example for figure~\ref{dslitg}.}

\item[E.]{Interacting particle quanta conforming to equation~\ref{phbarku} are  
identified as generalised wave-packets in  $A(x),\bv_n(x)$ subcomponent fields as well as in  
$g_{\mu\nu}(x)$  propagating as  minimal distortions in the gravitational field
 in solutions for equation~\ref{gfromavt} permitted within the 
constraints implicit in the generalised proper time basis.}

\end{itemize}

 While this framework provides a coherent conceptual framework
  for convolving general relativity with quantum theory
   it remains to be demonstrated, building upon sections~\ref{qugr5} 
and \ref{qugr6}, how the formalism of QFT and quantum mechanics in 
the appropriate limits can be fully recovered and new quantitative 
results derived.
 The explicit and rigorous mathematical expression for this theory 
may require novel techniques or methods adopted or adapted from other 
theories, such as we have described for the employment of linearised 
gravity in approaching the flat spacetime limit for HEP phenomena. 
Further such development will be needed to clarify, explain and 
possibly modify
 various elements of the theory in building upon the preliminary 
progress that has been made.  The questions to be addressed include 
the following considerations:

\begin{enumerate}

\item{Deriving from the symmetry breaking of equation~\ref{lpvn}, 
ultimately for the proposed form of generalised proper time of 
equation~\ref{lvto}, the limited number of possible elementary 
particle types is predicted to fully account for the Standard Model, building upon the progress that has been 
made~\cite{TimeE} as summarised in section~\ref{qugr3}, and in a 
manner consistent with the demands of the QFT limit discussed in the first half of  
subsection~\ref{qugr72} before figure~\ref{sldbbv}.
}

\item{Particle interactions from possible $A(x),\bv_n(x)$ field 
subcomponent exchanges associated with local degeneracies in
 minimal near-vacuum 
  geometric solutions for equation~\ref{gfromavt} as determined and 
restricted by the constraint  equations~\ref{gchift}--\ref{emcon},
 as provisionally exemplified in equations~\ref{ext0}--\ref{ksame} 
and figure~\ref{extofey}(a),
 need to be more rigorously 
 connected with the role of Lagrangian terms such as   
equations~\ref{lagsff}--\ref{lagdir} in a standard approach to QFT.}

\item{A pivotal question then concerns the full nature of the 
connection between the calculation of an event probability for the 
present theory based on internal field degeneracies underlying the 
local external geometry and that in QFT through a transition 
amplitude, as developed for equation~\ref{pdddds}
 in subsection~\ref{qugr52},
 and as also proposed to apply for the non-relativistic quantum 
mechanical limit with $P\propto \vert \Psi\vert^2$, as discussed in 
subsections~\ref{qugr73} and \ref{qugr74}.}

\item{The full nature of particle quanta and their interactions with 
each other and with detector apparatus is to be further explored, 
building upon the derivation of equation~\ref{phbarku} and the 
discussion of figures~\ref{sldbbv} and~\ref{dslitg}, including
 the interface and merging between microscopic and macroscopic 
systems and an understanding of the limits of validity of the methods 
of standard quantum theory such as the application of 
equation~\ref{Schro} for complex systems.}

\item{A full technical mathematical expression for this theory of 
quantum gravity for the general case, without the approximations of 
an almost flat background employed for the QFT and quantum mechanical 
limits, may be needed for a complete theoretical understanding of the 
more extreme regions of spacetime curvature, such as in the 
environment of the `singularities' of black holes or in the very 
early universe, as considered in subsection~\ref{qugr71}.}

\end{enumerate}

 This theory presents a rather fluid and organic picture of the world 
at the most elementary level, based upon the `slippery ambiguity' in 
the matter field components underlying the geometric construction of 
spacetime, as exemplified by the exchanges described for 
figure~\ref{extofey}(a) and associated with interaction processes 
such as those underlying the decay vertices in figure~\ref{sldbbv} or 
the fusing with experimental apparatus at $S$ and $I$ in 
figure~\ref{dslitg}.  
 Not only are the discrete properties and equations of motion of elementary particle quanta 
to be explained through the constraint 
equations~\ref{gchift}--\ref{emcon}, but also the emergence of the 
rigid, stable macroscopic structures of the everyday world with their 
classical physics properties, 
 from for example laboratory apparatus itself to
 the literally organic forms of biological life,  will need to be 
accounted for.

  A potential practical application of the theory, relating to `item 4' above and alluded to towards the end of 
both subsections~\ref{qugr72} and \ref{qugr74}, concerns going beyond elementary particle 
states, such as electrons, via hadronic states and through atoms and 
molecules to describe the limits of quantum superposition and the 
potential for constructing ever more macroscopic quantum states 
(which might be anticipated to fall very far short of the level of 
Schr\"{o}dinger's cat).
Exploration of the interface between the quantum and classical world 
more generally will shed light on the application of this theory.
 This could involve experiments for example in quantum interference 
which, beyond elementary particles and single atoms, has been 
observed for
  molecules as heavy as  
  fluorofullerene C$_{60}$F$_{48}$  (\cite{Hack}, \cite{Carl4} 
section~3).
  Experiments are also being developed involving ever larger 
structures including biological matter and even living 
organisms such as bacterial cells~\cite{Marl}.

The present theory could also have implications for the prospects of 
quantum computing as based upon the deployment of an entangled array 
of `qubits', each of which can be placed in a superposition of $\vert 
0 \rangle$ and 
 $\vert 1 \rangle$ states (see for example~\cite{VedP}). A quantum 
computation involves a unitary transformation in evolving from an 
initial input state to a final output state as registered by a single 
measurement to extract the result; with a capacity in principle to 
solve certain problems much faster than would be possible with a 
classical computer. However for quantum computation a `qubit', which 
may be the state of a single atom, is physically much smaller than a 
classical `bit', with superposed and entangled states being very 
susceptible to noise and decoherence. A combination of supercooling 
to $\sim 0.1\,$K, isolation from the environment and quantum error 
correction code (requiring many more qubits) is needed in aiming to 
build a fault-tolerant quantum computer. In developing this 
technology `we will discover a great deal of new physics involving 
entanglement, decoherence and the preservation of quantum 
superpositions' (\cite{VedP} section VII ending).

  Hence with the limits of quantum mechanics  still being 
investigated there is the potential for testing a new theory for 
which quantum mechanics itself is derived as a limiting case.  
 That is, experiments designed to augment the scale of quantum 
superposition or  develop the potential of quantum computing for 
example may themselves yield a new understanding in quantum mechanics 
that might provide a test of the present theory. 
Further laboratory experiments are also being considered with the 
potential to investigate aspects of quantum gravity specifically, as 
we noted in subsections~\ref{qugr24} and \ref{qugr71} with reference 
to the examples of~\cite{Haine,Faure,Howl,Kamp}.
  We have also noted that large scale cosmology could in principle 
provide an arena for observing implications of quantum gravity 
schemes as exemplified by~\cite{Ansel,Berg,Calc,Jacob}.

 To date the most significant success of the present theory has been 
seen in establishing explicit connections with the esoteric 
structures of the Standard Model, as briefly reviewed in 
section~\ref{qugr3} (\cite{Gener} section~3), which are obtained in a 
far more direct and unique manner than for models with extra spatial 
dimensions for example.
 While developments for the present theory beyond the Standard Model 
in the breaking of equation~\ref{lpvn} might prove predictive and 
testable in itself~(\cite{Gener} section~4, \cite{Ufield} section~6), the full 
potential to test the theory will require the full mathematical 
expression in the extended spacetime $M_4$ in order in principle to 
determine parameters of the Standard Model  such as
  the spectrum of particle masses and additional quantitative 
results.
 
 The further success for the theory in already  making  connections 
with QFT -- through the correlation with Lagrangian terms described 
for equations~\ref{lagsff}--\ref{emcon}, the relation to event rate 
calculations through equation~\ref{pdddds} and the elucidation of 
particle quanta culminating in equation~\ref{phbarku} -- supports the 
plausibility of the developments in sections~\ref{qugr5} and 
\ref{qugr6} towards achieving stand-alone calculations for HEP 
processes with this theory. However these various mathematical 
threads will require further development as well as tying together 
more closely in order to perform such calculations.   
  The identification of structures of the Standard Model, without 
needing to contrive these features, derives from the very local and 
necessary breaking of the full symmetry $\hG$ of equation~\ref{lpvn} 
over the \textit{local} external spacetime $M_4$, as described for 
equation~\ref{gbreak}.
 This success of `item 1' above hence 
  remains even if technical or new conceptual difficulties are 
encountered in addressing `items 2--5' of the above listed points 
concerning the investigation of physical structures in the 
\textit{extended} spacetime. In that case a modified or alternative 
means of proceeding within the framework of generalised proper time 
might  be considered while still retaining the successful elements.

  A main focus of this paper has been the success of the extent to 
which the seemingly contrary demands of quantum theory and general 
relativity, as reviewed in section~\ref{qugr2}, can be convolved both at a conceptual level and at an elementary 
mathematical level through
the development of the theory presented.
 This has been demonstrated through the explanatory power gained for
    well-established quantum mechanical experiments, such as 
described for figures~\ref{dslit} and \ref{dslitg}, and in particular 
for  HEP experiments, such as that discussed for figures~\ref{sldbb} 
and~\ref{sldbbv}, 
 providing in turn
  an arena where the advances of this theory might be tested.

  While non-trivial progress has been made for this theory
  in connecting with the empirical world
  clearly further developments are need on the mathematical side, 
building in particular upon sections~\ref{qugr5} and \ref{qugr6}. 
However, despite significant breakthroughs in recent decades, further
 technical progress will also be required of other approaches such as 
string theory and loop quantum gravity if they are to make any 
notable contact with the empirical world or improve in explanatory 
power. While tackling the ambition of quantising gravity or spacetime 
head-on as a technical mathematical problem such well-established  
approaches are still incomplete,
 as for the development of the present theory.
 However, unlike the scope of the present theory, the other 
approaches are  
  also neither able to account for the Standard Model nor the 
conceptual issues with quantum theory itself. Indeed it is not clear 
that any conceptual question, such as the `measurement problem', ever 
\textit{could} be resolved directly through technical mathematical 
developments without first stepping back to establish a firm 
conceptual footing capable in principle of addressing such 
foundational matters.

  Given the well-known technical difficulties with the more direct 
mathematical approaches to quantum gravity, here beginning rather 
with such a conceptual foundation provides a fresh perspective and a 
new opportunity to address these questions. The present theory is a 
more indirect approach to quantum gravity
 in that we \textit{begin} with a basis simply in the one dimension 
of time, an infinitesimal interval of which can be expressed 
identically in the general form of equation~\ref{lpvn}, 
\textit{before} constructing a \mbox{4-dimensional} spacetime 
structure capable of convolving the basic properties of both quantum 
theory and general relativity.
  While naturally there are still open questions as the theory 
develops  a number of pertinent issues in fundamental physics can 
already be resolved to a non-trivial degree. 
 The strengths of this framework include the simplicity of the basis, 
the establishment of empirical connections with HEP phenomena and the 
ability to address a series of conceptual issues in quantum theory by 
incorporating a significant and irreducible role for gravity in the 
laboratory. In all cases this progress has been achieved in a manner 
starkly contrasting the situation for most other approaches to 
quantum gravity and supports the proposal of 
 generalised proper time as the basis for a  comprehensive unified 
theory.

%\pagebreak

% fit references shorter space
%{\setlength{\baselineskip}{0.58\baselineskip}

{\small
{\setlength{\baselineskip}{0.65\baselineskip}

\par}
}

%Bibliography

%\pagebreak

%%
\par}% \linespread{1.0} for main text (28/11/15)
%match '{\setlength{\baselineskip}{0.625\baselineskip}' above

\end{document}